%% file: Manuscript.tex
\documentclass{article}

\usepackage{arxiv}
\usepackage{placeins}
\usepackage[utf8]{inputenc} 
\usepackage[T1]{fontenc}    
\usepackage{xr-hyper}
\usepackage{nameref,hyperref}
\usepackage{url}            
\usepackage{booktabs}       
\usepackage{amsfonts}       
\usepackage{nicefrac}       
\usepackage[protrusion = true, final]{microtype}      
\usepackage{lipsum}		
\usepackage{graphicx}
\usepackage[numbers]{natbib}
\usepackage{doi}
\usepackage{hyphenat}
\usepackage{amsmath, amssymb}
\usepackage{authblk}
\usepackage[symbol]{footmisc}
\usepackage{subfiles}

\usepackage{float}

\makeatletter

\usepackage[aboveskip=1pt,labelfont=bf,labelsep=period,justification=raggedright,singlelinecheck=off]{caption}

\usepackage[table]{xcolor}

\usepackage{array}
\newcolumntype{L}{>{\centering\arraybackslash}m{0.13\linewidth}}

\newcolumntype{+}{!{\vrule width 2pt}}

\newlength\savedwidth

\graphicspath{{Figures/}}

\newcommand*{\addFileDependency}[1]{
\typeout{(#1)}
\@addtofilelist{#1}
\IfFileExists{#1}{}{\typeout{No file #1.}}
}\makeatother

\title{MicroBundleCompute: Automated segmentation, tracking, and analysis of subdomain deformation in cardiac microbundles} 

\date{} 					

\author[1,2]{Hiba~Kobeissi}
\author[3]{Javiera~Jilberto}
\author[1,4,5]{M.~\c Ca\u gatay~Karakan}
\author[5,6,7]{Xining~Gao}
\author[3]{Samuel~J.~DePalma}
\author[5,6,7]{Shoshana~L.~Das}
\author[3]{Lani~Quach}
\author[8]{Jonathan~Urquia}
\author[3]{Brendon~M.~Baker}
\author[5,7]{Christopher~S.~Chen}
\author[3,9,10]{David~Nordsletten}
\author[1,2, *]{Emma~Lejeune}

\affil[1]{Department of Mechanical Engineering, Boston University, Boston, MA 02215, USA}
\affil[2]{Center for Multiscale and Translational Mechanobiology, Boston University, Boston, MA 02215, USA}
\affil[3]{Department of Biomedical Engineering, University of Michigan, Ann Arbor, MI 48109, USA}
\affil[4]{Photonics Center, Boston University, Boston, MA 02215, USA}
\affil[5]{Department of Biomedical Engineering, Boston University, Boston, MA 02215, USA}
\affil[6]{Harvard-MIT Program in Health Sciences and Technology, Institute for Medical Engineering and Science, Massachusetts Institute of Technology, Cambridge, MA 02139, USA}
\affil[7]{Wyss Institute for Biologically Inspired Engineering, Harvard University, Boston, MA 02215, USA}
\affil[8]{Department of Electrical and Computer Engineering, New York Institute of Technology, New York, NY 10023, USA}
\affil[9]{Department of Cardiac Surgery, University of Michigan, MI 48109, USA}
\affil[10]{Department of Biomedical Engineering, School of Imaging Sciences and Biomedical Engineering, King’s College London, King’s Health Partners, London SE1 7EH, United Kingdom}

\renewcommand{\shorttitle}{MicroBundleCompute}

\hypersetup{
pdftitle={MicroBundleCompute},
pdfsubject={q-bio.NC, q-bio.QM},
pdfauthor={Hiba~Kobeissi,Emma~Lejeune},
pdfkeywords={First keyword, Second keyword, More},
}

\begin{document}
\maketitle

\begin{abstract}
Advancing human induced pluripotent stem cell derived cardiomyocyte (hiPSC-CM) technology will lead to significant progress ranging from disease modeling, to drug discovery, to regenerative tissue engineering. Yet, alongside these potential opportunities comes a critical challenge: attaining mature hiPSC-CM tissues. At present, there are multiple techniques to promote maturity of hiPSC-CMs including physical platforms and cell culture protocols. However, when it comes to making quantitative comparisons of functional behavior, there are limited options for reliably and reproducibly computing functional metrics that are suitable for direct cross-system comparison. In addition, the current standard functional metrics obtained from time-lapse images of cardiac microbundle contraction reported in the field (i.e., post forces, average tissue stress) do not take full advantage of the available information present in these data (i.e., full-field tissue displacements and strains). Thus, we present ``MicroBundleCompute,'' a computational framework for automatic quantification of morphology-based mechanical metrics from movies of cardiac microbundles. Briefly, this computational framework offers tools for automatic tissue segmentation, tracking, and analysis of brightfield and phase contrast movies of beating cardiac microbundles. It is straightforward to implement, runs without user intervention, requires minimal input parameter setting selection, and is computationally inexpensive. In this paper, we describe the methods underlying this computational framework, show the results of our extensive validation studies, and demonstrate the utility of exploring heterogeneous tissue deformations and strains as functional metrics. With this manuscript, we disseminate ``MicroBundleCompute'' as an open-source computational tool with the aim of making automated quantitative analysis of beating cardiac microbundles more accessible to the community.
\end{abstract}

\keywords{cardiac tissue engineering \and mechanobiology \and open science \and microbundle \and hiPSC-CMs \and bright field imaging \and phase contrast microscopy}
\footnotetext{* Corresponding author: elejeune@bu.edu}

\newpage
\footnotetext{Email addresses: hibakob@bu.edu (Hiba Kobeissi), jilberto@umich.edu (Javiera Jilberto), karakan@bu.edu (M. Çağatay Karakan), xgao2@bu.edu (Xining Gao), samdep@umich.edu (Samuel J. DePalma), sldas@mit.edu (Shoshana L. Das), laniq@umich.edu (Lani Quach), jjurquia99@gmail.com (Jonathan Urquia), bambren@umich.edu (Brendon M. Baker), chencs@bu.edu (Christopher S. Chen), nordslet@umich.edu (David Nordsletten), elejeune@bu.edu (Emma Lejeune)}

\section*{Introduction}
\label{sec:intro}
\begin{figure}[!h]
\begin{center}
\includegraphics[width=0.95\textwidth]{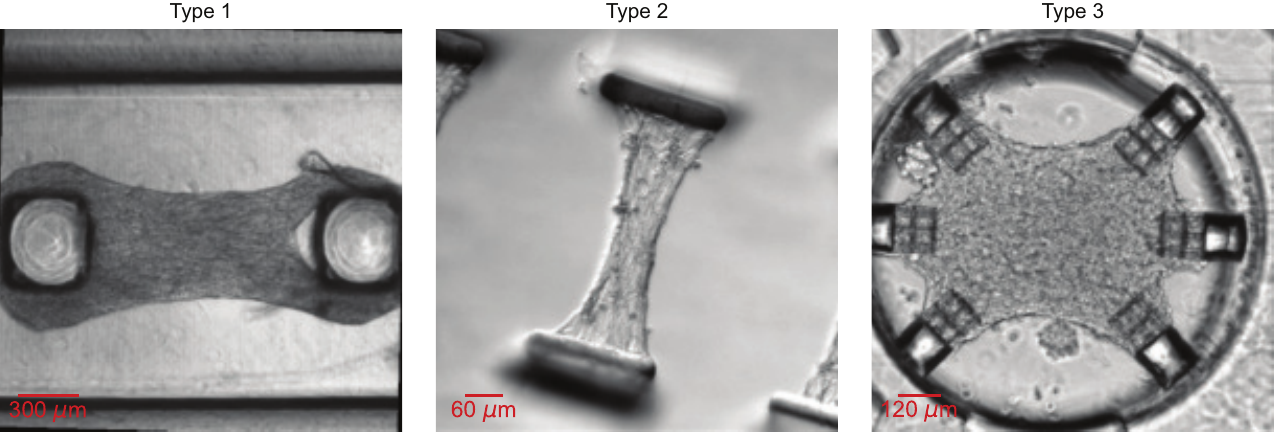}
\caption{{\bf Examples of $\mathbf{3}$ microbundle experimental testbeds of different types.}
We briefly note here that ``Type 1'' and ``Type 3'' represent 3D culture platforms while ``Type 2'' portrays a significantly thinner (almost 2D) platform. We elaborate more on each type in the ``\nameref{subsubsec:exp_data}'' Section and in \nameref{si:additional_ex}.}
\label{fig:diff_testbeds}
\end{center}
\end{figure}

Despite significant recent advances in cardiovascular disease prevention and diagnosis \cite{brush2022diagnostic, vardas2022year}, heart disease remains the leading cause of death among adults worldwide \cite{kaptoge2019world}. This is due, in part, to the fact that the native heart has a poor regenerative ability \cite{nguyen2021cardiac, bursill2022don}, and thus damage to the heart muscle during an adverse medical event such as a myocardial infarction is irreversible \cite{thygesen2007universal}. Cardiac tissue engineering is a promising approach to address this unmet societal need \cite{bursill2022don}. In particular, cardiac tissue engineering with human induced pluripotent stem cell derived cardiomyocyte (hiPSC-CM) based technology \cite{takahashi2007induction} is a promising approach to disease modeling \cite{brandao2017human, nakao2020applications, daly2021bioprinting, dou2022microengineered}, drug discovery \cite{bursill2022don, nakao2020applications, dou2022microengineered, hnatiuk2021human, ronaldson2018organs}, and regenerative tissue engineering \cite{hirt2014cardiac, masumoto2014human, karbassi2020cardiomyocyte}. However, the development of viable hiPSC-CM technology is very much ongoing. In particular, one major challenge is that differentiated hiPSCs initially resemble fetal cardiomyocytes - they are morphologically and functionally different compared to adult cardiomyocytes \cite{hnatiuk2021human}. Thus, developing technology to promote the maturation of hiPSC-CMs and, likewise, hiPSC-CM based tissue is an active area of research \cite{karbassi2020cardiomyocyte}. One impactful approach to promoting the maturation of hiPSC-CMs is the use of engineered tissue culture platform designs in both two \cite{depalma2021microenvironmental, batalov2021engineering} and three dimensions \cite{jayne2021direct, karakan2023direct, ronaldson2019engineering, javor2021pillar, zhao2019platform} (Fig \ref{fig:diff_testbeds}). Across these different platforms, there are multiple physical \cite{depalma2021microenvironmental, jayne2021direct, javor2021pillar, ruan2016mechanical}, electrical \cite{ jayne2021direct, karakan2023direct, zhao2019platform, ruan2016mechanical, feric2019engineered}, and chemical \cite{huang2020enhancement, lee2019nanoparticle, huebsch2022metabolically} knobs to tune to promote maturation and explore different physiological and pathological conditions. Even if we restrict our focus to microbundles (i.e., \textit{aligned, electromechanically coupled, microscale cardiac tissue bundles formed with hydrogel materials suspended between pillars}), there is massive variability across different experimental setups \cite{depalma2021microenvironmental, jayne2021direct, karakan2023direct, javor2021pillar}. 

Driven by this diversity in experimental approaches and the rapid growth of the field, the mechanical behavior of cardiac microbundles is challenging to compare across studies. Fundamentally, this challenge is driven by multiple factors, ranging from the high volume of data collected with these testbeds \cite{pointon2017cover}, to challenges associated with reproducing results when software and data are not shared under open-source licenses, or when extracting quantities of interest from data requires significant manual processing. To date, there have been multiple non-destructive image-based methods for quantifying the contractile action of cardiac microbundles \cite{ronaldson2019engineering, huebsch2015automated, ronaldson2018advanced, hansen2010development,tamargo2021millipillar, rivera2022contractility, oyunbaatar2016biomechanical, dostanic2020miniaturized, thavandiran2020functional, tsan2021physiologic, mery2023light}, often inspired by related approaches to assessing the contractile behavior of cardiomyocytes \cite{huebsch2015automated, steadman1988video,lim2008novel,hossain2010non, hayakawa2012noninvasive, ahola2014video, hayawaka2014image, rajasingh2015generation, czirok2017optical, shradhanjali2019spatiotemporal, scalzo2021dense, cheng2023quantification}. Broadly speaking, most of these tools can be grouped into four main categories: (1) edge detection systems \cite{ronaldson2018advanced, hansen2010development}, (2) pillar tracking-based methods \cite{tamargo2021millipillar, rivera2022contractility, oyunbaatar2016biomechanical, dostanic2020miniaturized, thavandiran2020functional, mery2023light}, (3) inter-frame pixel disparity methods \cite{ronaldson2019engineering, tsan2021physiologic}, and (4) optical flow-based tracking \cite{huebsch2015automated, mery2023light}. Each one of these methods has benefits and limitations that suit specific platforms, conditions, and research questions (some of these approaches are elaborated on in the ``\nameref{sec:materials&methods}'' Section). Despite this wide range of computational tools, many of which are available under open-source licenses \cite{ronaldson2019engineering, tamargo2021millipillar, rivera2022contractility, tsan2021physiologic, mery2023light, sala2018musclemotion}, few options can compare across multiple experimental testbeds and function on new datasets out of the box. In addition, the most popular approach to assessing microbundle contractile behavior – pillar tracking – does not necessarily capture the full richness of mechanical behavior in these systems. 

Given this research landscape, there is a clear need for an open-source computational tool to extract functional metrics from time-lapse images of microbundle contraction. To address this need, we have developed the ``MicroBundleCompute'' computational framework that we will present in this paper. In brief, ``MicroBundleCompute'' is disseminated as a Python package and is based on the Lucas-Kanade optical flow algorithm \cite{bouguet2001pyramidal} for computing full-field displacements, subdomain-averaged strains, and displacement and strain derived metrics. In this manuscript, we introduce the methods underlying the ``MicroBundleCompute'' framework, discuss our approach to validating the pipeline primarily via realistic synthetic data, and then show the results of implementing ``MicroBundleCompute'' on multiple testbeds, shown in Fig \ref{fig:diff_testbeds}. We not only portray the efficacy of our approach, but also show that examining full-field tissue deformation consistently reveals heterogeneous contractile behavior throughout the domain. Looking forward, our goal is to use this work as a starting point to move beyond previously developed ad hoc approaches to analyzing these data \cite{das2022mechanical}, and establish a computational foundation for performing cardiac microbundle assessment and quality control at scale. 

The remainder of this document is organized as follows. In the ``\nameref{sec:materials&methods}'' Section, we describe our methods to generate synthetic movies of beating cardiac microbundles with known ground truth to validate our computational pipeline, the diverse set of real examples on which we test the software, and finally, our code pipeline along with the main output metrics. Then in the ``\nameref{sec:results}'' Section, we present the main findings of our code validation on one synthetic example and show sample outputs on a few real movies. Finally, in the ``\nameref{sec:concl}'' Section, we share our final thoughts regarding ``MicroBundleCompute.'' Along with this document, we provide three supplementary documents where we explain in more details our validation pipeline (\nameref{si:pipeline_val}), share our complete set of real examples (\nameref{si:additional_ex}), and finally provide more information on basic pillar tracking (\nameref{si:pillar_track}), an additional feature of our computational tool. Overall, the intention of this manuscript is to outline our method and approach to making a tool of broad utility for the cardiac tissue engineering research community.

\section*{Materials and methods}
\label{sec:materials&methods}
In a broader sense, capturing full-field soft tissue deformation is critical for a number of research applications ranging from inverse material characterization \cite{raghupathy2011form, genovese2014digital, narayanan2021inverse, wang2021inference}, to high-fidelity biomechanical modeling of in vivo mechanisms \cite{weiss2020mechanics, Sree2019linking, kong2018finite}, and patient-specific modeling and procedure planning \cite{Joldes2019Suite,de2010imaging}. And recently, there has been keen interest in implementing image registration-based techniques widely used in the field of computer vision \cite{raghupathy2011form, wang2015digital} for full-field measurements, including digital image correlation (DIC) \cite{chu1985applications, zhang2004applications, annaidh2012characterization, blaber2015ncorr,yang2021fast}, 3D-DIC \cite{solav2018MultiDIC, murienne2016comparison}, digital volume correlation (DVC) \cite{bay1999digital, dall2014inevitable, acosta2018three, bersi2020multimodality}, and optical flow algorithms \cite{lucas1981iterative, boyle2014simple, boyle2019regularization}.

With regards to applications to cardiac microbundles in specific, available image-based techniques for computing the contractile behavior of microbundles, as categorized in the ``\nameref{sec:intro}'' into $4$ broad approaches, focus mainly on quantifying temporal profiles for the entire construct and as such, provide averaged results that lump the spatially heterogeneous tissue behavior into a single value per frame ($3$ of the $4$ approaches). 
However, tools developed based on optical flow ($4^{th}$ approach) can extract full-field, as well as directional outputs, such as directional displacement fields and strains.

To elaborate more, edge-detection approaches rely on quantifying the microbundle shape change between a relaxed (reference) state and a contracted (deformed) state. For example, in Ronaldson-Bouchard et al. \cite{ronaldson2018advanced}, tissue contractility was calculated by tracking the change in the tissue area while previously, in Hansen et al. \cite{hansen2010development}, the difference between the ends of the tissue was used to measure contraction. Yet, these tools were custom-developed and are not readily available online for the broader research community. As for pillar tracking-based methods, there are currently a number of available tools \cite{tamargo2021millipillar, rivera2022contractility, mery2023light} out of the identified implementations that were developed and kept in-house \cite{oyunbaatar2016biomechanical, dostanic2020miniaturized, thavandiran2020functional}. In these approaches, the pillar or cantilever head deflection is estimated to generate contraction waveforms and extract contraction kinetics including contraction frequency, force, as well as the time to achieve $10\%$, $50\%$, or $90\%$ of the peak contraction or relaxation. 

Available tissue tracking methods via pixel intensity disparity have been implemented based on different approaches. For example, in ``MUSCLEMOTION'' \cite{sala2018musclemotion}, which is offered as an ImageJ \cite{bourne2010imagej} plugin, the absolute difference in pixel intensity between a reference frame and a frame of interest is calculated, whereas the MATLAB-based \cite{MATLAB} ``CardiacContractileMotion'' \cite{ronaldson2019engineering} identifies the tissue region in a relaxed baseline state and tracks, within this region of interest, changes in pixel motion with time. Another MATLAB-based \cite{MATLAB} tool, ``ContractQuant'' \cite{tsan2021physiologic}, which was specifically developed for implementation with micron-scale 2D cardiac muscle bundles, uses cross-correlation to track pixel features and find the best match for a specified region of interest across consecutive frames. Outputs from these software are in general similar to those extracted with pillar tracking approaches and include contraction and relaxation profiles and velocities, as well as contraction and relaxation times. 

Overall, these approaches are suitable when only averaged values, such as the mean value and the time rate or velocity of tissue shortening, shrink or contraction, are enough. However, tools developed based on optical flow can provide richer outputs. For example, Huebsch et al. \cite{huebsch2015automated} implemented block-matching optical flow \cite{ghanbari1990cross} methods in MATLAB \cite{MATLAB} to estimate absolute as well as directional full-field contractility. And very recently, a particle image velocimetry toolbox in MATLAB \cite{MATLAB} was utilized in \cite{mery2023light} to calculate displacement vectors of sub-regions or patches identified within a manually selected region of interest in the ``reference'' frame using a cross-correlation approach. And from these displacement fields, strain maps were subsequently derived. To the best of our knowledge, these methods appear to be robust and relatively versatile, yet at present they lack automation.

Within this scope, we present here our high-throughput optical flow-based computational framework to extract full-field deformation metrics from lab-grown cardiac microbundles.
In the Sections that follow, we describe both the data used for testing and validating the ``MicroBundleCompute'' software, and the details of our computational methods. First, in the ``\nameref{subsec:data}'' Section, we introduce the two general categories of cardiac microbundle data that we have used in developing the software: synthetic data with a known ground truth and experimental data. Next, in the ``\nameref{subsec:code}'' Section, we explain the details of the code pipeline and describe core functionalities and output metrics.

\subsection*{Data} 
\label{subsec:data}
To validate our pipeline, we first invest significant effort into creating labeled data with a known ground truth. In the ``\nameref{subsubsec:syn_data_gen}'' Section, we outline our process for synthetic data generation. As a brief note, this is complemented with additional information in \nameref{si:pipeline_val} where we show comparisons to another form of synthetic data and manually labeled experimental data. In the ``\nameref{subsubsec:exp_data}'' Section, we then show three distinct classes of microbundle experimental testbeds that we will use to showcase the function and versatility of our software.

\subsubsection*{Synthetic data generation} 
\label{subsubsec:syn_data_gen}
\begin{figure}[!h]
\begin{center}
\includegraphics[width=0.95\textwidth]{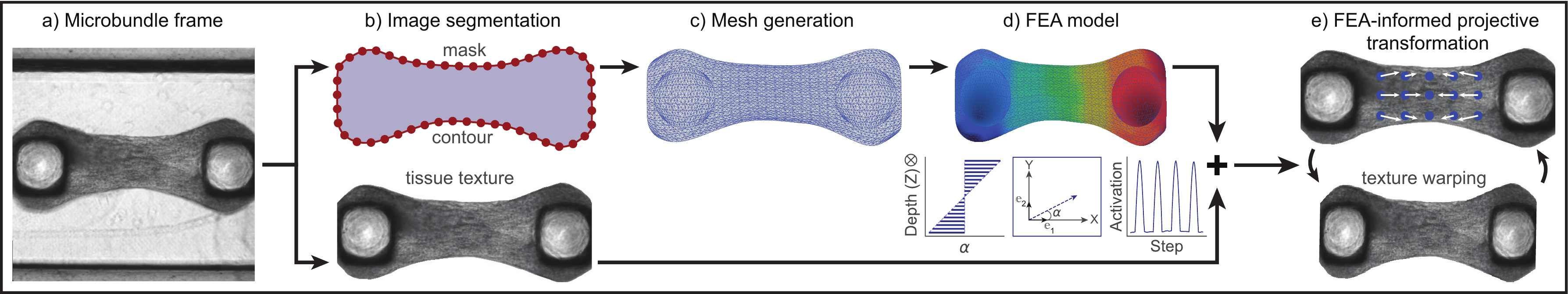}
\caption{{\bf Schematic representation of the synthetic data generation pipeline.}
Illustrations of the main elements of the synthetic data generation pipeline in order of implementation: (a) an image of a microbundle movie frame; (b) a mask of the microbundle, extracted contour coordinates, and a segmented tissue texture; (c) a three-dimensional mesh for the FE model; (d) main variables within the FE simulations in order from left to right: profile of a linearly variable fiber direction with respect to depth, an illustration of the tissue depth direction, a uniform time series activation; (e) extracted surface displacement results and a warped image texture based on an estimated projective transformation.}
\label{fig:Syn_Data_Pipeline}
\end{center}
\end{figure}

Here, we describe the steps to generate realistic synthetic brightfield movies of beating cardiac microbundles based on examples from ``Type 1,'' as described in the ``\nameref{subsubsec:exp_data}'' Section. 
In Fig \ref{fig:Syn_Data_Pipeline}, we summarize the main steps of this pipeline. 
For each synthetic data example, we begin with a frame from an experimental movie (Fig \ref{fig:Syn_Data_Pipeline}a). As the first step of this pipeline, we manually trace the tissue region in a relaxed valley frame, and use the traced region to obtain both tissue geometry and image texture. From the traced region, we extract the coordinates of the external contour to generate Finite Element (FE) simulation geometry, and isolate the tissue texture that will be warped following the FE simulation results (Fig \ref{fig:Syn_Data_Pipeline}b).

To inform our FE simulations, we first generate a simplified three-dimensional microbundle geometry based on the contour coordinates of a mask extracted from the single representative valley frame (Fig \ref{fig:Syn_Data_Pipeline}c). Specifically, we extrude the 2D surface created by connecting these contour coordinates to a thickness of $400 \mu$m, a reasonable microbundle thickness given our target experimental setups \cite{javor2021pillar}. To approximate the pillars, we implement the geometry and dimensions detailed in \cite{javor2021pillar}, which matches one of the main platforms used in our experimental dataset (``Type 1''). 
To create the FE mesh, we use Gmsh $4.10.5$ \cite{geuzaine2009gmsh}, where the final mesh consists of $205,524$ tetrahedral elements which was deemed sufficient for our purpose, following a mesh refinement study. We provide a detailed schematic of the three-dimensional mesh geometry in \nameref{si:pipeline_val}.

In Fig \ref{fig:Syn_Data_Pipeline}d, we briefly summarize the main components of the FE model as implemented in FEniCS $2019.1.0$ \cite{alnaes2015fenics, logg2012automated}. Following popular recent work in the field of soft tissue biomechanics \cite{pezzuto2014orthotropic, gurev2015high, finsberg2018efficient}, we model the cardiac tissue as a nearly-incompressible transversely isotropic hyperelastic material where deformation is driven by periodic activation \cite{pezzuto2014orthotropic,finsberg2018efficient,land2015verification}. Of note, we create synthetic data with heterogeneous deformation fields (specifically, (1) a fully actively contracting tissue domain, (2) a passive circular inclusion at the center of the actively contracting tissue), and we vary the direction of contractile alignment (1) through the thickness or depth of the tissue or (2) along its length. We model the poly(dimethylsiloxane) (PDMS) pillars as passive Neo-Hookean material, and treat the interface between the cardiac tissue and the pillars as perfectly bonded.
For each FE simulation, we extract, with respect to time, the $X$, $Y$, and $Z$ positions of the mesh cell centers at the top surface of the microbundle for each step and save the results as text files. This simplified FE model serves a single purpose: to computationally generate synthetic data of realistically beating microbundles with known ground truth deformation. 
Then, following the schematic illustration in Fig \ref{fig:Syn_Data_Pipeline}e, we estimate a projective transformation based on the initial and deformed positions of the mesh cell centers and warp the image texture accordingly using the \verb|warp| transform function in the scikit-image $0.19.3$ Python library \cite{scikit_image}. To enable a heterogeneous transformation, we subdivide the image domain and perform subdomain specific warping. Additional details are available in \nameref{si:pipeline_val}.

Overall, our main synthetic dataset consists of $60$ generated movies of beating experimentally derived image textures. 
To obtain these $60$ examples, we use $15$ different base texture images extracted from $5$ experimental movies of ``Type 1'' as described in the ``\nameref{subsubsec:exp_data}'' Section. We then deform these extracted textures with FE results obtained from $4$ different FE simulations run under the variable conditions specified above. To perform quantitative evaluation, we extract a $90\times90$ pixel region from each domain center and make all direct comparisons based on this domain. 

In addition, we perform additional validation against a single computationally generated synthetic example of ``Type 2" data based on a more sophisticated tissue-specific FE model described in detail in Jilberto et al. \cite{jilberto2023computational} and \nameref{si:pipeline_val}.
For more information on implementing the Finite Element model, image warping, and the addition of Perlin noise \cite{perlin1985noise}, we refer the reader to \nameref{si:pipeline_val}. We also make the entire synthetic dataset prior to the addition of Perlin noise available along with all the Python code and files that are necessary to re-generate our dataset available on GitHub (\href{GitHub} {https://github.com/HibaKob/SyntheticMicroBundle}).

\subsubsection*{Experimental data} 
\label{subsubsec:exp_data}
Our experimental dataset can be systematically categorized into $3$ different types (Fig \ref{fig:diff_testbeds}). ``Type 1'' includes movies obtained from standard experimental microbundle  strain gauge devices \cite{boudou2012microfabricated, xu2015microfabricated, bielawski2016real}. We refer to data collected from non-standard platforms termed FibroTUGs \cite{depalma2023microenvironmental} as ``Type 2'' data. As for ``Type 3,'' they represent data obtained from a highly versatile experimental platform \cite{jayne2021direct, karakan2023direct} and as such, include the most diverse examples in this collection. 

Specifically, “Type 1” examples were prepared as previously detailed in \cite{das2022mechanical}. Briefly, PDMS (Dow Silicones Corporation, Midland, MI) microbundle devices were first cast from 3D printed molds (Protolabs). Each device contains $6$ wells, each with two pillars with rectangular cross sections and spherical caps, where the cardiac microbundles are seeded. Up to $2$ days before seeding, the devices were sequentially treated with $0.01\%$ poly-l-lysine (ScienCell) and then with $0.1\%$ glutaraldehyde (EMS) to promote cell attachment to the caps. On the day of seeding, devices were cleaned with $70\%$ ethanol and ultraviolet (UV) sterilized. Next, the device wells were incubated with a small volume of $2\%$  Pluronic F-127 (Sigma) to prevent cell attachment at the base of the well. hiPSC-CMs, differentiated and purified as described by Lian et al.  \cite{lian2013directed}, were seeded with human ventricular cardiac fibroblasts in a Matrigel (Corning) and fibrin (Sigma) extracellular matrix (ECM) solution. Microbundles were maintained in growth medium, with replacement every other day. Time-lapse videos of tissue contractions were acquired $5$-$7$ days after seeding at $30$ Hz using a $4\times$ objective on a Nikon Eclipse Ti (Nikon Instruments Inc.) with an Evolve EMCCD camera (Photometrics), while maintaining a temperature of $37^{\circ}$C and $5\%$ CO\textsubscript{2} .

As for the second type, FibroTUG microbundles were fabricated as described previously \cite{depalma2021microenvironmental, depalma2023microenvironmental}. First, arrays of PDMS cantilevers were fabricated by soft lithography as detailed in \cite{depalma2023microenvironmental}. Then, fiber matrices, suspended between pairs of these cantilevers, were generated by selective photo-crosslinking of electrospun dextran vinyl sulfone (DVS) fiber matrices deposited onto the microfabricated PDMS cantilevers \cite{depalma2021microenvironmental, depalma2023microenvironmental}. Matrix and cantilever stiffnesses were tuned by adjusting photoinitiator concentrations and cantilever height, respectively, while matrix alignment was controlled by altering the translation speed of collection substrates during fiber deposition \cite{depalma2021microenvironmental, depalma2023microenvironmental}. Following functionalization of the electrospun fiber matrices with cell adhesive cRGD peptides, iPSC-CMs, differentiated and purified \cite{depalma2021microenvironmental}, were patterned onto matrices using microfabricated seeding masks cast from 3D-printed molds. Finally, time-lapse videos of the microbundle’s spontaneous contractions were acquired at $\sim$ $65$ Hz on Zeiss LSM800 equipped with an Axiocam 503 camera while maintaining a temperature of $37 ^{\circ}$C and $5\%$ CO\textsubscript{2} .

Examples from ``Type 3'' were generated using the protocol previously described in \cite{jayne2021direct, karakan2023direct}. In brief, a combination of soft lithography and two-photon direct laser writing (DLW, Nanoscribe Photonic Professional GT+) was used to fabricate the seeding platforms. The process involves printing negative master molds using DLW, casting PDMS onto the molds, followed by sandwiching, curing and demolding. This results in $0.5$ - $0.6$ mm-thick PDMS devices with embedded microfluidic channels and deformable seeding wells. As a final step, cage-like microstructures were printed using DLW on the sides of the wells of the demolded PDMS devices to facilitate cell attachment. After device fabrication, differentiated hiPSC-CMs as per the procedures described in \cite{jayne2021direct}, were seeded into the wells with human mesenchymal stem cells in a collagen ECM solution, with the growth medium changed every other day. Time-lapse videos of the tissue contractions were acquired $4$-$9$ days after seeding at $30$ Hz using $4 \times$ or $10 \times$ objectives on a Nikon Eclipse Ti (Nikon Instruments Inc.) with an Evolve EMCCD camera (Photometrics) equipped with a temperature and CO\textsubscript{2} equilibrated environmental chamber.

In total, we include in this framework $24$ real experimental data, $11$ examples from ``Type 1,'' $7$ from ``Type 2,'' and $6$ from ``Type 3.'' This diverse pool of examples allows us to not only demonstrate the adaptability of our computational pipeline to different input examples, but also gain valuable insight about the heterogeneous contractile action of cardiac tissue by extracting and observing relevant mechanical metrics, such as full-field displacements, subdomain-averaged strains, and displacement and strain-derived outputs, as shown in the ``\nameref{res:exp_data}'' Section and in \nameref{si:additional_ex}. We note that details about each specific experimental example are provided in \nameref{si:additional_ex} as well as on Dryad \cite{microbundle2023data} (\href{Dryad}{https://doi.org/10.5061/dryad.5x69p8d8g}) where the whole dataset, ``Microbundle Time-lapse Dataset,'' is made available.

\subsection*{Code} 
\label{subsec:code}
\begin{figure}[!h]
\begin{center}
\includegraphics[width=0.95\textwidth]{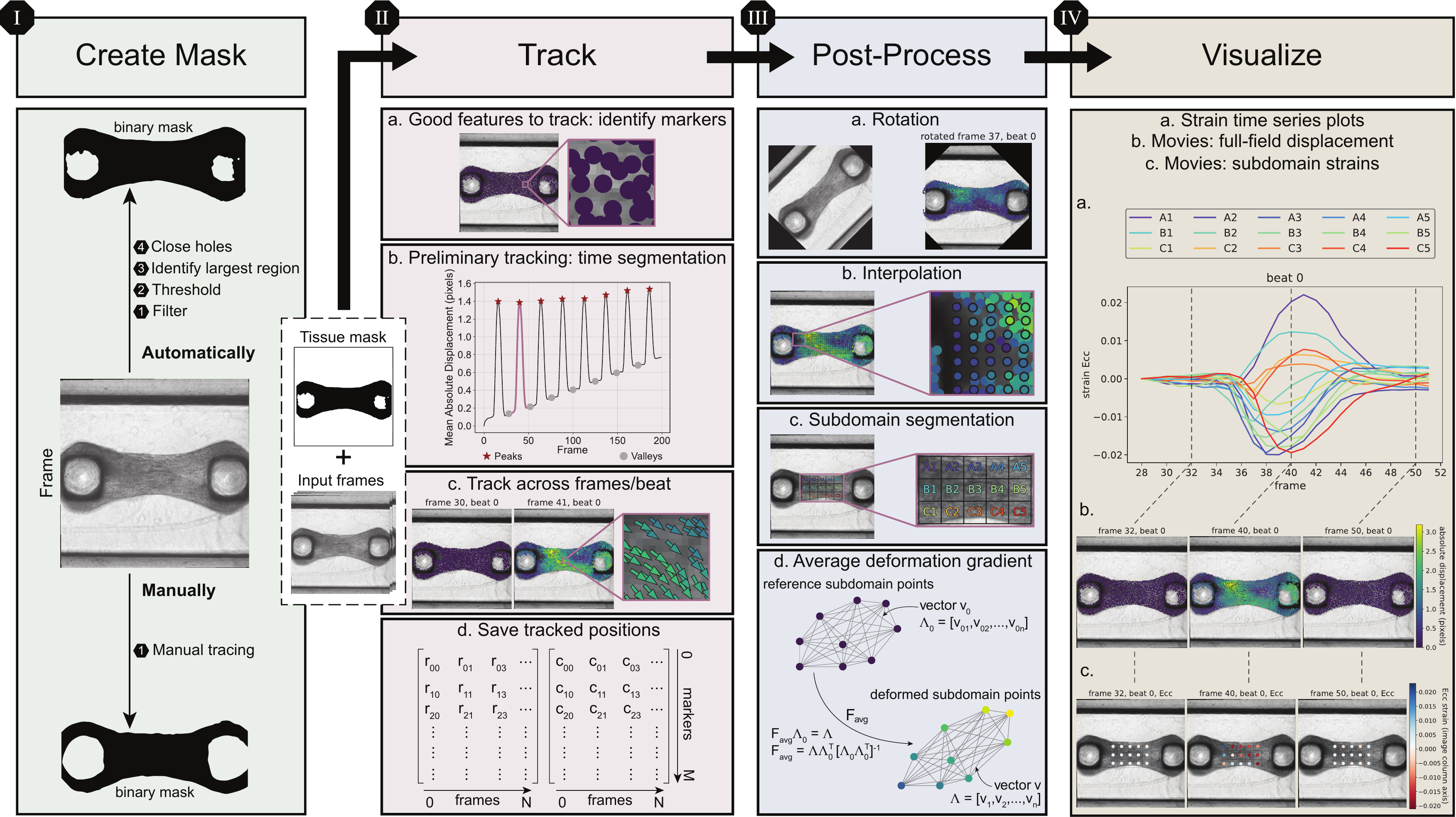}
\caption{{\bf Schematic illustration of the ``MicroBundleCompute'' computational framework.} 
Two main inputs are required: I) a binary mask generated either automatically using the software or manually and consecutive movie frames of the beating microbundle. II) For tracking, (a) marker points are identified on the first frame and (b) tracked across all frames to identify individual beats and perform time segmentation. (c) This allows us to perform the analysis per beat and correct for the observed drift as discussed in the ``\nameref{subsub:time_seg}'' Section. (d) Finally, we save the row and column positions of the tracked markers per a single beat and use these saved outputs to compute full-field displacements and derive strain results. III) Post-processing functionalities include (a) rotation of the images and tracking results, (b) interpolation of the results at query points, and (c) segmentation into subdomains for which (d) average deformation gradients and subsequently, subdomain-averaged strains are calculated. IV) Finally, to visualize the results, the software outputs (a) time series plots per beat and movies of (b) full-field and (c) subdomain-averaged results.}
\label{fig:code_pipeline}
\end{center}
\end{figure}

In this Section, we describe the main working components of our ``MicroBundleCompute'' software for the automatic analysis of deformation in brightfield and phase contrast movies of cardiac microbundles. Because our goal is to implement an approach with simplicity, versatility, and adaptability in mind, our pipeline is structured with four modular components: (1) image pre-processing and mask creation, (2) deformation tracking, (3) post-processing (e.g., rotation, interpolation, strain analysis), and (4) visualization.  In Fig \ref{fig:code_pipeline}, we provide a graphical summary of the major functionalities included in this pipeline and the computational workflow. As a brief note, the software GitHub repository (\href{GitHub} {https://github.com/HibaKob/MicroBundleCompute}) contains instructions on how to install and run the code, detailed explanations of each main function, and a more thorough description of the formatting of output files. 

The implementation of these methods is divided across four Python files: \verb|create_tissue_mask|, \verb|image_analysis|, \verb|strain_analysis|, and \verb|optional_preprocessing|.
The bulk of core functionality is contained in \verb|image_|  \verb|analysis|, and all functions of the code are designed to be modular when possible such that they can be replaced in the future if a need arises. In addition, many of the specific post-processing steps are optional, and adaptation to compute additional quantities of interest should be straightforward. As we describe our pipeline, we will specify which Python file a given function is located in. Essential functions included in our pipeline are as follows, following the order illustrated in Fig \ref{fig:code_pipeline}.

\subsubsection*{Tracking region mask generation} 
\label{subsub:tissue_mask}
Within \verb|create_tissue_mask|, we provide two different options for creating a binary mask of the tissue region: manual or automatic (Fig \ref{fig:code_pipeline} I). At present, we provide $3$ basic segmentation functions for automatic microbundle mask creation: 1) a straightforward threshold-based mask, 2) a threshold-based mask that is applied after a Sobel filter, and 3) a threshold-based mask that is applied to either the minimum or the maximum (specified by the user as an input) of all movie frames. We implement our tissue segmentation pipeline mainly based on the \verb|threshold_otsu| and \verb|sobel| functions provided within the \verb|filters| module in scikit-image $0.19.3$ Python library \cite{scikit_image}.

In addition to programming alternative automated mask generation functions, software users can provide an externally generated tissue mask (e.g., a manually generated mask) by including a file named ``tissue\_mask.txt'' where the mask is a two-dimensional array in which the tissue domain is denoted by ``1'' and the background domain is denoted by ``0'' to allow for the analysis of domains that may fall outside the original scope of this endeavor. 

\subsubsection*{Sparse optical flow algorithm for tracking} 
\label{subsub:optical_flow}
Within \verb|image_analysis|, we provide all of the essential functions to automatically run the tracking algorithm illustrated in Fig \ref{fig:code_pipeline} II. Our tracking pipeline is built on OpenCV’s \cite{bradski2008learning} pyramidal implementation of the Lucas-Kanade sparse optical flow algorithm \cite{bouguet2001pyramidal} where we identify markers as Shi-Tomasi ``good features to track'' corner points \cite{shi1994good} that fall within the specified tissue mask. In brief, these corner points are points where a slight shift in location leads to ``large'' changes in intensity along both the horizontal and vertical axes. 
There are two key parameters required to tune OpenCV's \verb|goodFeaturesToTrack| function: \verb|qualityLevel|, a minimum score to measure if a feature can be tracked well, and \verb|minDistance|, the minimum permitted Euclidean distance between two identified corners. We initialize \verb|minDistance| = $3$ and \verb|qualityLevel| = $0.1$. Then, we iteratively decrease \verb|qualityLevel| until \verb|coverage| $> 40$, where \verb|coverage| is defined as the average number of pixels associated with each tracking point, for up to $15$ iterations. 
As for \verb|minDistance|, we define a \verb|local_coverage| measure, which is the \verb|coverage| computed on $20 \times 20$ pixel subdivisions. The \verb|minDistance| is automatically incremented by $1$ as long as the largest $3$ \verb|local_coverage| values are less than or equal to $50$ and the number of iterations does not exceed $2$. We note that we adjust \verb|qualityLevel| and \verb|minDistance| simultaneously. 

For running OpenCV’s \cite{bradski2008learning} Lucas-Kanade optical flow \cite{bouguet2001pyramidal} function (\verb|calcOpticalFlowPyrLK|), we automatically tune the parameter \verb|winSize|, which dictates the size of the integration window.
Crucially, the window size in both horizontal and vertical directions, $w_x$ and $w_y$, should be larger than the maximum tracked pixel motion between frames. To specify \verb|winSize|, we adopt a pragmatic approach where we initialize \verb|winSize| = $5$, perform a preliminary tracking step, calculate the maximum absolute displacement, and compare its magnitude to the initial window size. If the calculated displacement is larger, we increase \verb|winSize| by $5$, and continue to iterate until the condition is met or the number of iterations exceeds $15$. Critically, keeping \verb|winSize| from being larger than necessary reduces error during tracking. 
In the remainder of this Section, we will describe our methods for leveraging this basic sparse optical flow algorithm to effectively and automatically analyze microbundle domains. 

\subsubsection*{Temporal segmentation} 
\label{subsub:time_seg}
After automatic identification of features and tracking algorithm parameters, we run a preliminary tracking step. Representative results of preliminary tracking are shown in Fig \ref{fig:code_pipeline} II.b as a plot of mean absolute displacement vs. frame number. We use these preliminary tracking results to perform temporal segmentation where we delineate individual tissue beats.
To accomplish this, we use the SciPy signal processing library function \verb|find_peaks| \cite{2020SciPy}. The \verb|find_peaks| input parameters, \verb|distance|, the minimum horizontal distance between neighbouring peaks, and \verb|prominence|, minimum prominence for a perturbation to be recognized as a peak, are identified automatically. 
Specifically, we initialize \verb|distance| and \verb|prominence| to values of $20$ and $0.1$ respectively. Then \verb|distance| is updated to take a value equal to $1.5 \times$ the horizontal distance separating two consecutive intersection points between the time series and its mean. For the \verb|prominence| parameter, we keep it constant at a value of $0.1$, which we found to be suitable for all the example videos that we have analyzed to date. 
After peaks are identified, we define valleys as the midpoints of two consecutive peaks. As such, to be able to identify a pair of valleys, or in other words a single beat, a minimum of $3$ beats should be present in any movie to enable automatic analysis and an accurate approximation of beat period. Temporal segmentation into individual beats is then performed based on the temporal location of each valley. 

Beyond the determination of microbundle time-related properties (e.g., beat period), temporal segmentation is an essential part of our pipeline as we use it to work around the tracked feature drift observed over the duration of the movie (see Fig \ref{fig:code_pipeline} II.b for an illustration of drift). 

After identifying individual beats, we split, based on the first and last beat frame numbers, the main two arrays storing column (horizontal) and row (vertical) locations of marker points obtained during preliminary tracking into multiple arrays corresponding to each segmented beat. Likewise, instead of frame $0$ being the reference configuration for the whole tracking duration, the fiducial marker positions in the first frame of each beat become the baseline for all future output calculations within the beat.  

\subsubsection*{Optional rotation and interpolation} 
\label{subsub:rot_interp}
After the tracking step is complete, we include two optional features for post-processing: sample rotation and fiducial marker displacement interpolation. 
First, we include an option to rotate both the images and the tracking results based on a specified center of rotation and desired horizontal axis vector. 
The center of rotation and desired horizontal axis vector can be either specified manually or identified automatically based on the geometry of the tissue mask.
As a brief note, rotation is performed after tracking as the process involves interpolation which can lead to loss of image resolution.
Also, we automatically rotate the tissue domain before performing strain subdomain calculations to match the global row (vertical) and column (horizontal) coordinate system. 
The second optional feature, interpolation, allows the user to interpolate the tracking results returned at the automatically identified fiducial marker points to user-specified locations, on a structured grid for example. This step can be performed after tracking and optional rotation, and will output the interpolated displacement fields to specified sampling points for either visualization or downstream analysis.

\subsubsection*{Subdomain spatial segmentation} 
\label{subsub:subdomain_seg}
In contrast to standard approaches to analyzing cardiac microbundles \cite{sala2018musclemotion}, our approach is unique in that we compute full-field quantities of interest over the tissue domain. In order to better analyze these full-field results and reliably post-process displacement fields to compute strain, we perform spatial segmentation to define tissue subdomains over which we can report average strain quantities \cite{Franck2016mean_def, Lejeune2021sarcgraph}.
This subdomain spatial segmentation is implemented in the \verb|strain_analysis| file within ``MicroBundleCompute.'' We provide two options to specify the subdomain extents defined as a rectangle: 1) automatic subdomain generation via clipping the input tissue mask or 2) manually providing subdomain extent coordinates. Given rectangular subdomain extents, we then delineate individual subdomain tiles by specifying either the target number of tiles in each column and row or by specifying the target tile dimensions in pixels. Representative subdomain segmentation results are illustrated in Fig \ref{fig:code_pipeline} III.c. 

\subsubsection*{Strain computation} 
\label{subsub:strain_calc}

With these defined subdomain regions, we then compute the average deformation gradient $\mathbf{F_{avg}}$ of each subdomain and use it to compute relevant strain metrics. As stated previously, we compute average subdomain strain rather than full-field strain due to: (1) the desire to reduce the influence of imaging artifacts and noise, and (2) increased ease of comparison between samples.  
Within each subdomain, we define the standard continuum mechanics deformation gradient $\mathbf{F}$ as follows: 
\begin{align}
\mathbf{F}d\mathbf{X}= d\mathbf{x}
\end{align}
where $\mathbf{F}$ maps a vector $d\mathbf{X}$ in the initial or reference configuration to its deformed configuration $d\mathbf{x}$ \cite{holzapfel2002nonlinear}. To apply this within the context of our tracking pipeline, we define a set of $n$ vectors, $\mathbf{\Lambda_0}$, that connect each potential pair of fiducial markers that lie within the extents of each subdomain in the reference configuration. We define the reference configuration with respect to each cardiac tissue beat as the first frame in the segmented beat frame. Thus, $\mathbf{\Lambda_0}$ is defined in frames that represent the most relaxed tissue state. We then compute $\mathbf{\Lambda}$ following the same structure as  $\mathbf{\Lambda_0}$ for each subsequent movie frame, where the updated fiducial marker positions capture the subdomain deformed configuration. With this definition, we can set up the over-determined system of equations:
\begin{align}
\mathbf{F_{avg}}\mathbf{\Lambda_0}= \mathbf{\Lambda} \quad \text{where} \quad \mathbf{\Lambda_0} = [\mathbf{v_{01}}, \mathbf{v_{02}},...,\mathbf{v_{0n}}] \quad \text{and} \quad \mathbf{\Lambda} = [\mathbf{v_{1}}, \mathbf{v_{2}},...,\mathbf{v_{n}}]
\end{align}
where $\mathbf{F_{avg}}$ is a $2 \times 2$ matrix, and $\mathbf{\Lambda_0}$ and $\mathbf{\Lambda}$ are $2 \times n$ matrices of vectors in the initial (reference) and current (deformed) configurations, respectively. Of note, when the initial and current frames are identical, $\mathbf{\Lambda_0}$ = $\mathbf{\Lambda}$, $\mathbf{F_{avg}}$ = $\mathbf{I}$, a $2 \times 2$ identity matrix. To solve this over-determined system, we can use the normal equation to find the best fit average deformation gradient as:

\begin{align}
\mathbf{F_{avg}} = \mathbf{\Lambda} \mathbf{\Lambda^T_0}\left[ \mathbf{\Lambda_0}\mathbf{\Lambda^T_0}\right]^{-1} \, .
\end{align}

We schematically illustrate our method to compute the mean deformation gradient in Fig \ref{fig:code_pipeline} III.d. With the computed $\mathbf{F_{avg}}$, finding the average Green-Lagrange strain tensor is straightforward: 

\begin{align}
\mathbf{E_{avg}} = \frac{1}{2}\left(\mathbf{C_{avg}} - \mathbf{I}\right) \quad \text{where} \quad \mathbf{C_{avg}} = \mathbf{F^T_{avg}}\mathbf{F_{avg}}
\end{align}

and we can then compute strain on a per subdomain per beat basis to obtain subdomain time series results for $\mathbf{F_{avg}}$ and subsequently $\mathbf{E_{avg}}$.

\subsubsection*{Data structure preparation}
\label{subsub:data_prep}

Tracking all identified fiducial markers for an extended number of frames produces a large quantity of output data for each movie. Thus, we selectively save output results such that they are both comprehensible and easily accessible for downstream data analysis.
First, we save information regarding the column (horizontal) and row (vertical) positions of the tracked marker points per beat. Specifically, we store one row-position text file and one col-position text file for each beat formatted as a $M\times N$ array where $M$ is the number of markers that were tracked and $N$ is the number of frames in the beat.
Similarly, we output mean deformation gradient results as text files saving the column and row positions of the center for each subdomain and the $4$ components of the $2\times2$ mean deformation gradient per subdomain per beat.

\subsubsection*{Data analysis, key metrics, and visualization} 
\label{subsub:data_analysis}

In Fig \ref{fig:code_pipeline} IV, we show key output metrics and their visualizations. In brief, we provide tools to visualize full-field displacement and average subdomain strain, and provide key quantities of interest such as maximum strain, beat period, and synchrony. 
In all cases, we build on the popular matplotlib package \cite{hunter2007matplotlib} for producing all visualizations.
In addition to the metrics directly enabled by our novel pipeline, we provide other commonly pursued relevant outputs including beat frequency, beat mean amplitude, and tissue width at the domain center. To convert the numerical outputs from dimensions of pixels and frame numbers to physical units, the user can specify: 1) frames per second and 2) the length scale in units of $\mu$m/pixel.

\subsubsection*{Quality checking and rejection of unsuitable examples} 
\label{subsub:quality_check}

In our extensive experience testing our code on different synthetic and real examples, we have identified three main instances that negatively influence the fidelity of our outputs and decrease our confidence in analysis results: 
\begin{itemize}
    \item[1.] blurred input movie frames that prevent effective identification of corner points for the tracking described in the ``\nameref{subsub:optical_flow}'' Section.
    \item[2.] movies that start from a contracted tissue position that confounds the temporal segmentation step described in the ``\nameref{subsub:time_seg}'' Section.
    \item[3.] movies where all displacement is on \textit{sub}-pixel length scales lead to large relative errors given our choice of tracking algorithm. 
\end{itemize}
The specific influence of these conditions are explored in \nameref{si:pipeline_val}. 
To address these conditions, we provide both warnings in the code when we detect that these conditions arise and functions to correct these scenarios (e.g., removing initial movie frames to address case 2).

\subsubsection*{Note on pillar tracking} 
\label{subsub:pillar_tracking}

In the broader cardiac microbundle and microtissue literature \cite{tamargo2021millipillar, mery2023light}, pillar force and tissue stress are two commonly computed metrics. Broadly speaking, pillar force is computed by tracking pillar displacement and converting displacement to force via cantilever beam equations where the pillars are treated as beams with known elastic modulus and geometry \cite{legant2009microfabricated}. Stress is then computed from pillar force by dividing force by tissue cross sectional area where tissue width is measured from the in-plane images \cite{rivera2022contractility} and tissue depth is assumed based on typical values observed via three dimensional imaging modalities \cite{kabadi2015into}. We are able to adapt our framework to readily track pillar displacement by simply specifying pillar regions as the area to track rather than the tissue domain. Due to lack of novelty, we do not emphasize this functionality in this paper. However, we do provide additional details on this topic in \nameref{si:pillar_track}.

\section*{Results and discussion}
\label{sec:results}

In this Section, we show a summary of our main results. In the ``\nameref{res:syn_data}'' Section, we present representative validation examples from validating our pipeline on an example from the synthetic dataset of ``Type 1'' with known ground truth behavior, and in the ``\nameref{res:exp_data}'' Section, we share examples of implementing our computational framework on $3$ different experimental platforms (i.e., $3$ different data types as explained in the ``\nameref{subsubsec:exp_data}'' Section). As a brief note, we provide a more comprehensive set of results in \nameref{si:pipeline_val} on validation, \nameref{si:additional_ex} on additional real data examples, and \nameref{si:pillar_track} on pillar tracking. 

\subsection*{Synthetic data examples} 
\label{res:syn_data}
\begin{figure}[p]
\begin{center}
\includegraphics[width=0.9\textwidth]{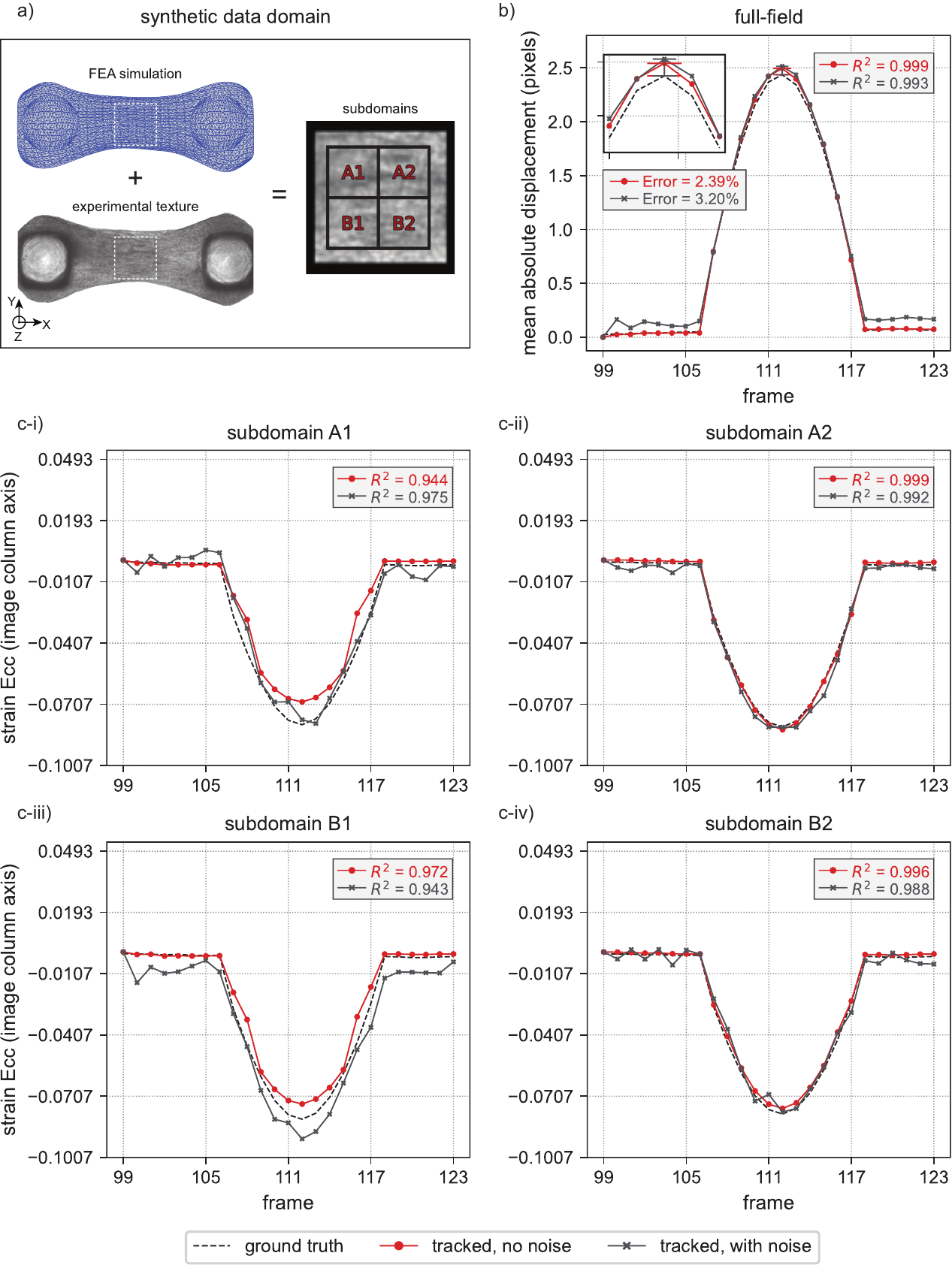}
\caption{{\bf Validation against a single synthetic example.}
Validation results against synthetic example ST\_1 in its original (no noise) state as well as after the addition of Perlin noise (magnitude ratio of $12\%$ and octaves of $40$): (a) a schematic representation of the synthetic data domain and subdomain divisions for strain calculations; (b) tracked and FEA-extracted (ground truth) mean absolute displacement of the full-field data for a single beat (beat $4$), where maximum error for synthetic data with noise is observed; (c-i)--(c-iv) tracked and FEA-extracted (ground truth) subdomain-averaged $E_{cc}$ strain for beat $4$ for each subdomain. Overall, the $R^2$ values indicate relatively good agreement between the tracking output and ground truth data. For more information on validation with synthetic data, refer to \nameref{si:pipeline_val}.}
\label{fig:syn_error}
\end{center}
\end{figure}

Here, we briefly summarize the results of our validation studies. In Fig \ref{fig:syn_error}, we show the performance of our pipeline on a single representative synthetic example ST\_1. In \nameref{si:pipeline_val}, we provide a comprehensive summary of all $416$ synthetic validation examples based on data of ``Type 1'' as well as a single synthetic example of ``Type 2'' data. In our validation studies, we compare ``MicroBundleCompute'' displacement and strain outputs against the known ground truth for synthetic data as originally generated (without noise) and for the same examples with added Perlin noise of different magnitude ratios and octaves. As a brief note, other types of noise such as shot noise \cite{shot1960} would be appropriate alternatives to Perlin noise for the purpose of this investigation. In preliminary studies, we found that Perlin noise with a range of magnitudes and octaves was a sufficiently general choice to provide a robust challenge for our framework. The code required to reproduce all synthetic data examples of ``Type 1'' can be found on GitHub (\href{GitHub} {https://github.com/HibaKob/SyntheticMicroBundle}). 

\FloatBarrier

In Fig \ref{fig:syn_error}a, we specify the synthetic data domain for which the validation studies were performed, a $90 \times 90$ domain warped with FEA-informed displacements as briefly described in the ``\nameref{subsubsec:syn_data_gen}'' Section and presented in more details in \nameref{si:pipeline_val}. In Fig \ref{fig:syn_error}b, we plot the mean absolute displacement with respect to frame number for beat $4$, the beat with the maximum error at the peak displacement (full contraction) for the synthetic example with Perlin noise of magnitude ratio of $12\%$ and octaves of $40$. We specifically highlight these Perlin noise parameters because they mimic, to a great extent, noise artifacts found in the real experimental data. A direct comparison between both tracked and ground truth mean absolute displacements per beat revealed that the percentage error at peak displacement is less than $2.5\%$ for the example shown in Fig \ref{fig:syn_error}b, with the addition of Perlin noise slightly raising this error to $3.2\%$. In addition, we assessed the ability of our computational framework to ``predict'' the mean absolute displacement by calculating the coefficient of determination ($R^2$) and found good agreement. Overall, considering all validation examples in \nameref{si:pipeline_val} prior to the addition of Perlin noise, peak mean displacement errors were found to be less than $7\%$ for \textit{super}-pixel displacements (\nameref{si:pipeline_val} Table S1\_1 and Fig S1\_5) whereas \textit{sub}-pixel displacements had higher errors that reach up to $15\%$ (\nameref{si:pipeline_val} Table S1\_1 and Fig S1\_6) with a minimum $R^2$ value of $0.939$. Furthermore, it is evident that the addition of Perlin noise produces higher shifts within the tracked outputs, with higher errors reported for higher magnitude ratios and lower octaves (\nameref{si:pipeline_val} Figs S1\_23 and S1\_24). Specifically, as shown in \nameref{si:pipeline_val}, Table S1\_4 and Fig S1\_23, the percentage error at the peak mean absolute displacement increases to around $14.4\%$ for \textit{super}-pixel displacement examples. For the \textit{sub}-pixel displacement examples, mean absolute displacement errors become unreasonably high in the majority of the synthetic examples and are on the order of $10^3 - 10^4$ (\nameref{si:pipeline_val} Fig S1\_24). And, in some of these \textit{sub}-pixel displacement examples where the added Perlin noise is unrealistically extreme (higher magnitude ratios and lower octaves), the code fails to produce meaningful outputs as indicated by missing data points in \nameref{si:pipeline_val}, Fig S1\_24.

In Fig \ref{fig:syn_error}c-i--iv, we show the error on the column-column direction (i.e., horizontal) Green-Lagrange strain ($E_{cc}$) per subdomain for beat $4$. We refer to this direction as the ``column-column'' direction to be consistent with the row (vertical) and column (horizontal) directions defined by our input images. Here, the reported $R^2$ values reveal that the minimum value of $0.943$ occurs in subdomain \verb|B1|. And, in general, these plots indicate that strain magnitudes tend to be underestimated. Furthermore, errors on $E_{cc}$ follow the observed trend for displacement errors, where synthetic examples of \textit{sub}-pixel displacements exhibit higher errors with some cases having negative $R^2$ values (\nameref{si:pipeline_val} Tables S1\_2 and S1\_3 and Fig S1\_8).

In \nameref{si:pipeline_val}, we share the validation results for all synthetic examples. We note that strain outputs are sensitive to subdomain divisions, specifically subdomain size. Determining a suitable subdomain size is a delicate process that is governed by two main opposing factors. The subdomain size should be large enough such that each subdivision contains an appropriate number of automatically identified fiducial markers to ensure that the computations are less sensitive to noise. On the contrary, the subdomain size should be small enough to avoid the loss or reduction of information due to averaging the heterogeneous deformation, or put more explicitly, obtaining attenuated or zero strain values due to lumping regions that are experiencing opposing deformations, for example extension versus compression. In general, based on our comprehensive experience with implementing the code on a number of synthetic and real examples, we recommend a subdomain side length that is between $30$ and $40$ pixels. From a methodological perspective, we propose that the user observes the displacement field and avoids having subdomains that span regions where the change in displacement flips sign. We recommend that manual examination of subdomain size be carried out once per dataset, for example, select a single movie from a batch of $100$ to confirm that the subdomain division is appropriate. In the future, we plan to investigate the approach adopted in \cite{boyle2014simple} and \cite{boyle2019regularization} where strains are directly informed from affine warping functions optimized via the Lucas-Kanade algorithm \cite{lucas1981iterative} without computing displacement fields.

As described in this Section and in \nameref{si:pipeline_val}, we have performed extensive validation studies for the ``MicroBundleCompute'' computational framework against a total of $416$ synthetic examples of ``Type 1'' ($16$ examples generated under baseline conditions and the remaining $400$ created by adversely altering the original $16$ examples via the addition of Perlin noise) and a single synthetic example of ``Type 2'' with known ground truth. These validation studies: 1) corroborate the output of the software against FEA-labelled data, 2) test its performance against realistic synthetic examples, and 3) evaluate its robustness against challenging examples with excessive noise artifacts. The obtained results reveal that for \textit{super}-pixel displacements, our software is quite robust to all tested cases with relatively low magnitudes and high octaves, or in other words, when the Perlin noise patterns resemble speckle noise rather than pronounced textures (see \nameref{si:pipeline_val} Fig S1\_4). However, for examples with entirely \textit{sub}-pixel displacements, the performance of the software degrades. In summary, ``MicroBundleCompute'' breaks down when the synthetic data has noise artifacts that appear similar to the original texture, and for \textit{sub}-pixel displacements. Yet, given that real experimental examples of beating microbundles generally produce displacements that exceed a single pixel and that Perlin noise examples with higher octaves more faithfully represent naturally occurring noise in real microbundle data than lower values, we anticipate that ``MicroBundleCompute'' will output reliable mechanical metrics in real experimental settings on condition that the natural contrast of the microbundle textures is visibly present and in focus in the time-lapse videos. 

\subsection*{Experimental data examples} 
\label{res:exp_data}

We provide here a summary of implementing ``MicroBundleCompute'' on the experimental dataset described in the ``\nameref{subsubsec:exp_data}'' Section and \nameref{si:additional_ex}, and shared on Dryad (\href{Dryad}{https://doi.org/10.5061/dryad.5x69p8d8g}) under a Creative Commons CC0 $1.0$ Universal Public Domain Dedication. Specifically, the ``Microbundle Time-lapse Dataset’’ \cite{microbundle2023data} contains all raw videos in ``.tif’’ format for the $24$ experimental time-lapse images of beating cardiac microbundles, $23$ of which are brightfield videos, while the remaining single example is a phase contrast video. Besides the raw videos and the experimental metadata describing the conditions under which they were obtained, we include the tissue mask used for each example, whether generated automatically via our computational pipeline or manually via tracing in ImageJ \cite{bourne2010imagej}. These time-lapse videos and masks were used to generate the results shown here and in \nameref{si:additional_ex}. We note briefly that it is only possible to develop automatic mask segmentation functions for examples where there is imaging consistency and when we have an ample number of examples to identify a pattern. Future extensions of this framework will include automatic mask functions tailored to specific experimental needs.

\begin{figure}[H]
\begin{center}
\includegraphics[width=0.95\textwidth]{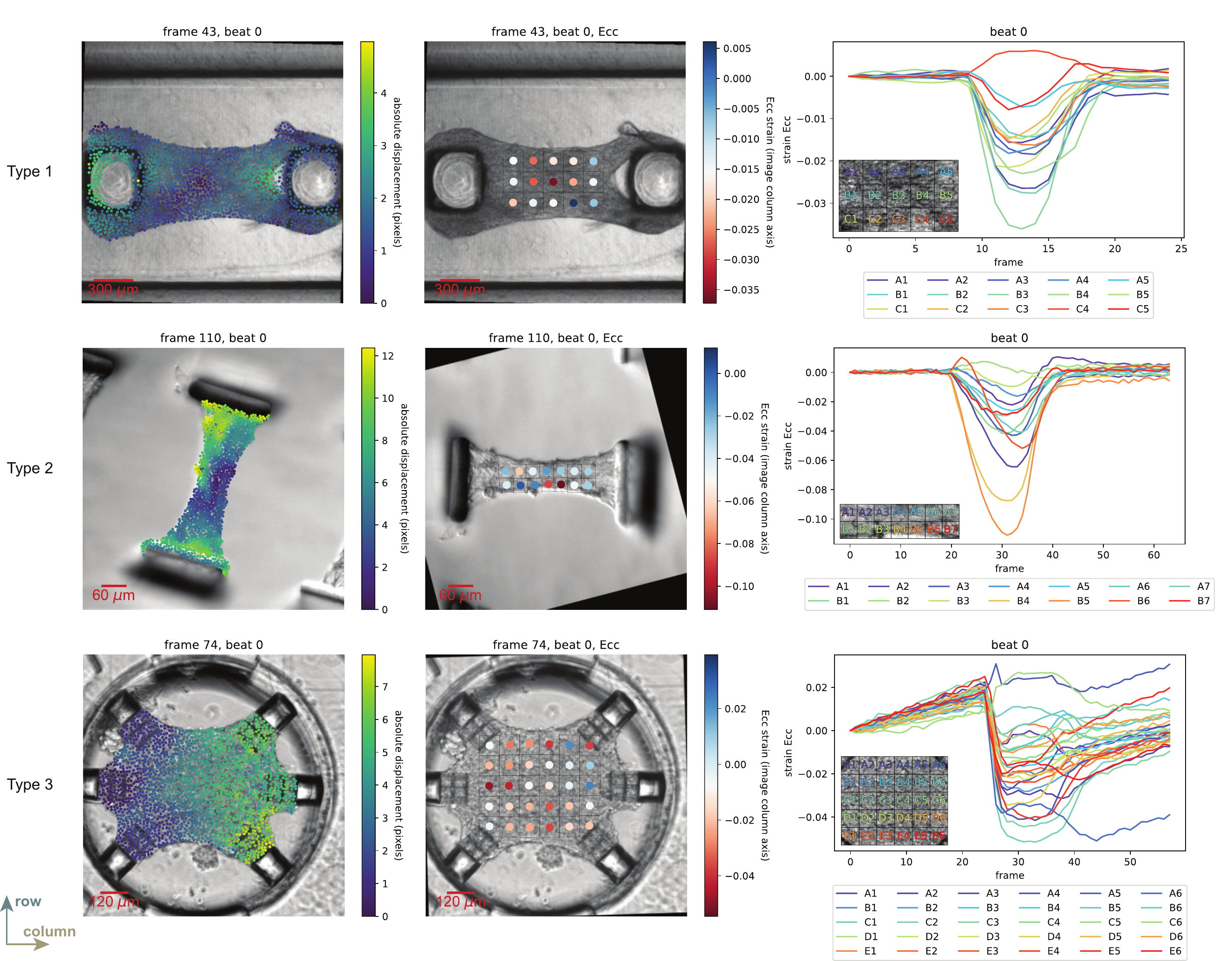}
\caption{{\bf Examples of ``MicroBundleCompute'' outputs.}
We show sample outputs of the code when run on $3$ examples from the $3$ different data types, specifically Example $9$ from ``Type 1,'' Example $5$ from ``Type 2,'' and Example $1$ from ``Type 3'' as shared on Dryad \cite{microbundle2023data}. Note that the strain output for ``Type 2'' data is automatically rotated such that $E_{cc}$ aligns with the column-column direction. Additionally, the ``Type 3'' example shows an actuated microbundle at 0.5 Hz by applying sawtooth pressure waves with $\sim$ - 6 kPa peak amplitude (equivalent to $\sim$ 2.5\% strain) on the right side using a microfluidic pump (Elveflow OB1), and hence, the considerable discrepancy of the $E_{cc}$ time series plot from $0$ at the end of beat $0$.}
\label{fig:res_real_examples}
\end{center}
\end{figure}

In Fig \ref{fig:res_real_examples}, we show visualizations of output results generated by running ``MicroBundleCompute'' on $3$ experimental examples, each from a different data type. Of note, we only visualize $3$ of many potential software outputs: (1) full-field absolute displacement, (2) spatially distributed subdomain-averaged Green-Lagrange strain $E_{cc}$ (automatically rotated to align with the column-column horizontal direction and plotted at beat $0$ strain peak), and (3) time series plot of $E_{cc}$ strains with respect to beat $0$ frames. These are representative examples from the comprehensive list of outputs described in the software GitHub repository (\href{Github} {https://github.com/HibaKob/MicroBundleCompute}). While displacement and strain visualizations reveal spatial information on tissue heterogeneous behavior and spatial contraction patterns, time series strain plots highlight this spatial synchrony (or lack of synchrony) of subdomain contraction across each beat.  

\FloatBarrier

To complement the results in Fig \ref{fig:res_real_examples}, we show, in \nameref{si:additional_ex}, representative results obtained for all $24$ real examples following the same format. Critically, all of the results presented in the supplementary document, for all $3$ of the data types, were obtained without the need for any parameter tuning. As is evident in \nameref{si:additional_ex}, Fig S2\_1, we can group data from ``Type 1'' into $2$ different categories: 1) examples $1$--$6$ which constitute relatively challenging examples to the code since the imaging brightness/contrast is not optimized to fully accentuate the natural tissue texture, and 2) examples $7$--$11$ where a texture is clearly visible in the microbundle region, making them a set of optimal examples for running the code. Despite the fact that ``Type 1'' data comprises the most consistent and well studied experimental platform, the full-field absolute displacement plots reveal wide variation in spatially distributed behavior. However, magnitude ranges are relatively consistent across samples with mean absolute displacement values at the peaks varying between $1.2$ and $2.3$ pixels ($4.8$ and $9.2$ $\mu$m). We anticipate that this observation might be attributed to variations in pillar-tissue attachment behavior across different examples (see \nameref{si:pillar_track} Fig S3\_1). However, further investigations are required to develop a systematic way to assess the effect of this pillar-tissue interaction on microbundle contractility and analyze the extracted metrics in this context. 

Similar spatial heterogeneity is observed for the "Type 2" samples (\nameref{si:additional_ex} Fig S2\_2), where examples $1$--$5$, which were prepared under the same experimental conditions (soft, aligned matrix), exhibit discrepant displacement distributions but fairly consistent magnitude ranges (mean absolute displacement values at the maximum contraction vary between $5.0$ and $7.9$ pixels or equivalently, $4.54$ and $7.17$ $\mu$m). For examples $6$ and $7$, which are prepared with stiff matrices, the results reveal lower contraction magnitudes where an aligned stiff matrix (example $6$) shows higher contractions than the randomly distributed one (example $7$) \cite{depalma2023microenvironmental}.
Of note, by examining all results of cardiac microbundle ``Type 1'' and ``Type 2'' data, a noticeable diagonal contraction pattern is apparent, especially in examples $2$, $7$, $9$, and $10$ from ``Type 1'' (\nameref{si:additional_ex} Fig S2\_1) and examples $2$, $3$, and $5$ from ``Type 2'' (\nameref{si:additional_ex} Fig S2\_2). This observation, enabled by full-field tracking, indicates that future study to investigate the association between microbundle contraction, fiber alignment, and emergent load paths between the pillars, would be meaningful future work.

The diversity of ``Type 3'' experimental data prevents direct comparison between samples. However, the time series strain visualizations (\nameref{si:additional_ex} Fig S2\_3) reveal that cardiac microbundles grown on these experimental constructs typically contract more synchronously than microbundles of ``Type 1'' and ``Type 2,'' where the strain time series plots in \nameref{si:additional_ex}, Figs S2\_1 and S2\_2 respectively, show aspects of temporal heterogeneity for which peak contractions do not occur at the same frame within all subdomains per beat. Finally, the visualized average subdomain $E_{cc}$ strains, as well as the remaining two strain components (row-row direction Green-Lagrange strain $E_{rr}$ and column-row direction Green-Lagrange strain $E_{cr}$) that are computed and saved but not included in the set of representative outputs that we visualize here, give insight about the regions within the microbundle that are contracting or bulging in a given direction across each beat.  

\begin{figure}[!h]
\begin{center}
\includegraphics[width=0.95\textwidth]{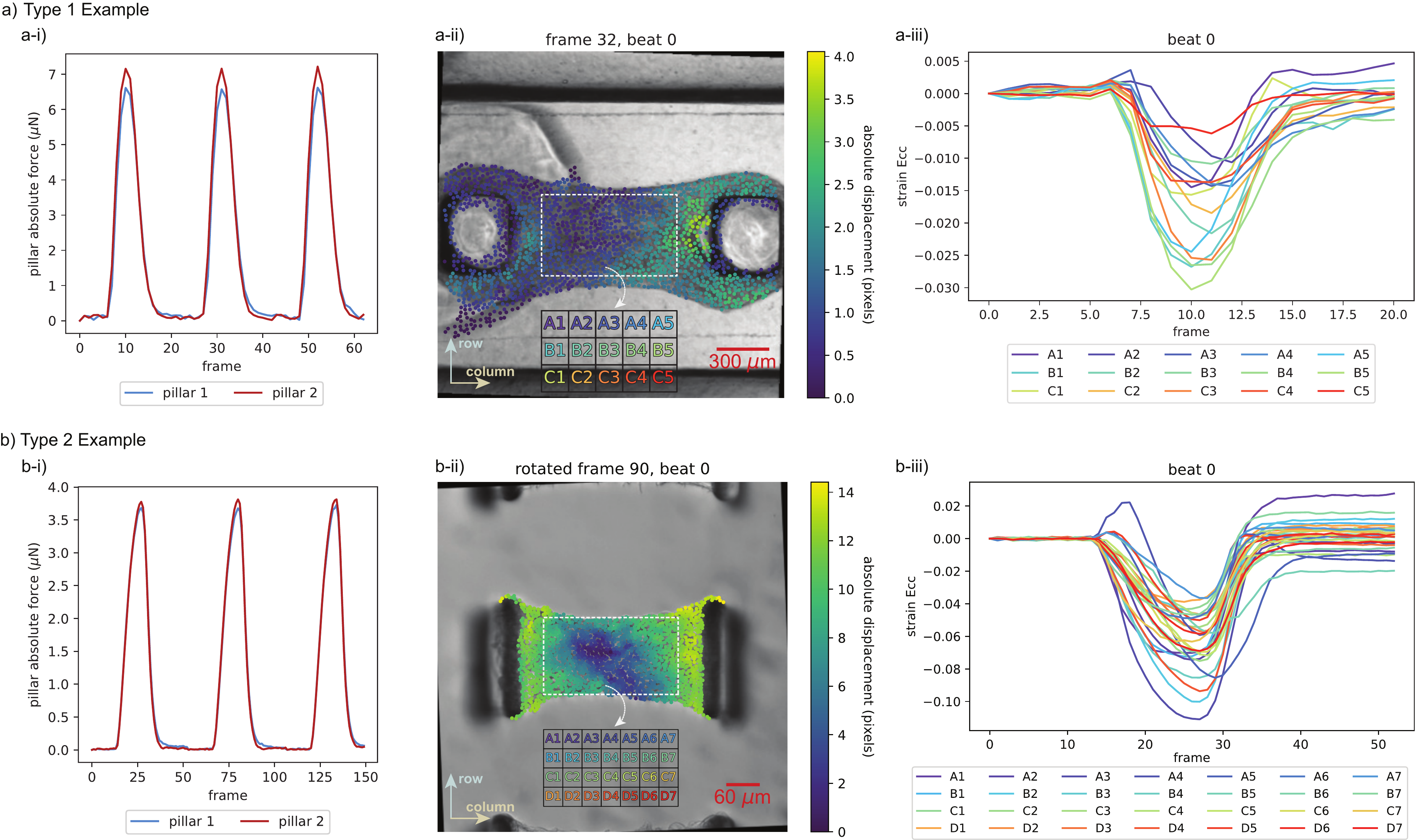}
\caption{{\bf Direct comparison of pillar tracking and tissue tracking.}
Implementing the pillar tracking functionality of ``MicroBundleCompute'' on (a) Example $10$ from ``Type 1'' data, (a-i) we obtain the pillar displacement based tissue force to describe the microbundle beating behavior, whereas full-field tissue tracking reveals the heterogeneous (a-ii) full displacement field as well as (a-iii) subdomain-averaged strains computed within the region marked by the dashed white box in (a-ii) where marker points within this region only are considered. In (b) we show the same outputs for Example $3$ from ``Type 2.'' We note that, for this example, we show the rotated displacement output to be consistent with the subdomain segmentation orientation. To view the original non-rotated displacement results, refer to \nameref{si:additional_ex}, Fig S2\_2.}
\label{fig:Pillar_vs_full}
\end{center}
\end{figure}

As we mention in the ``\nameref{subsub:pillar_tracking}'' Section and describe in more detail in \nameref{si:pillar_track}, it is straightforward to implement our computational framework to track pillar displacements, adding to the versatility of the ``MicroBundleCompute'' software framework. In Fig \ref{fig:Pillar_vs_full}, we show, side by side, the outputs obtained via pillar tracking (Figs \ref{fig:Pillar_vs_full}a-i and b-i) and via tracking the entire tissue domain (Figs \ref{fig:Pillar_vs_full}a-ii--iii and b-ii--iii) for an example from ``Type 1'' and ``Type 2'' data respectively. Pillar tracking enables the calculation of an average absolute or directional pillar displacement based tissue force, which can be used to infer an average tissue stress given approximate tissue width and thickness as described in more detail in \nameref{si:pillar_track}. On the other hand, tissue tracking reveals abundant information regarding the inherent heterogeneous nature of microbundle beating. For example, full-field displacement fields (Figs \ref{fig:Pillar_vs_full}a-ii and b-ii) show regions where maximum or minimum tissue contractions are taking place. Furthermore, while subdomain-averaged strains also underline the spatial heterogeneity of the tissue contractions, visualizing them with respect to time (or frame number) reveals the nature of temporal synchrony across the different subdivisions of the beating microbundle (Figs \ref{fig:Pillar_vs_full}a-iii and b-iii). We note that a similar comparison can be carried out for all ``Type 1'' and ``Type 2'' data for which pillar tracking results are included in \nameref{si:pillar_track}, while tissue tracking results are shared in \nameref{si:additional_ex}.

\begin{figure}[!h]
\begin{center}
\includegraphics[width=0.95\textwidth]{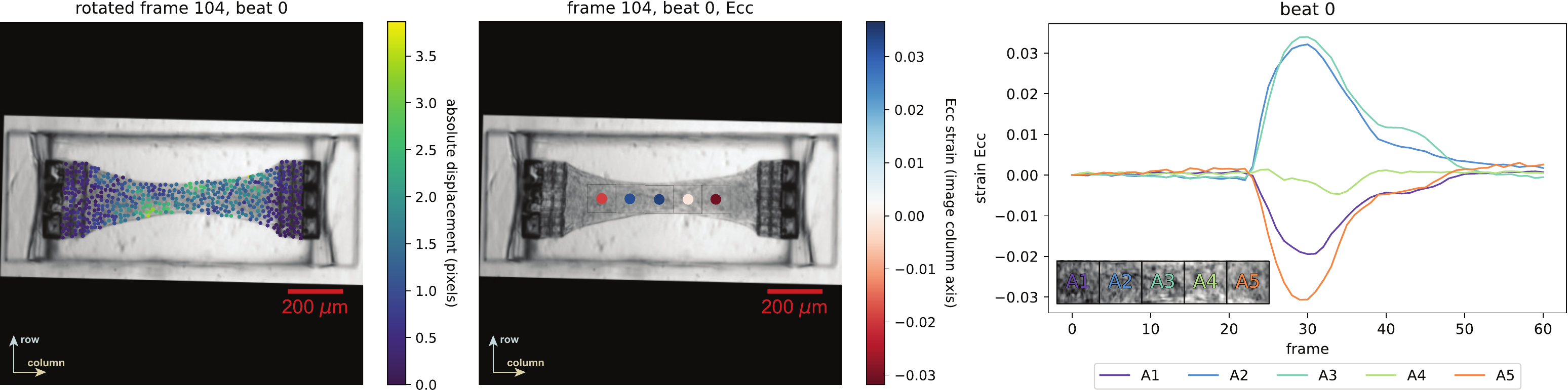}
\caption{{\bf Outputs of ``MicroBundleCompute'' run on Example $\mathbf{6}$ from ``Type 3'' reveal regions of tissue extension at the center.}
We briefly note that for displacement outputs, rotating the results is optional, while for strain outputs, the frames are rotated such that the major axis of the microbundle automatically aligns with the column (i.e., horizontal) axis. To view the original orientation of this example, refer to \nameref{si:additional_ex}, Fig S2\_3.}
\label{fig:type3_ex6}
\end{center}
\end{figure}

Finally, we include in Fig \ref{fig:type3_ex6} an example of ``Type 3'' data which clearly indicates that the microbundle is experiencing positive $E_{cc}$ strains at the center, specifically in subdomains \verb|A2| and \verb|A3| as indicated in the inset legend on the lower left corner of the strain time series plot. According to Wang et al. \cite{wang2013necking}, this observation suggests that a necking instability is forming on this tissue over time, which is also supported by the thinning tissue width at the center. Within this context, being able to extract and examine strain distributions and magnitudes enables further investigations into understanding and determining the factors, such as pillar stiffness and ECM density \cite{wang2013necking}, that produce microbundles that are more stable against necking. 

Based on the results shown in this Section and in \nameref{si:additional_ex}, conventional metrics comprising tissue force and tissue stress obtained via basic pillar tracking offer insightful yet lumped information about the microbundle behavior. Complemented with reliable and reproducible full-field data, such as displacement distributions and subdomain-averaged strains, as in \cite{das2022mechanical} for example, these metrics become more useful to assessing the highly heterogeneous cardiac tissue behavior and understanding the complex underlying mechanisms driving this behavior. Furthermore, this spatial information would allow us to study injury models of tissue \cite{das2022mechanical}, as well as further investigate mechanical observations from pillar tracking, for example, the unbalanced pillar forces noted in \nameref{si:pillar_track}, Figs S3\_1 and S3\_2. Finally, with sufficient collected data, this approach would enable the development of appropriate statistical models of the whole tissue. These endeavors are planned as part of our future work.

\section*{Conclusion}
\label{sec:concl}
In this work, we describe our approach to creating a computational tool for analyzing brightfield and phase contrast movies of beating microbundles. In brief, we describe our process for converting movies of beating microbundles to meaningful quantitative metrics, validating our approach against synthetically generated data, and testing it on a diverse pool of real experimental examples. In addition to ease of use, limited requirements for user intervention, relatively short run time, and no parameter tuning, ``MicroBundleCompute'' is easy to implement out of the box on new experimental datasets. Because it relies solely on the natural contrast of brightfield and/or phase contrast microbundle images and the resulting intensity gradients, it is also straightforward to integrate with existing experimental workflows. To enable broad adoption, we share the software under open-source license and look forward to receiving feedback from different users on how to adapt the code to tailor to their specific needs and enhance the overall user experience.  

Looking forward, we aim to constantly improve the code. Future extensions will include automated quality control and pre-processing of input data, as well as enhancements to current functionality such as automatic adjustment of input movies that do not start from a fully relaxed frame. As viable alternatives arise, we also plan to benchmark it against available cardiac microbundle tissue analysis tools. In addition, we plan to continue developing our pipeline to address alternative quantitative metrics and different imaging modalities such as calcium imaging \cite{guatimosim2011imaging} and integrate these results with structural information such as sarcomere geometry and alignment \cite{Lejeune2021sarcgraph, Mohammadzadeh2023}. From the results shown in the ``\nameref{res:exp_data}'' Section as well as \nameref{si:additional_ex} and \nameref{si:pillar_track}, it is also clear that there is significant variation across individual microbundle behavior both within and across testbeds. One key motivation for applying the ``MicroBundleCompute'' framework to these data moving forward is that it will make it possible to better understand and analyze this heterogeneity. In addition, extracting comparable mechanical metrics reliably and reproducibly across different testbeds allows for the identification of the favorable configurations and conditions that promote hiPSC-CM based tissue maturation, and ultimately, converge to an optimum system. Overall, our intention is for other researchers to directly benefit from disseminating this work. As a final note, here we demonstrate the utility and functionality of “MicroBundleCompute” for a particular highly used engineered cardiac tissue format: cardiac microbundles. In concurrent work \cite{rios2023mechanically}, we leverage the fundamental core of this computational framework and make some minor modifications and extensions to extract mechanical metrics from actuated 2D muscle sheets. Looking forward, we will continue to generalize our framework to provide noninvasive, label-free, and high-throughput tools to facilitate contraction measurements across different engineered contractile tissue platforms to benefit the tissue engineering research community.

\section*{Acknowledgments}
We gratefully acknowledge the collaborative opportunities facilitated by the CELL-MET Engineering Research Center. We especially thank the administrative team at CELL-MET for coordinating the Research Experiences for Undergraduates (REU) program. We would also like to acknowledge the staff at Boston University libraries for providing advice regarding data dissemination practices.

\appendix 

\newpage
\clearpage
\FloatBarrier
\section{Pipeline Validation Appendix} 
\label{si:pipeline_val}
 \renewcommand{\thefigure}{S1\_\arabic{figure}}
 \setcounter{figure}{0}

 \renewcommand{\thetable}{S1\_\arabic{table}}
 \setcounter{table}{0}
 \subfile{S1_Appendix}
\newpage
\FloatBarrier
\section{Additional Examples Appendix} 
\label{si:additional_ex}
 \renewcommand{\thefigure}{S2\_\arabic{figure}}
 \setcounter{figure}{0}

 \renewcommand{\thetable}{S2\_\arabic{table}}
 \setcounter{table}{0}
\subfile{S2_Appendix}

\newpage
\FloatBarrier
\section{Pillar Tracking Appendix} 
\label{si:pillar_track}
\renewcommand{\thefigure}{S3\_\arabic{figure}}
 \setcounter{figure}{0}

 \renewcommand{\thetable}{S3\_\arabic{table}}
 \setcounter{table}{0}

\subfile{S3_Appendix}

\FloatBarrier
\newpage
\bibliographystyle{unsrtnat}
\typeout{}
\bibliography{references}

\end{document}

%% file: S1_Appendix.tex
In this Supplementary Document, we elaborate on the ``Synthetic data generation'' Section and provide further details on our approach to validate the output of our cardiac microbundle tracking software. In the ``\nameref{si1:methods}'' Section, we describe the steps to generate Finite Element Analysis (FEA)-informed synthetic movies of beating microbundles with a known ground truth. We also provide details on our approach to collect ground truth data via manual tracking performed on real examples. Then in the ``\nameref{si1:res}'' Section, we present the results obtained by comparing the tracked output generated by the software and the ground truth data, and comment on the accuracy and robustness of our computational framework.

\section*{Methods} 
\label{si1:methods}

\subsection*{Synthetic dataset} 
\label{subsi1:syn_data}
In this Section, we describe additional details of our approach to creating realistic brightfield movies of beating cardiac microbundles, as depicted in Fig 2 of the main document. We note that our synthetic dataset is primarily derived from ``Type 1'' examples, since ``Type 1'' cardiac microbundle data is a more common approach in the literature.

For the ``Type 1'' synthetic data, we first start by identifying frames from which we can extract tissue textures. We systematically choose frames that represent valley and peak positions of the beating microbundle (i.e., the most relaxed or most contracted states) in order to sample multiple distinct textures (Fig 2a). We then manually trace the microbundle element in a valley frame. From the obtained mask, we extract the coordinates of the mask contour and crop out the background of the movie frame to obtain an image that contains just the tissue texture (Fig 2b).

\begin{figure}[h]
\begin{center}
\includegraphics[width=0.75\textwidth]{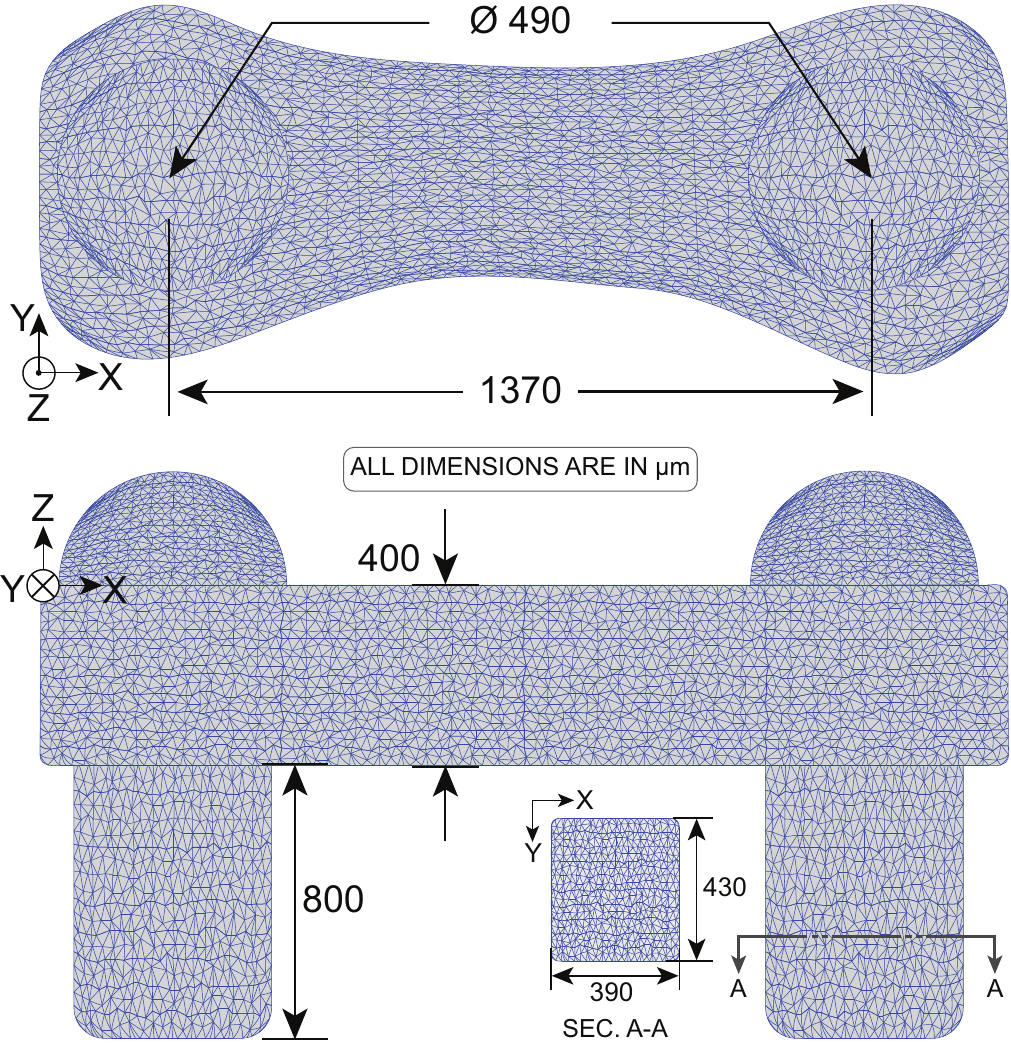}
\caption{\label{fig:Syn_Data_Mesh}Schematic representation of the ``Type 1'' microbundle mesh implemented in our Finite Element simulations.}

\end{center}
\end{figure}

For the Finite Element (FE) simulations, we generate a microbundle model based on the contour coordinates of a mask extracted from a single representative valley frame using Gmsh $4.10.5$ \cite{geuzaine2009gmsh} (Fig 2c \& Fig \ref{fig:Syn_Data_Mesh}). By connecting the points representing the mask contour, we obtain a 2D tissue surface, which we then extrude to a representative thickness for our real ``Type 1'' data. To simulate pillars, we follow the design and dimensions detailed in \cite{javor2021pillar}, which corresponds to a commonly implemented platform design. The generated mesh consists of $205,524$ tetrahedral elements.

In Fig 2d, we briefly summarize the main components of the FE model as implemented in FEniCS $2019.1.0$ \cite{alnaes2015fenics, logg2012automated}. Specifically, we model the cardiac microbundle as a nearly-incompressible hyperelastic material, following recent work in the literature \cite{pezzuto2014orthotropic, gurev2015high, finsberg2018efficient}. We describe the incompressible hyperelastic material with the strain energy density function $\Psi$ written as: 
\begin{align}
\Psi = \Psi(\mathbf{F}) - p(J-1) \label{incomp_psi}
\end{align}
where $\mathbf{F}$ is the deformation gradient tensor, $p$ is a Lagrange multiplier to impose the incompressibility constraint, and $J = det(\mathbf{F})$ is the volume ratio and should be equal to unity for full incompressibility \cite{holzapfel2002nonlinear}. The scalar $p$ acts as a reaction hydrostatic pressure resisting any volume change and can only be determined from the equilibrium equations and the boundary conditions. 

To relax the incompressibility constraint, we adopt the standard volumetric-isochoric decomposition of the deformation gradient tensor to obtain a more robust mathematical formulation of the constitutive model equations and avoid numerical difficulties such as  mesh locking that arise when finite element methods are implemented in the analysis of incompressible materials \cite{pezzuto2014orthotropic, weiss1996finite}. This representation can be derived following the multiplicative decomposition of the deformation gradient tensor into purely isochoric $\mathbf{F_{iso}}$ and purely volumetric $\mathbf{F_{vol}}$ components where $\mathbf{F} = \mathbf{F_{iso}}\mathbf{F_{vol}}$. As such we can write $\mathbf{F_{iso}} = J^{-1/3}\mathbf{I}$ and $\mathbf{F_{vol}} = J^{1/3}\mathbf{I}$.

The two components of the deformation gradient tensor contribute in an additive manner to the strain energy density function, where we can write: 
\begin{align}
\Psi(\mathbf{F}) = \Psi_{iso}(\mathbf{F_{iso}}) + \Psi_{vol}(\mathbf{F_{vol}}) \label{decomp_psi}
\end{align}
The isochoric component of the strain energy density function can be defined as any hyperelastic model. Here, we consider a Neo-Hookean material model:
\begin{align}
\Psi_{iso} = \frac{1}{2} \mu \big[tr\left(\mathbf{C_{iso}}\right) -3\big] \label{neo_hook_eq}
\end{align}
where $\mu$ is the shear modulus and $\mathbf{C_{iso}} = \mathbf{F^{T}_{iso}}\mathbf{F_{iso}} = J^{-2/3} \mathbf{F^{T}}\mathbf{F}$ with $\mathbf{C}$ being the right Cauchy-Green tensor. As for the volumetric strain energy, several forms are cited in the literature. In our simulation, we define $\Psi_{vol} = \frac{\kappa}{2} ln(J)^2$ where $\kappa$ is the bulk modulus of the material, which is orders of magnitude higher than the nearly-incompressible material's shear modulus. Finally, for nearly-incompressible formulation, incompressibility is implemented by the constraint relating the Lagrange multiplier $p$ and $\Psi_{vol}$ where $p = -d\Psi_{vol}(J)/dJ$.

In the numerical setting, we employ a mixed formulation finite element method with Taylor-Hood tetrahedral elements \cite{hood1974navier}, where the displacement field is represented by a continuous piecewise quadratic Lagrange shape function ($P2$) while the pressure field is represented by a continuous piecewise linear Lagrange shape function ($P1$).

Finally, to model the active properties of the cardiac tissue and its ability to actively contract and generate force without external loads, we implement the active strain approach \cite{ambrosi2012active}. In brief, this approach is based on a multiplicative decomposition of the deformation gradient tensor $\mathbf{F}$ into an active component $\mathbf{F_a}$ and an elastic component $\mathbf{F_e}$ \cite{pezzuto2014orthotropic, finsberg2018efficient}. For a transversely isotropic activation, $\mathbf{F_a}$ is defined as:
\begin{align}
\ \mathbf{F_a} = \left(1-\gamma \right)\mathbf{f_0} \otimes \mathbf{f_0} + \frac{1}{\sqrt{1-\gamma}}\left(\mathbf{I}- \mathbf{f_0} \otimes \mathbf{f_0}\right) \label{activ_Fa}
\end{align}
where $\gamma$ is the activation in the fiber direction and $\mathbf{f_0}$ is the fiber axis. The elastic part $\mathbf{F_e}$ is defined as $\mathbf{F_e} = \mathbf{F}\mathbf{F^{-1}_a}$.
Applying this to the isochoric deformation gradient, we get $\mathbf{(F_{iso})_e} = \mathbf{F_{iso}} \mathbf{F^{-1}_a}$. Finally, the strain energy density will be a function of $\mathbf{F_e}$ only, and following our previous decomposition, the Neo-Hookean material model can be written as: 
\begin{align}
\Psi_{iso} = \frac{1}{2} \mu \left[tr\left(\mathbf{C_{iso}}\right)_\mathbf{e} -3\right] \label{neo_elastic_psi_eq}
\end{align}
where $\left(\mathbf{C_{iso}}\right)_\mathbf{e} = \mathbf{(F^{T}_{iso})_e} \mathbf{(F_{iso})_e}$.

We define a uniform periodic time series activation ranging between $0$ and a maximum activation value (see Fig 2d - lower right inset) along the fiber axis as: 

\begin{equation}
\gamma = 
\begin{cases}
-A \cos(\omega t) \qquad & \gamma \ge 0.008\\
0 & \gamma < 0.008
\end{cases}\ \qquad  \text{with} \quad 0 \leq \omega \leq 2\pi
\end{equation}

To introduce slight variability to the periodic function, we add correlated Brownian (red) noise \cite{zhivomirov2018method}. The lower left and middle insets in Fig 2d provide a general depiction of the defined fiber direction. In our simulations, we linearly vary the fiber angle $\alpha$ defined with the global $X$-axis (horizontal) from a lower value of $9.33^{\circ}$ to a maximum of $15.33^{\circ}$ in two different tissue dimensions: 1) with respect to tissue depth (global $Z$-axis, into the page), where the fiber angle is $9.33^{\circ}$ on the top surface of the tissue ($Z = 0$) and becomes $15.33^{\circ}$ at the tissue's bottom ($Z = -400 \mu$m) and 2) with respect to tissue length (global $X$-axis, horizontal), where the fiber angle is $9.33^{\circ}$ on the left side of the tissue ($X = 0$) and becomes $15.33^{\circ}$ at the tissue's right end.
 
We also simulate spatially heterogeneous activation by including a circular passive region in the middle of the tissue where the activation is zero and increases gradually with distance from the periphery of the circle to reach the maximum defined activation value at the current step. We implement this by defining an inclusion ratio as: 

\begin{equation}
r = 
\begin{cases}
1 \qquad & d < r_{min}\\
\frac{r^2_{min}}{\beta r^2_{min} + (1-\beta)d } \qquad & d \ge r_{min}
\end{cases}\ \qquad  \text{with} \quad d = S_{\alpha}(\|P - p_1\|,...,\|P - p_n\|)
\end{equation}

where $r_{min}$ is the radius of the circular inclusion, $\alpha = -80$, $\beta = 0.9$, $P$ is the center of the inclusion, $p_1,...,p_n$ are the tissue spatial mesh coordinates, $\| \quad \|$ denotes the Euclidean distance, and $S_{\alpha}$ is the smooth maximum approximation function given by:
\begin{equation}
S_{\alpha}(x_1,...x_n) = \frac{\sum^n_{i=1}x_i e^{\alpha x_i}}{\sum^n_{i=1} e^{\alpha x_i}} \qquad \text{where} \quad S_{\alpha} \rightarrow min \quad \text{as} \quad \alpha \rightarrow -\infty
\end{equation}

and calculate the heterogeneous activation as $\gamma_{h} = \gamma(1 - r)$.

As for the pillars,  we define a passive compressible Neo-Hookean constitutive model as: 
\begin{equation}
\Psi=\frac{1}{2} \, \mu \,  \big[\textbf{F}:\textbf{F}-3-2\ln(\det\textbf{F})\big]+\frac{1}{2} \, \lambda \, \big[\frac{1}{2}[(\det\textbf{F})^{2}-1]-\ln(\det\textbf{F})\big]
\end{equation}
where $\textbf{F}$ is the deformation gradient, and $\mu$ and $\lambda$ are the Lam{\'e} parameters equivalent to Young's modulus $E$ and Poisson's ratio $\nu$ as $E = \mu \,  (3\lambda+2\mu)/(\lambda+\mu)$ and $\nu=\lambda/(2(\lambda+\mu))$. 
We specify a value of $0.47$ for the Poisson's ratio, and a value of $1.6$ MPa for the Young's modulus.

\begin{figure}[h]
\begin{center}
\includegraphics[width=1\textwidth]{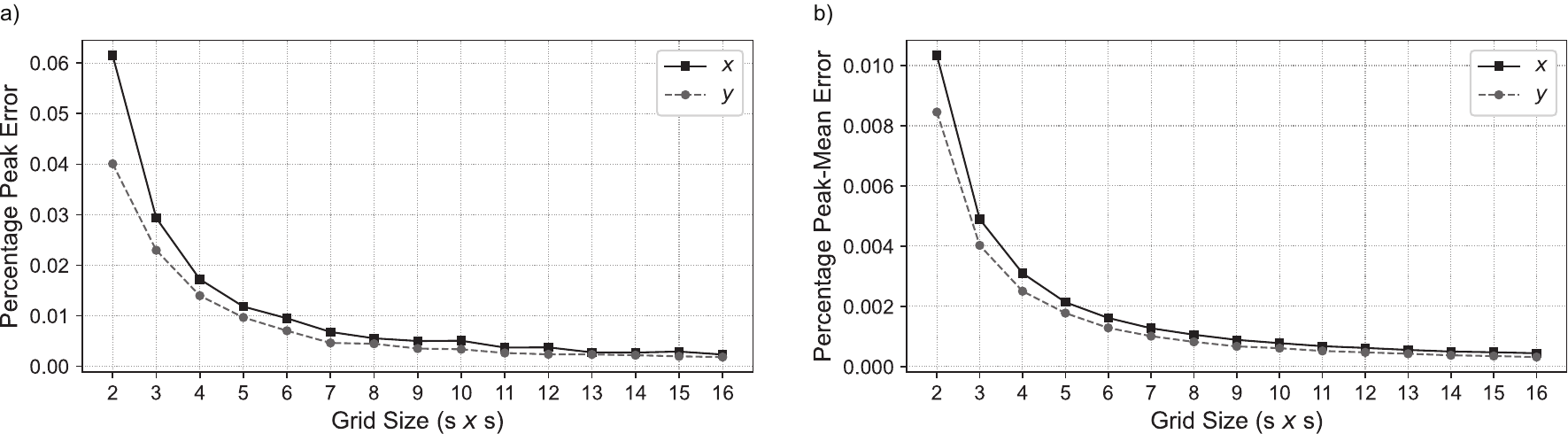}

\caption{\label{fig:conv_grid_size}Convergence of the percentage error of the $x$ and $y$ positions of the mesh cell centers in comparison to a ground truth with respect to grid size: (a) The peak (maximum) percentage error in all the $s \times s$ grids for each respective grid size; (b) The peak (maximum) of the mean percentage error in each grid of the total $s \times s$ grids.}
\end{center}
\end{figure}

From these Finite Element simulations, we extract the $X$, $Y$, and $Z$, positions of the mesh cell centers at the top surface of the microbundle for each step. We then estimate a projective transformation based on the initial and deformed positions of the mesh cell centers and warp the image texture accordingly using the ``warp'' transform function in the scikit-image $0.19.3$ Python library \cite{scikit_image}. To deform the synthetic texture with a heterogeneous transformation, we divide the domain into smaller slightly overlapping subdomains and deform each one according to a projective transformation calculated specifically for the subdomain. We perform a convergence study to decide on an appropriate number of subdomains and choose a grid size that results in the lowest error between the deformation gradient mapping obtained from FE simulations $\mathbf{F_{FEA}}$ and its equivalent that is obtained from approximating a projective transformation $\mathbf{F_{projective}}$ (Fig \ref{fig:conv_grid_size}). In our final implementation, we divide each synthetic texture domain into $16 \times 16$ subdomains. 

Overall, our synthetic dataset consists of $60$ generated movies of beating microbundle textures. We extract $90 \times 90$ pixel regions for the tissue textures from 3 different frames per movie ($2$ valleys and $1$ peak) and employ for this purpose, $5$ experimental movies from ``Type 1.'' Thus, we obtain $3 \times 5 = 15$ different base texture images. We then deform these extracted textures with FE results from $4$ different FE simulations: 1) homogeneous activation across the whole microbundle domain with the fiber direction varying linearly in the $X$ direction, 2) homogeneous activation across the whole microbundle domain with the fiber direction varying linearly in the $Z$ direction, 3) heterogeneous activation where the active microbundle domain has a passive inclusion in the middle with the fiber direction varying linearly in the $X$ direction, and finally, 4) heterogeneous activation where the active microbundle domain has a passive inclusion in the middle with the fiber direction varying linearly in the $Z$ direction. We briefly note that all the files required to reproduce our complete synthetic dataset of ``Type 1'' along with the dataset files are made available on Github (\href{GitHub} {https://github.com/HibaKob/SyntheticMicroBundle}).

As a final step, we add spatially correlated Perlin noise \cite{perlin1985noise}, to both make the dataset more realistic and to create more challenging examples to test the robustness of our computational framework. We add Perlin noise at varying levels of magnitude defined as a ratio of the maximum image intensity, and with a varying number of octaves as shown in Fig \ref{fig:perlin_syn}. From these examples, we see that the addition of noise to the synthetic images leads to accrued drift that is similar to what we observe when tracking displacements in real data. 

\begin{figure}[h]
\begin{center}
\includegraphics[width=1\textwidth]{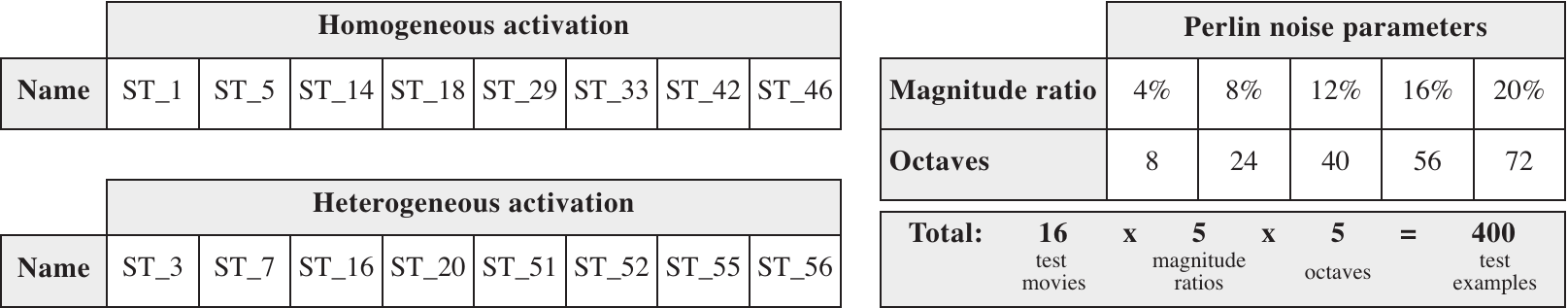}
\caption{\label{fig:data_syn}A tabulated summary of the synthetic data along with the added Perlin noise parameters implemented for code validation. Here, ``ST'' refers to ``synthetic texture.''}    
\end{center}
\end{figure}

For code testing and validation, we select a subset of the entire synthetic dataset consisting of $16$ original synthetic movies, with half of this testing data corresponding to homogeneous activation and the other half to heterogeneous activation, as depicted in Fig \ref{fig:data_syn}. For each of these $16$ examples, we add Perlin noise of $5$ different magnitude ratios and $5$ different octaves (Fig \ref{fig:data_syn}). In total, we run our code on $400$ noisy test examples with known ground truth. We provide the results in the ``\nameref{si1:res}'' Section. 

Finally, we include a single synthetic example with known ground truth based on ``Type 2'' data. The main steps to generate this example are similar to the ones followed when generating ``Type 1'' synthetic data; that is: 1) extracting the microbundle texture, 2) running a FE simulation mimicking the microbundle behavior, and finally, 3) warping the extracted texture with displacement results obtained from the FE simulations. These FE simulations are based on a tissue-specific model that has a detailed representation of fibers and cardiomyocytes as described in Jilberto et al. \cite{jilberto2023computational}, which makes the kinematic behavior similar to that of real tissues. We provide this synthetic video with the supplementary material as ``S1 Movie'' and the performance of our framework on this example in the ``\nameref{si1:res}'' Section.

\newpage
\FloatBarrier
\begin{figure}[ht!]
\begin{center}
\includegraphics[width=1\textwidth]{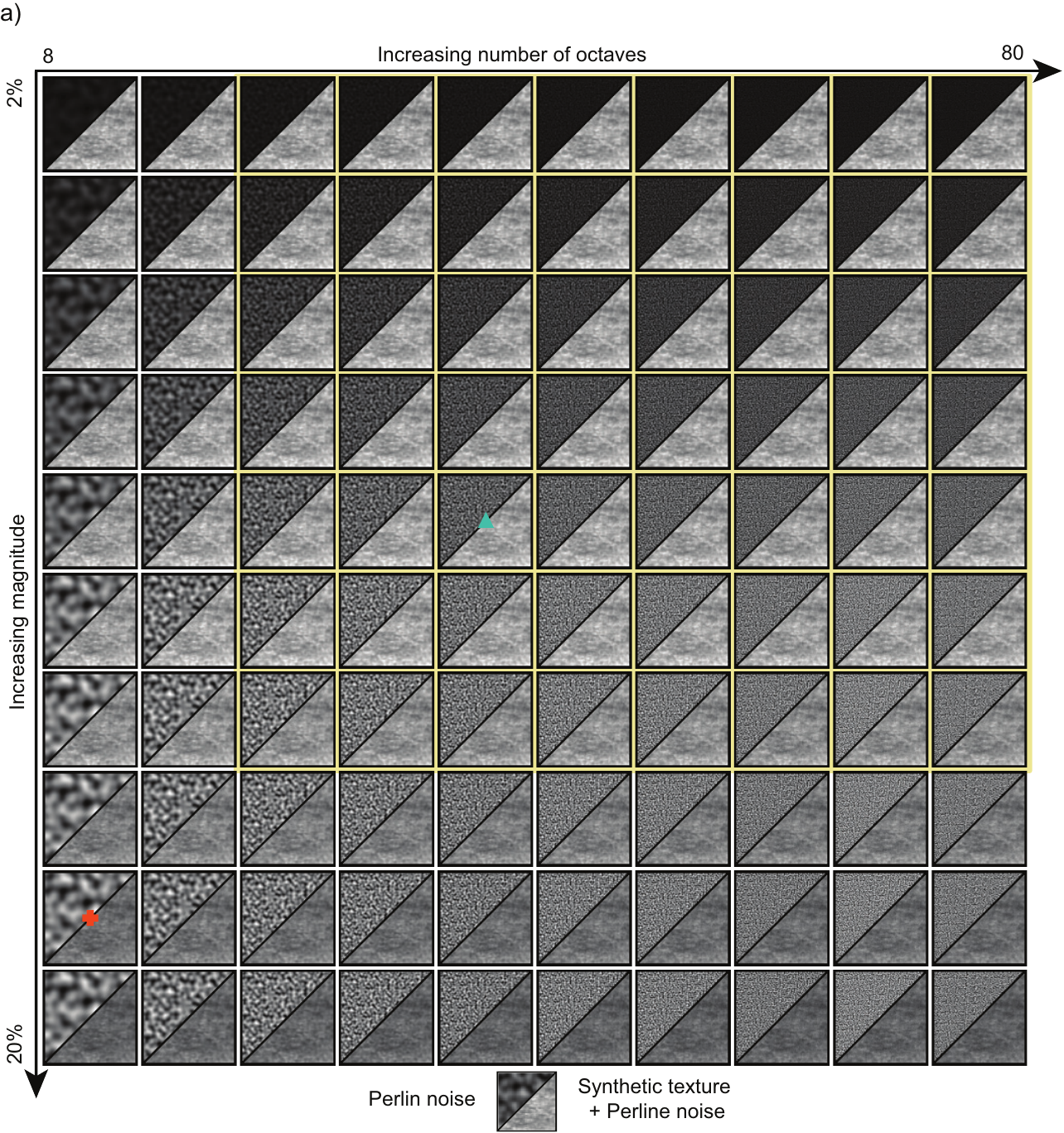}
\end{center}
\end{figure}

\newpage
\FloatBarrier
\begin{figure}[ht!]
\begin{center}
\includegraphics[width=1\textwidth]{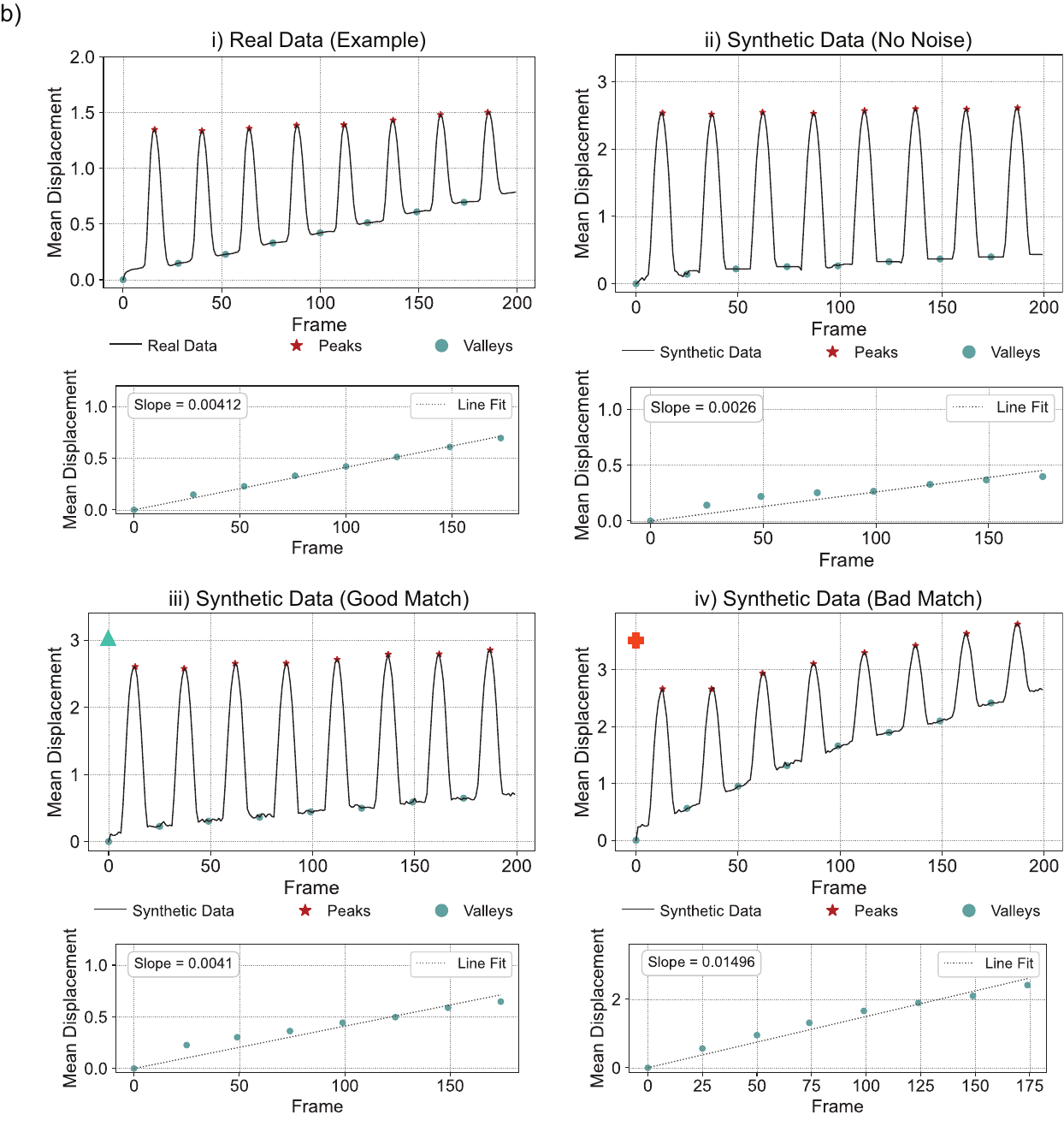}
\end{center}
\end{figure}

\newpage
\FloatBarrier
\clearpage
\begin{figure}[ht!]
\begin{center}
\includegraphics[width=0.98\textwidth]{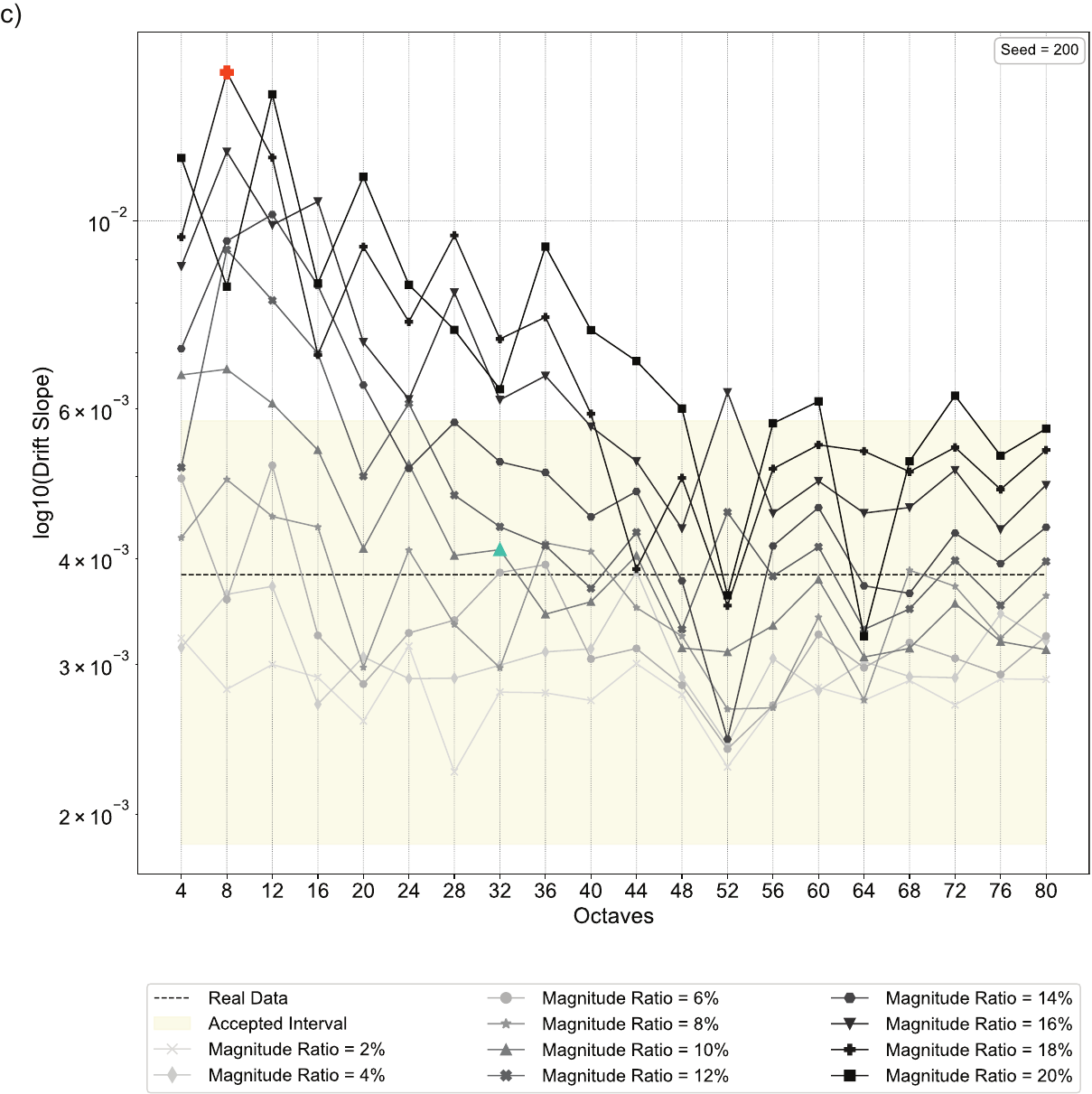}
\caption{\label{fig:perlin_syn}To make our synthetic dataset more challenging and more representative of real experimental data, we add Perlin noise to the synthetic microbundle textures: (a) the upper triangle shows an illustration of Perlin noise (a function of magnitude and octave), and the lower triangle shows the Perlin noise superimposed on the synthetic microbundle texture; (b) time series plots of (i) an example of tracked real data showing the drift in valleys along with the value of the slope of the linear fit of the valleys, (ii) an example of synthetic data without the addition of Perlin noise where it is clear that the drift in valleys is negligible, (iii) noisy synthetic data generated with $32$ octaves and $10\%$ magnitude ratio for which the slope of the drift is a good match to real data, (iv) an example of excessive distortion to the time series caused by the addition of Perlin noise generated with $18$ octaves and $8\%$ magnitude ratio; (c) a quantitative comparison of the drift in valleys (measured by the base $10$ logarithm of the slope of the drift) with respect to the octave and magnitude ratio. From this analysis, we observe that there are a range of Perlin noise parameters (highlighted with a yellow background) that lead to drift behavior that is consistent with the average observed in $5$ real examples of “Type 1” data. From this investigation, we: 1) show that adding Perlin noise to our synthetic data leads to synthetic data that better recapitulates the challenges associated with the real experimental data, and 2) identify the characteristics of the Perlin noise that we will use in our validation dataset. 
}
\end{center}
\end{figure}

\subsection*{Manual tracking} 
\label{subsi1:manual_track}
Another approach to validate our computational pipeline is to track the displacements of manually selected points and compare the results to those obtained by our tracking software. Specifically, we track the positions of $30$ points across a single beat. We perform the manual tracking on two different examples of ``Type 2'' data, where each example is tracked by $2$ different users (Fig \ref{fig:man_track} ``Case 1'' \& ``Case 2''). We include the results of this manual tracking validation the ``\nameref{si1:res}'' in Section.

\section*{Results} 
\label{si1:res}

\subsection*{Validation against synthetic data} 
\label{subsubsi1:syn_res}
\begin{figure}[p]
\begin{center}
\includegraphics[width=0.9\textwidth]{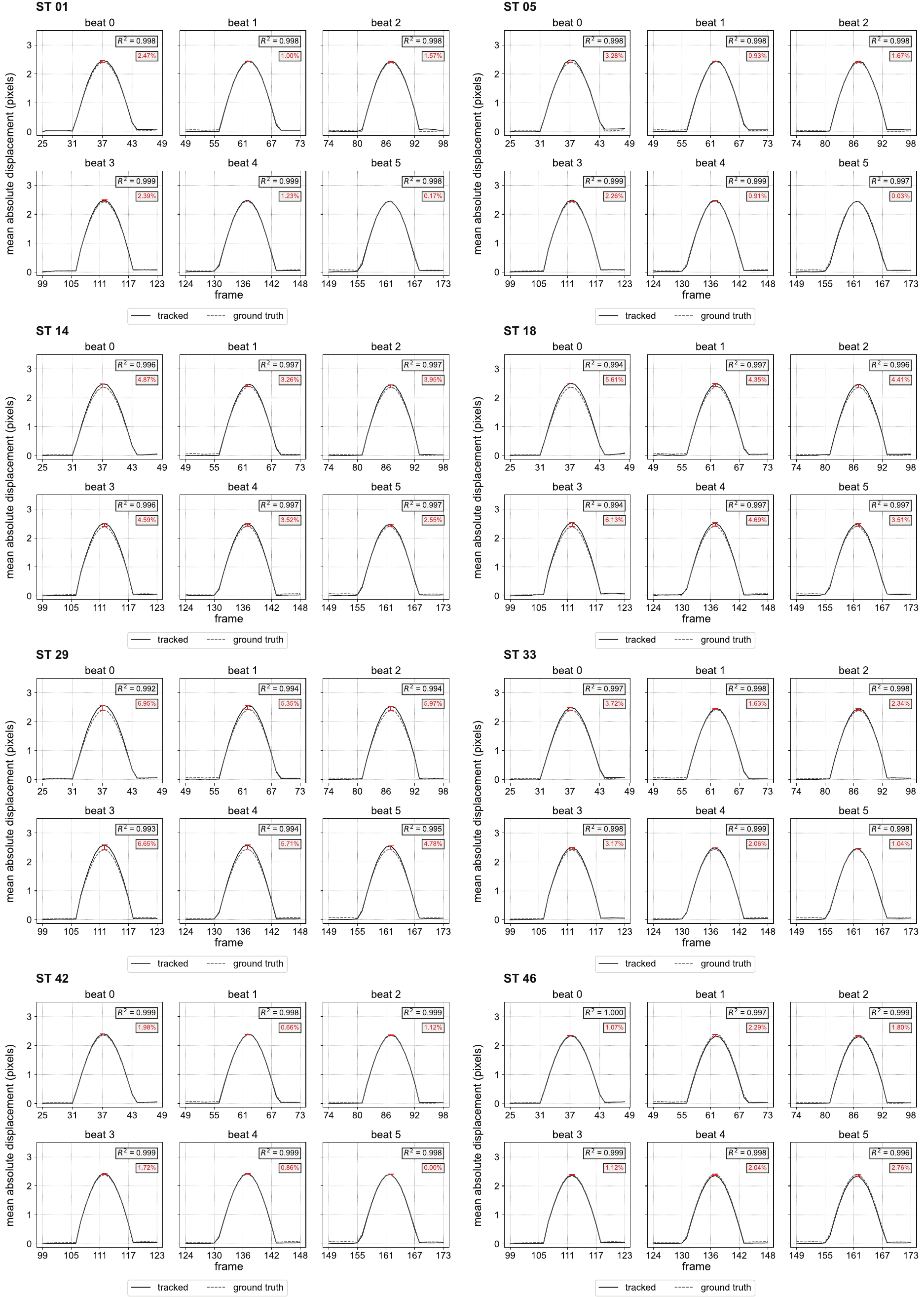}
\caption{\label{fig:homog_MAD}Error in mean absolute displacement for synthetic data of ``Type 1'' based on FE simulations with homogeneous activation.}    
\end{center}
\end{figure}

\begin{figure}[p]
\begin{center}
\includegraphics[width=0.9\textwidth]{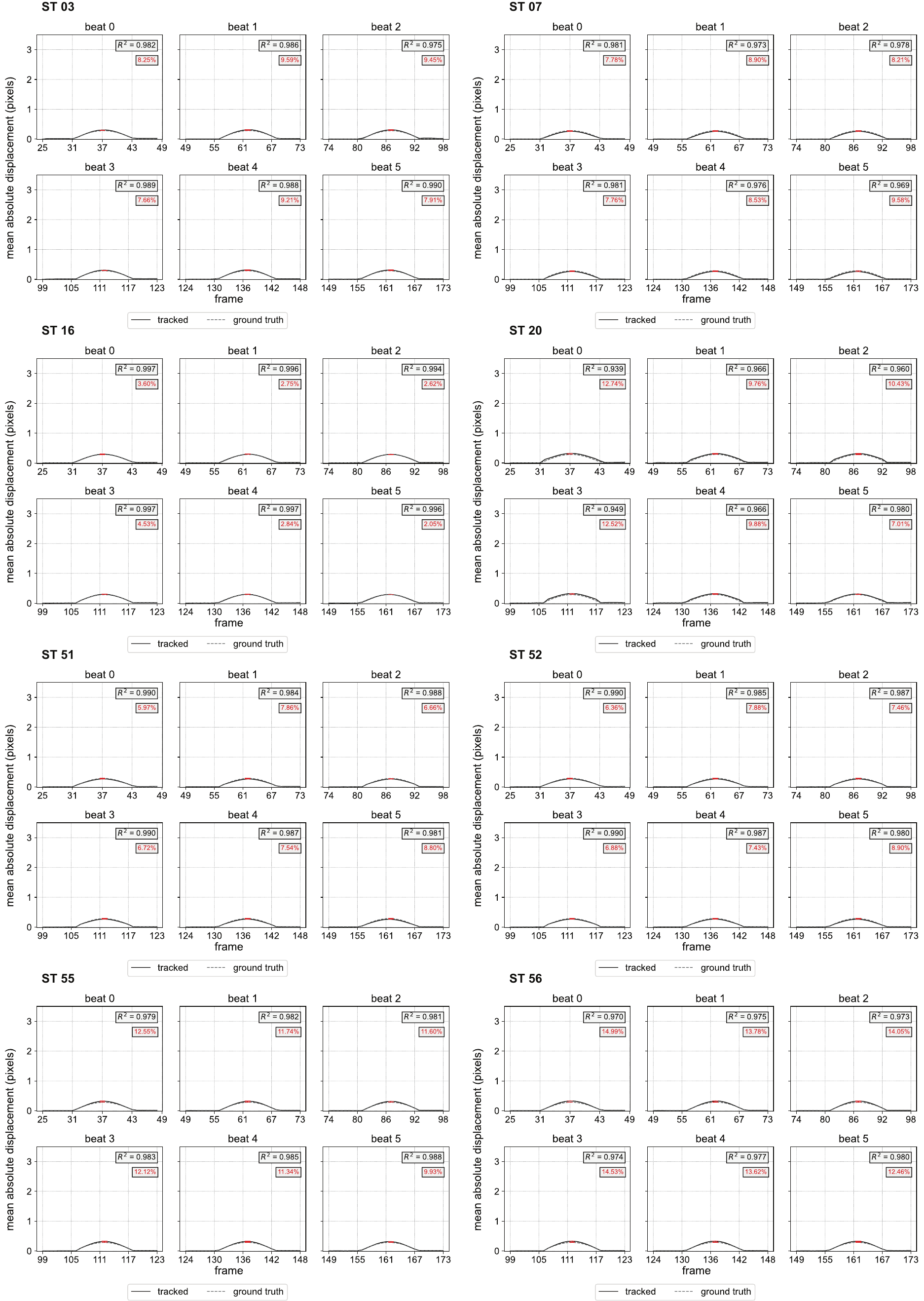}
\caption{\label{fig:hetero_MAD}Error in mean absolute displacement for synthetic data of ``Type 1'' based on FE simulations with heterogeneous activation.}    
\end{center}
\end{figure}

\begin{figure}[p]
\begin{center}
\includegraphics[width=0.9\textwidth]{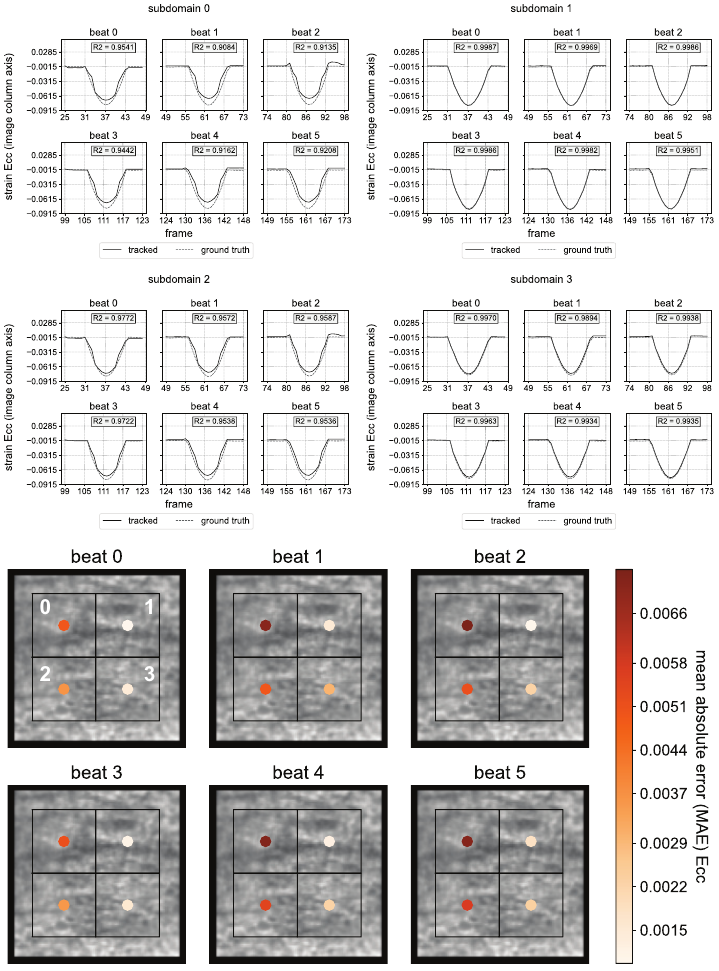}
\caption{\label{fig:ST1_error}Error in $E_{cc}$ strain for {\fontfamily{pcr}\selectfont``SyntheticTextures\_1''}. The numbers $0, 1, 2, 3$ shown on beat 0 MAE plot indicate the subdomain number.}    
\end{center}
\end{figure}

\begin{figure}[p]
\begin{center}
\includegraphics[width=0.9\textwidth]{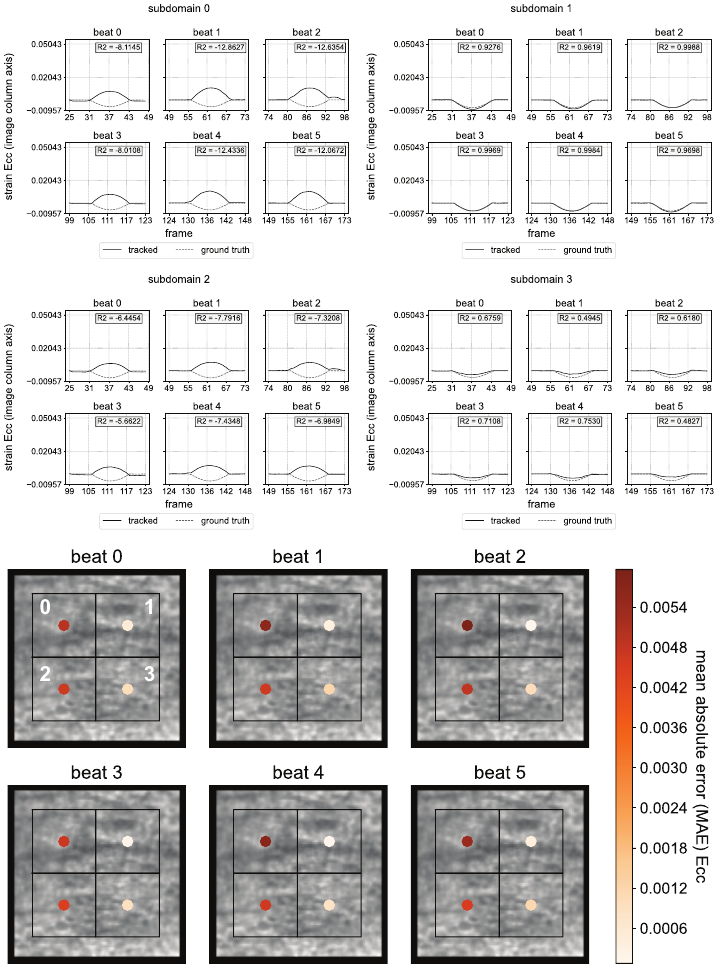}
\caption{\label{fig:ST3_error}Error in $E_{cc}$ strain for {\fontfamily{pcr}\selectfont``SyntheticTextures\_3''}. The numbers $0, 1, 2, 3$ shown on beat 0 MAE plot indicate the subdomain number.}    
\end{center}
\end{figure}

\begin{figure}[p]
\begin{center}
\includegraphics[width=0.9\textwidth]{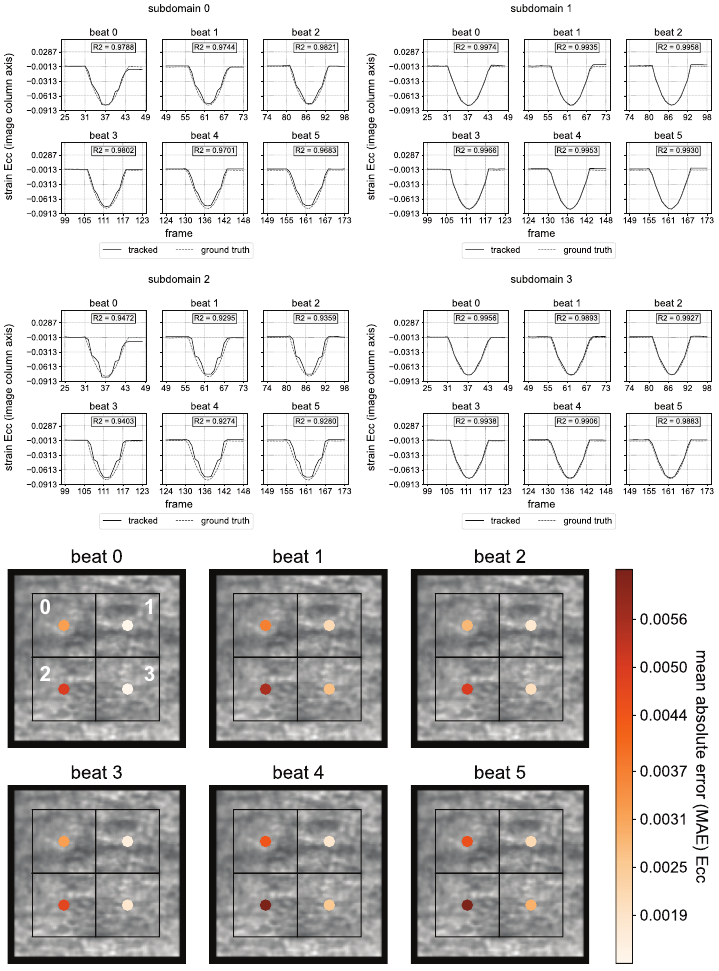}
\caption{\label{fig:ST5_error}Error in $E_{cc}$ strain for {\fontfamily{pcr}\selectfont``SyntheticTextures\_5''}. The numbers $0, 1, 2, 3$ shown on beat 0 MAE plot indicate the subdomain number.}    
\end{center}
\end{figure}

\begin{figure}[p]
\begin{center}
\includegraphics[width=0.9\textwidth]{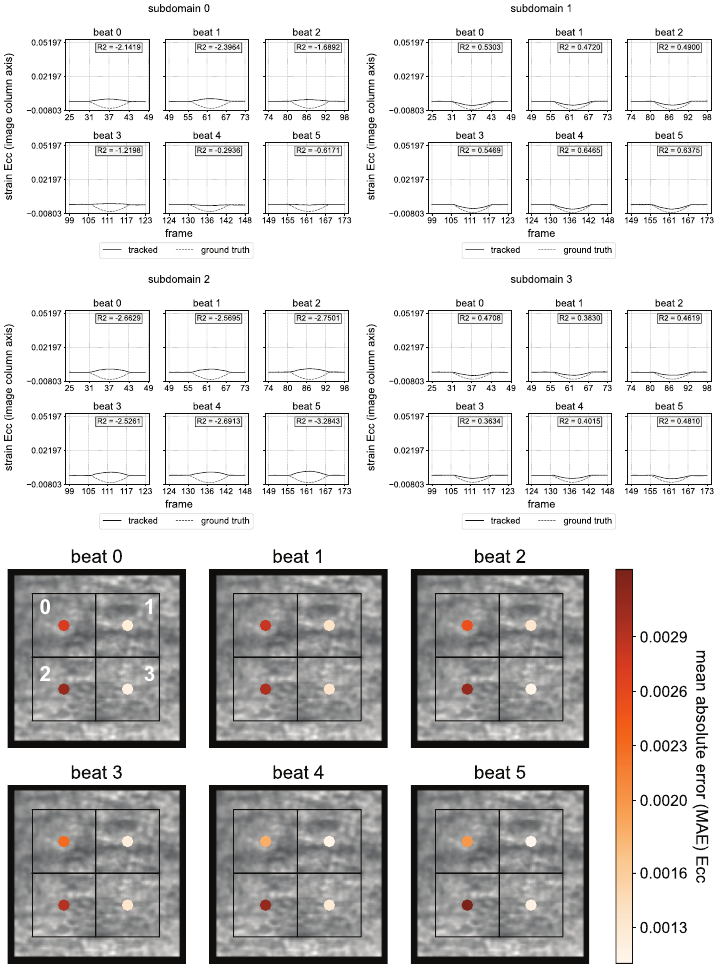}
\caption{\label{fig:ST7_error}Error in $E_{cc}$ strain for {\fontfamily{pcr}\selectfont``SyntheticTextures\_7''}. The numbers $0, 1, 2, 3$ shown on beat 0 MAE plot indicate the subdomain number.}    
\end{center}
\end{figure}

\begin{figure}[p]
\begin{center}
\includegraphics[width=0.9\textwidth]{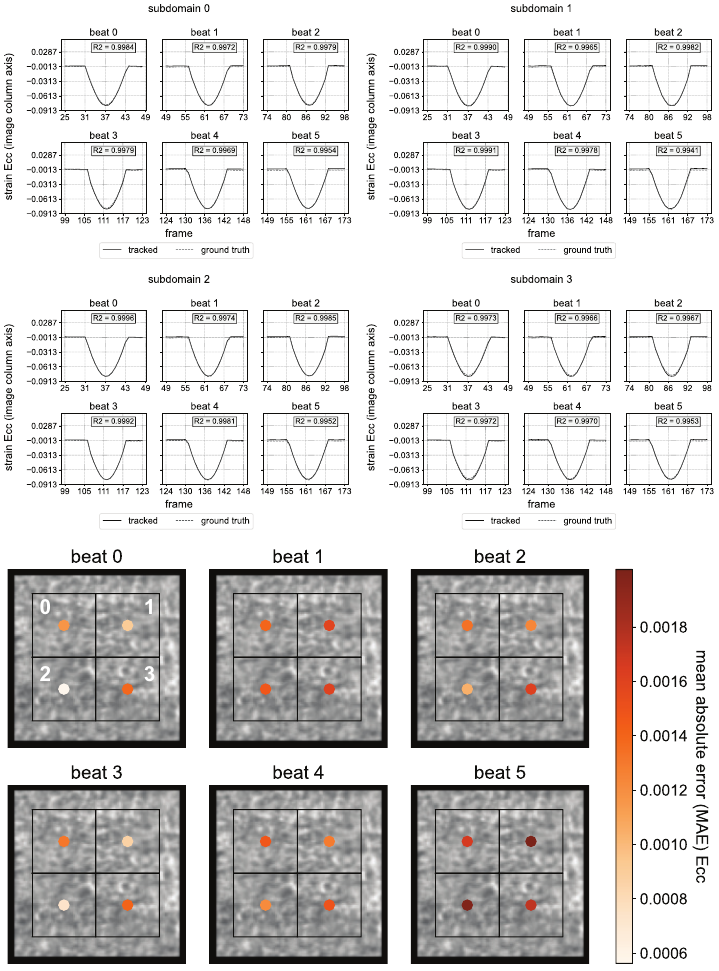}
\caption{\label{fig:ST14_error}Error in $E_{cc}$ strain for {\fontfamily{pcr}\selectfont``SyntheticTextures\_14''}. The numbers $0, 1, 2, 3$ shown on beat 0 MAE plot indicate the subdomain number.}    
\end{center}
\end{figure}

\begin{figure}[p]
\begin{center}
\includegraphics[width=0.9\textwidth]{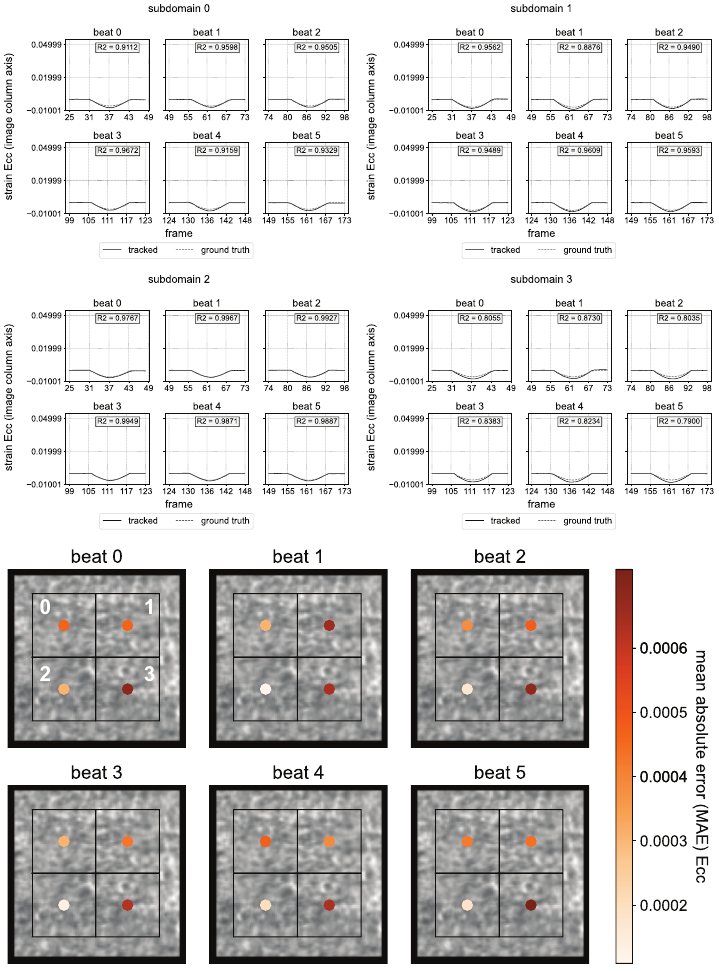}
\caption{\label{fig:ST16_error}Error in $E_{cc}$ strain for {\fontfamily{pcr}\selectfont``SyntheticTextures\_16''}. The numbers $0, 1, 2, 3$ shown on beat 0 MAE plot indicate the subdomain number.}    
\end{center}
\end{figure}

\begin{figure}[p]
\begin{center}
\includegraphics[width=0.9\textwidth]{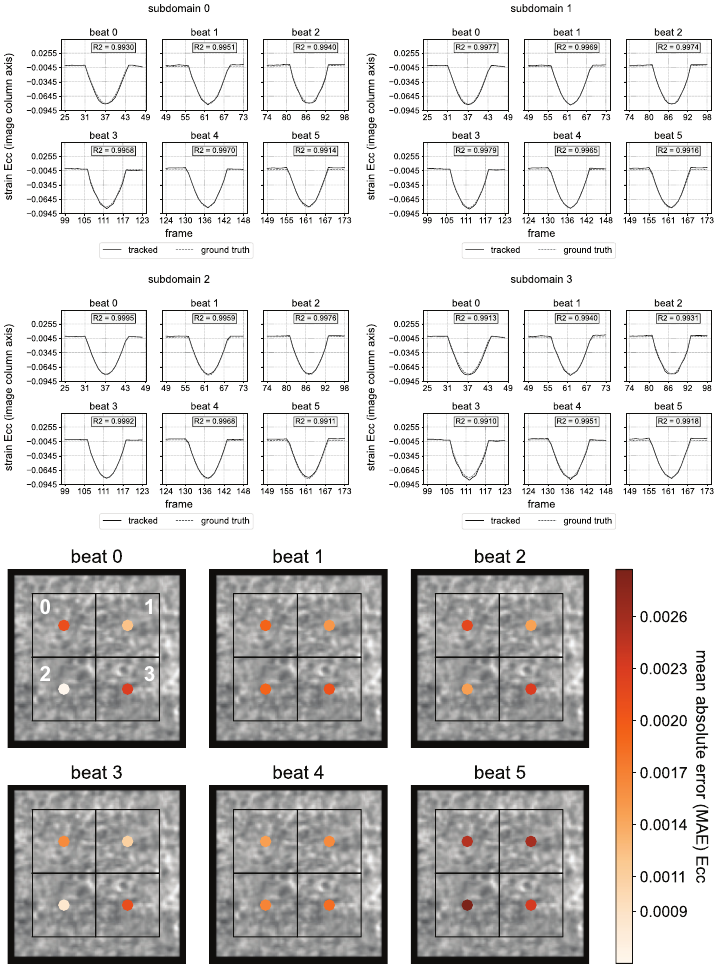}
\caption{\label{fig:ST18_error}Error in $E_{cc}$ strain for {\fontfamily{pcr}\selectfont``SyntheticTextures\_18''}. The numbers $0, 1, 2, 3$ shown on beat 0 MAE plot indicate the subdomain number.}    
\end{center}
\end{figure}

\begin{figure}[p]
\begin{center}
\includegraphics[width=0.9\textwidth]{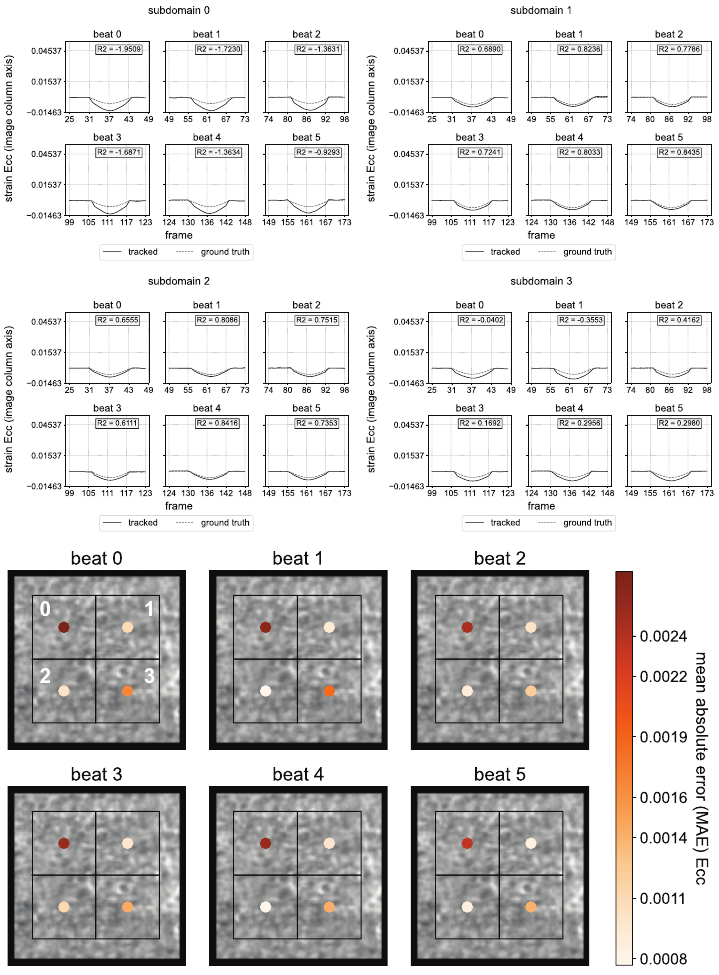}
\caption{\label{fig:ST20_error}Error in $E_{cc}$ strain for {\fontfamily{pcr}\selectfont``SyntheticTextures\_20''}. The numbers $0, 1, 2, 3$ shown on beat 0 MAE plot indicate the subdomain number.}    
\end{center}
\end{figure}

\begin{figure}[p]
\begin{center}
\includegraphics[width=0.9\textwidth]{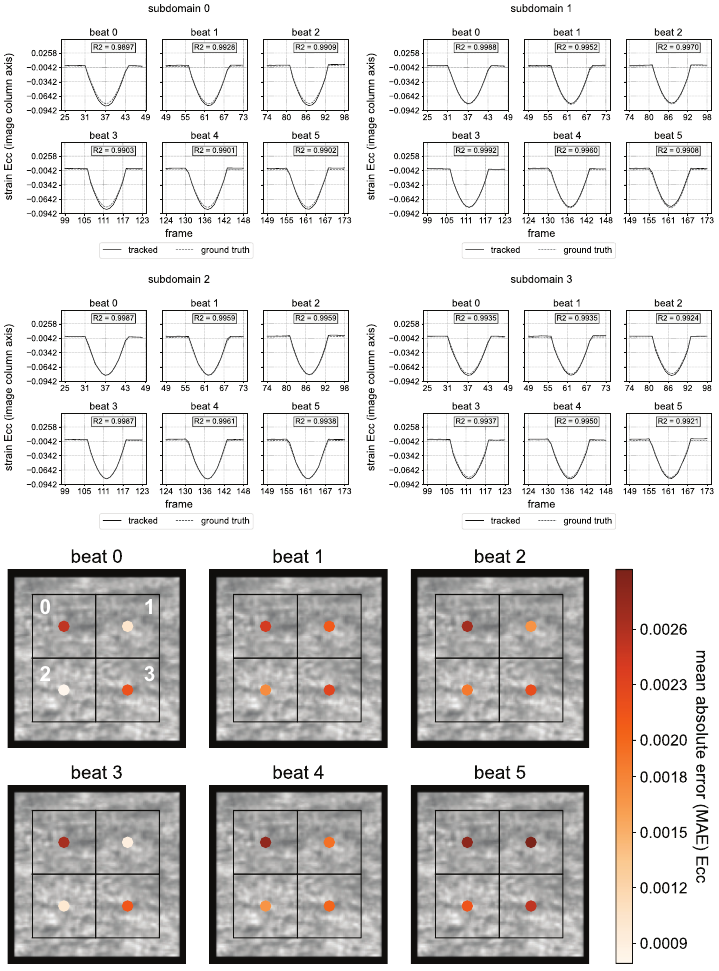}
\caption{\label{fig:ST29_error}Error in $E_{cc}$ strain for {\fontfamily{pcr}\selectfont``SyntheticTextures\_29''}. The numbers $0, 1, 2, 3$ shown on beat 0 MAE plot indicate the subdomain number.}    
\end{center}
\end{figure}

\begin{figure}[p]
\begin{center}
\includegraphics[width=0.9\textwidth]{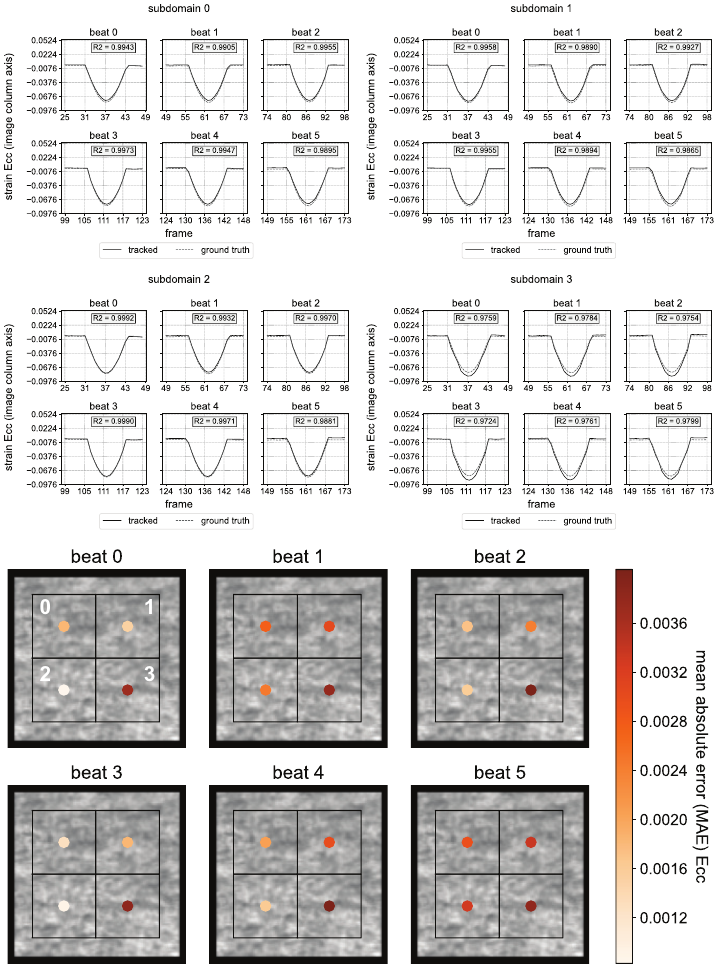}
\caption{\label{fig:ST33_error}Error in $E_{cc}$ strain for {\fontfamily{pcr}\selectfont``SyntheticTextures\_33''}. The numbers $0, 1, 2, 3$ shown on beat 0 MAE plot indicate the subdomain number.}    
\end{center}
\end{figure}

\begin{figure}[p]
\begin{center}
\includegraphics[width=0.9\textwidth]{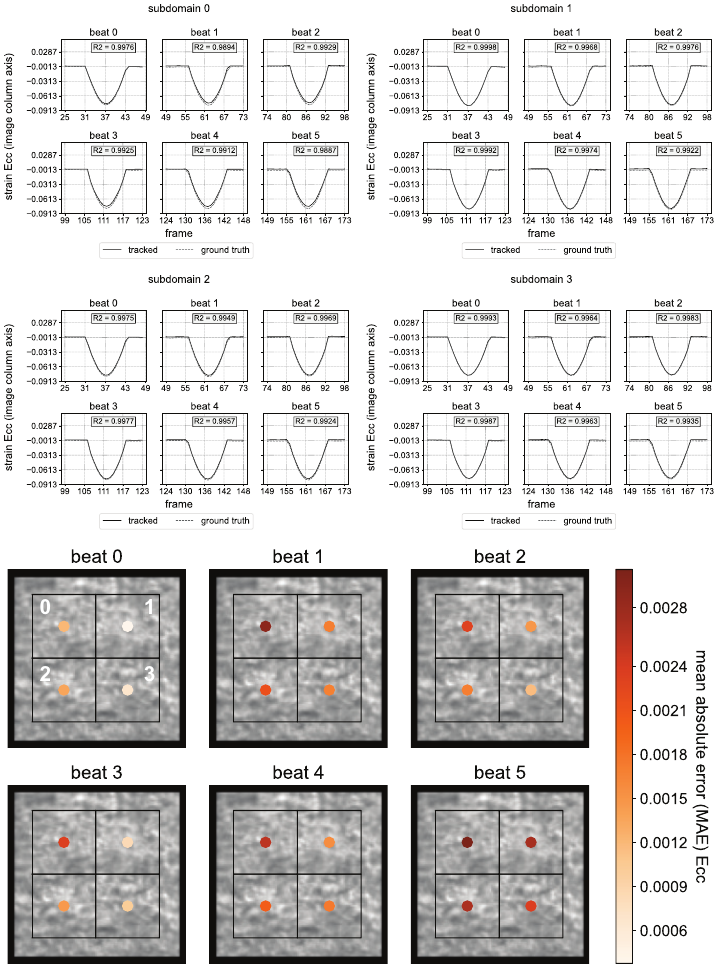}
\caption{\label{fig:ST42_error}Error in $E_{cc}$ strain for {\fontfamily{pcr}\selectfont``SyntheticTextures\_42''}. The numbers $0, 1, 2, 3$ shown on beat 0 MAE plot indicate the subdomain number.}    
\end{center}
\end{figure}

\begin{figure}[p]
\begin{center}
\includegraphics[width=0.9\textwidth]{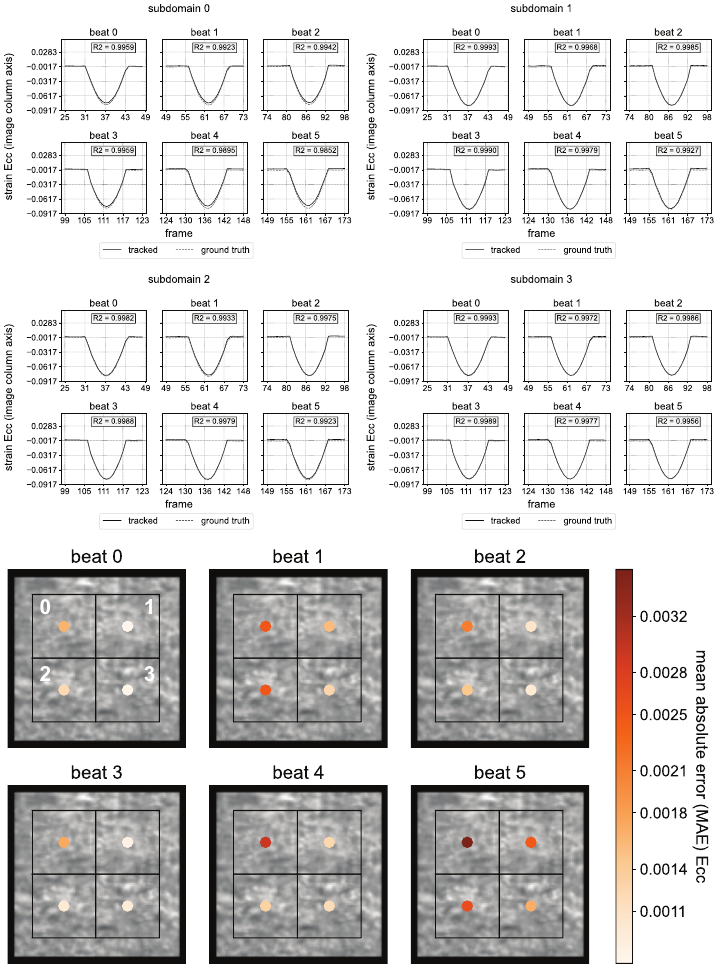}
\caption{\label{fig:ST46_error}Error in $E_{cc}$ strain for {\fontfamily{pcr}\selectfont``SyntheticTextures\_46''}. The numbers $0, 1, 2, 3$ shown on beat 0 MAE plot indicate the subdomain number.}    
\end{center}
\end{figure}

\begin{figure}[p]
\begin{center}
\includegraphics[width=0.9\textwidth]{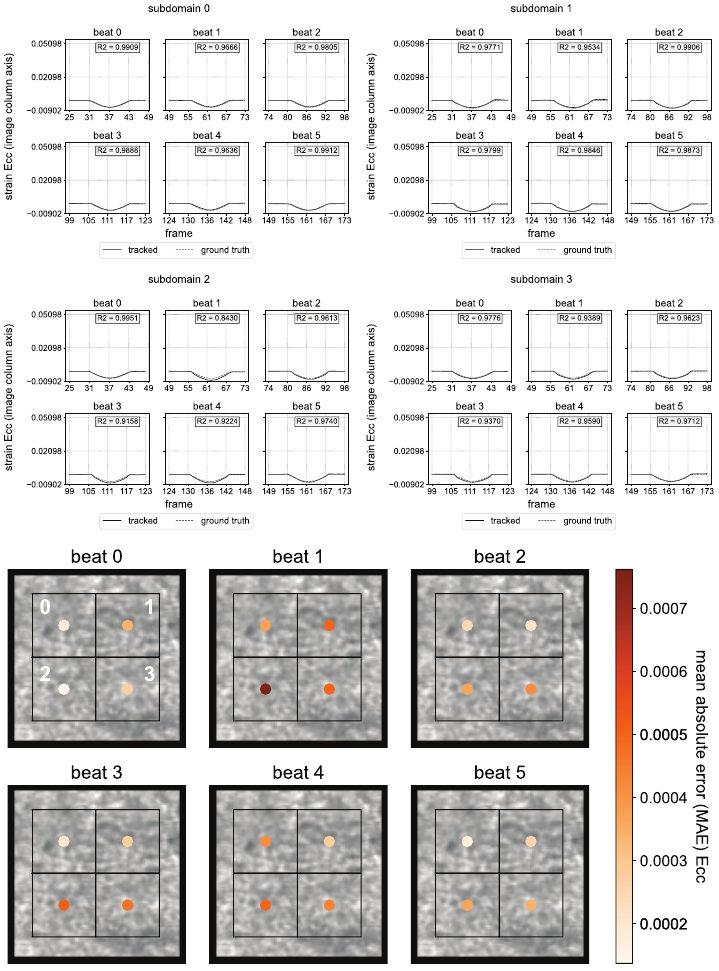}
\caption{\label{fig:ST51_error}Error in $E_{cc}$ strain for {\fontfamily{pcr}\selectfont``SyntheticTextures\_51''}. The numbers $0, 1, 2, 3$ shown on beat 0 MAE plot indicate the subdomain number.}    
\end{center}
\end{figure}

\begin{figure}[p]
\begin{center}
\includegraphics[width=0.9\textwidth]{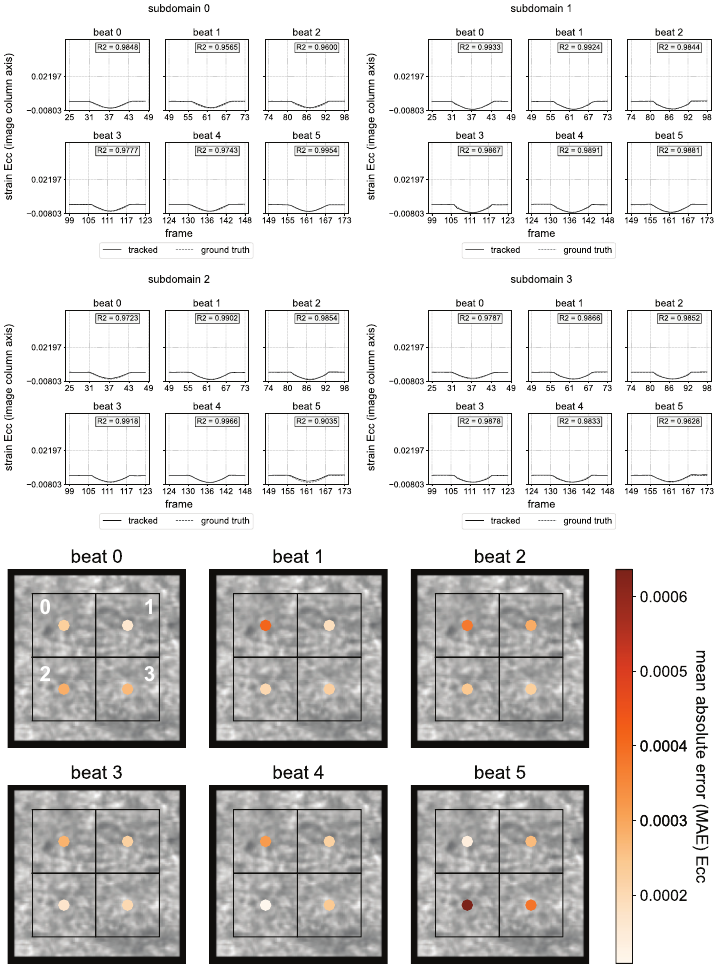}
\caption{\label{fig:ST52_error}Error in $E_{cc}$ strain for {\fontfamily{pcr}\selectfont``SyntheticTextures\_52''}. The numbers $0, 1, 2, 3$ shown on beat 0 MAE plot indicate the subdomain number.}    
\end{center}
\end{figure}

\begin{figure}[p]
\begin{center}
\includegraphics[width=0.9\textwidth]{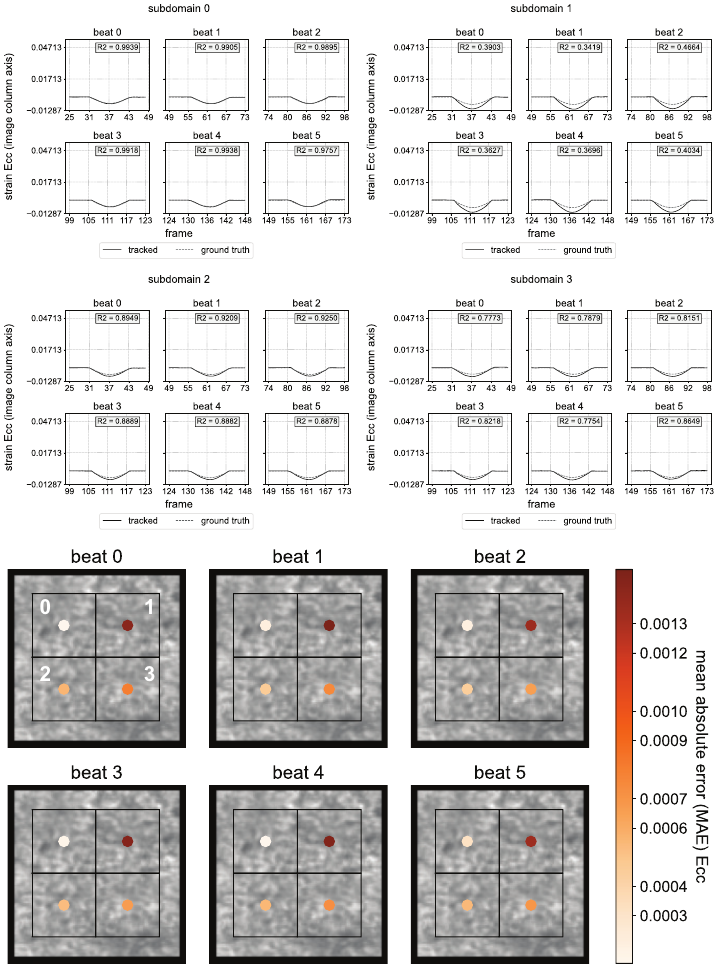}
\caption{\label{fig:ST55_error}Error in $E_{cc}$ strain for {\fontfamily{pcr}\selectfont``SyntheticTextures\_55''}. The numbers $0, 1, 2, 3$ shown on beat 0 MAE plot indicate the subdomain number.}    
\end{center}
\end{figure}

\begin{figure}[p]
\begin{center}
\includegraphics[width=0.9\textwidth]{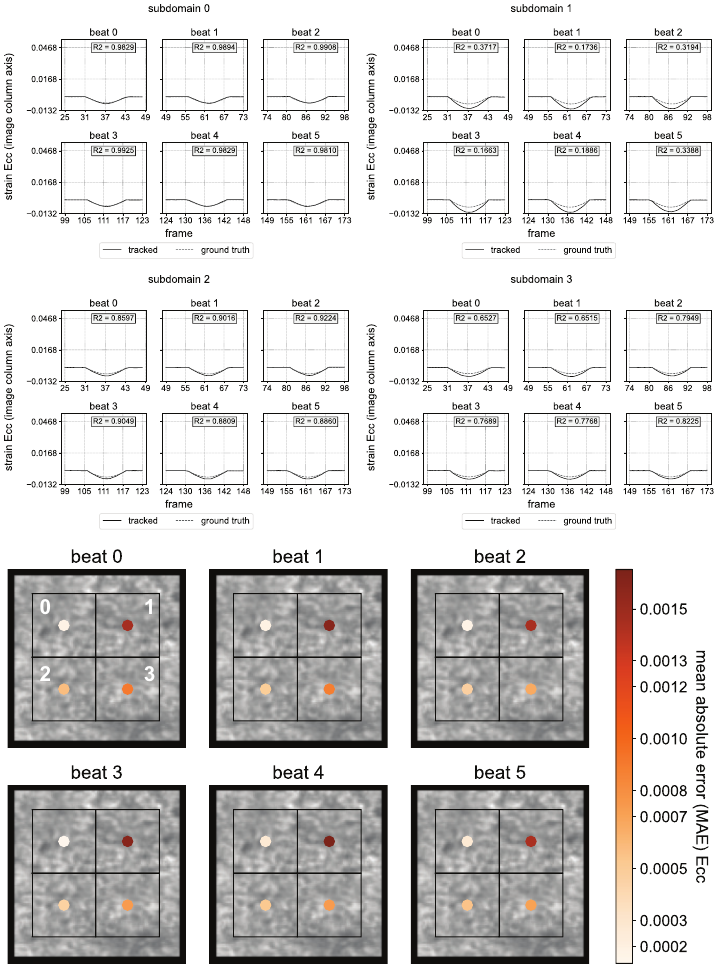}
\caption{\label{fig:ST56_error}Error in $E_{cc}$ strain for {\fontfamily{pcr}\selectfont``SyntheticTextures\_56''}. The numbers $0, 1, 2, 3$ shown on beat 0 MAE plot indicate the subdomain number.}    
\end{center}
\end{figure}

Here, we provide the results of validating ``MicroBundleCompute'' against synthetic data of beating microbundles. 
For the generated data based on ``Type 1'' data without any added Perlin noise, we show the computed errors for both mean absolute displacement (MAD) (Figs \ref{fig:homog_MAD} and \ref{fig:hetero_MAD}) and subdomain-averaged $E_{cc}$ strain (strain in the column-column or horizontal-horizontal direction) (Table \ref{tab:full_MAE} and Figs \ref{fig:ST1_error} -- \ref{fig:ST56_error}) obtained by comparing the tracking software output to the known ground truth. For the $16$ original validation examples, we base our error analysis on $4$ basic metrics: 1) percentage error at the peak MAD, 2) coefficient of determination ($R^{2}$) between tracked and ground truth MAD for each beat, 3) coefficient of determination ($R^{2}$) between tracked and ground truth subdomain-averaged $E_{cc}$ for each beat per subdomain, and 4) the mean absolute error (MAE) in $E_{cc}$ per subdomain per beat. 

For synthetic examples that are based on homogeneous activation simulations, where the resulting microbundle contractions exceed a single pixel, the percentage errors at peak MAD fall between $2\%$ and $7\%$ for all examples (Table \ref{tab:sum_MAD}). Overall, as shown in Table \ref{tab:sum_MAD}, the $R^2$ values indicate good agreement between the tracked and ground truth mean absolute displacement profiles, with $0.992$ being the lowest $R^2$ value at the maximum error. For \textit{sub}-pixel displacements, however, the peak MAD error range rises to $5\%$ -- $15\%$ and the lowest $R2$ value drops to $0.939$ (Table \ref{tab:sum_MAD}). As such, we consider tracked \textit{sub}-pixel displacements to be of lower fidelity and advise the user of ``MicroBundleCompute'' software to interpret results derived from \textit{sub}-pixel displacements with caution.  

For the $E_{cc}$ strain outputs, the computed errors are subdomain dependent. Overall, consistent with the behavior observed for MAD errors, synthetic examples based on homogeneous activation simulations exhibit significantly lower strain errors, with a maximum MAE of $9.2\%$ of the peak ground truth $E_{cc}$ in the corresponding subdomain versus an extreme value of $98\%$ for \textit{sub}-pixel displacement simulations (Table \ref{tab:sum_MAE}). In Table \ref{tab:full_MAE}, we present the complete summary of the MAE $E_{cc}$ strain validation performed on all synthetic examples.

\begin{table}[ht]
\begin{center}
\caption{\label{tab:sum_MAD}\textbf{Summary of the mean absolute displacement validation errors for synthetic data without any added Perlin noise.}} 
\includegraphics[width=0.6\textwidth]{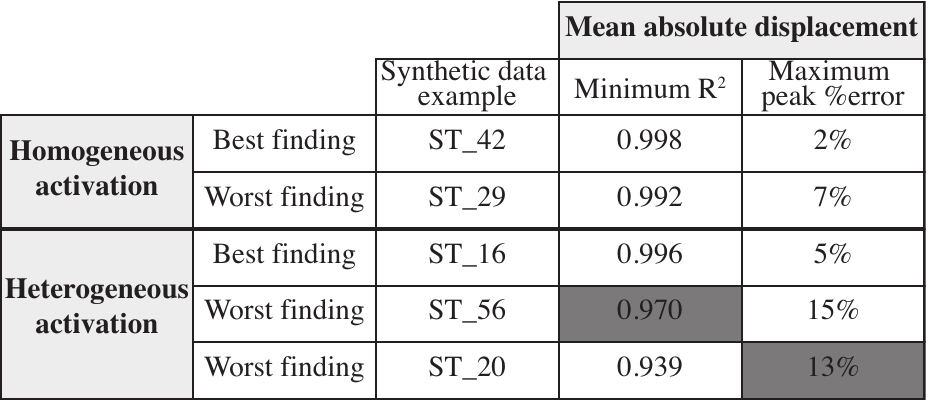}
\end{center}
\end{table}

\begin{table}[ht]
\begin{center}
\caption{\label{tab:sum_MAE}\textbf{Summary of the mean absolute $E_{cc}$ strain validation errors for synthetic data without any added Perlin noise.}}
\includegraphics[width=1\textwidth]{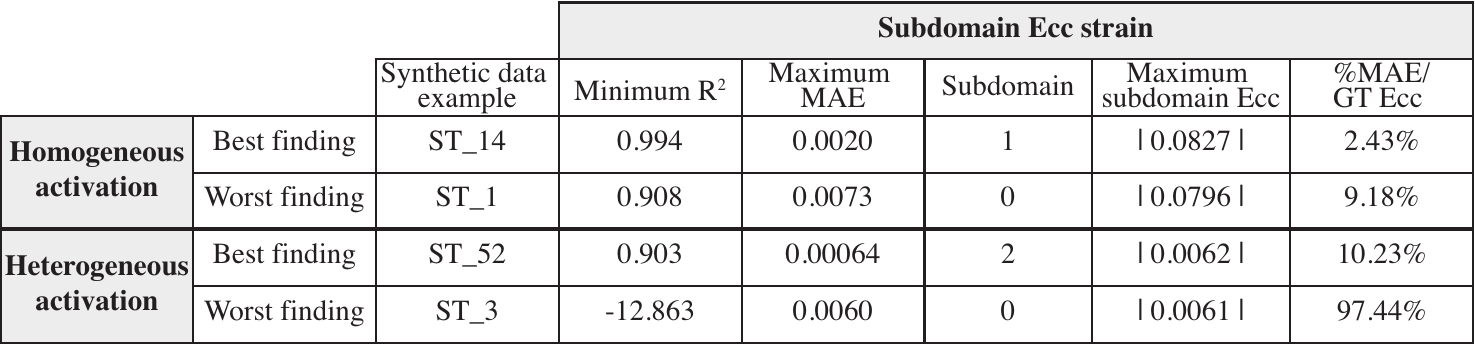} 
\end{center}
\end{table}

\begin{table}[ht]
\begin{center}
\caption{\label{tab:full_MAE}{\bf Summary of the mean absolute $E_{cc}$ strain validation errors for all synthetic data without any added Perlin noise. (Corresponds to Figs \ref{fig:ST1_error}--\ref{fig:ST56_error}.)}}
\begin{tabular}{|L|L|L|L|L|L|}
\hline
Synthetic data example & Maximum MAE & R2       & Subdomain & Maximum subdomain Ecc & \%MAE/ GT Ecc \\ \hline
ST\_1                      & 7.30E-03    & 0.9135   & 0         & $| 7.96\text{E-}02 |$              & 9.18\%          \\ \hline
ST\_3*                      & 5.96E-03    & -12.6354 & 0         & $| 6.12\text{E-}03 |$              & 97.44\%         \\ \hline
ST\_5                      & 6.22E-03    & 0.9274   & 2         & $| 8.25\text{E-}02 |$              & 7.53\%          \\ \hline
ST\_7*                      & 3.27E-03    & -3.2843  & 2         & $| 6.51\text{E-}03 |$              & 50.23\%         \\ \hline
ST\_14                     & 2.01E-03    & 0.9941   & 1         & $| 8.27\text{E-}02 |$              & 2.43\%          \\ \hline
ST\_16*                     & 7.22E-04    & 0.7900     & 3         & $| 6.02\text{E-}03 |$              & 11.99\%         \\ \hline
ST\_18                     & 2.87E-03    & 0.9911   & 2         & $| 8.23\text{E-}02 |$              & 3.49\%          \\ \hline
ST\_20*                     & 2.72E-03    & -1.9509  & 0         & $| 5.95\text{E-}03 |$              & 45.70\%         \\ \hline
ST\_29                     & 2.92E-03    & 0.9908   & 1         & $| 8.21\text{E-}02 |$              & 3.56\%          \\ \hline
ST\_33                     & 4.04E-03    & 0.9754   & 3         & $| 7.83\text{E-}02 |$              & 5.15\%          \\ \hline
ST\_42                     & 3.06E-03    & 0.9887   & 0         & $| 8.11\text{E-}02 |$              & 3.77\%          \\ \hline
ST\_46                     & 3.53E-03    & 0.9852   & 0         & $| 8.11\text{E-}02 |$              & 4.35\%          \\ \hline
ST\_51*                     & 7.61E-04    & 0.8430    & 2         & $| 6.46\text{E-}03 |$              & 11.78\%         \\ \hline
ST\_52*                     & 6.36E-04    & 0.9035   & 2         & $| 6.22\text{E-}03 |$              & 10.23\%         \\ \hline
ST\_55*                     & 1.49E-03    & 0.3419   & 1         & $| 7.17\text{E-}03 |$              & 20.73\%         \\ \hline
ST\_56*                     & 1.65E-03    & 0.1886   & 1         & $| 7.12\text{E-}03 |$              & 23.20\%         \\ \hline
\end{tabular}
\end{center}
Examples marked by an asterisk ``*'' indicate  heterogeneous activation functions. We note that the $R^2$ values reported here correspond to the subdomain and beat for which the highest MAE for $E_{cc}$ is observed.
\end{table}

As anticipated, the introduction of Perlin noise to ``Type 1'' synthetic dataset results in higher tracking errors as shown in Figs \ref{fig:homog_syn_res} and \ref{fig:hetero_syn_res} and summarized in Tables \ref{tab:sum_MAD_noisy} and \ref{tab:sum_MAE_noisy} for the homogeneous activation cases. Again, movies with \textit{sub}-pixel displacements are adversely affected by noise addition to a point where the code fails to identify beats to track. Such cases are indicated by missing data points in Fig \ref{fig:hetero_syn_res}. Overall, the errors that arise from the addition of Perlin noise decrease as the number of octaves increases, and the magnitude ratio decreases. For homogeneously activated data, the maximum MAD remains less than $14.5\%$ (Table \ref{tab:sum_MAD_noisy}) for all tested cases, even those marked as extremely noisy compared to realistic data, as shown in Fig \ref{fig:perlin_syn}. However, this is not the case with heterogeneously activated examples where the MAD errors become unreasonably high in most cases (on the order of $10^3 - 10^4$).

In general, for $E_{cc}$, the MAE is below $36\%$ but higher than $15\%$ of the peak ground truth $E_{cc}$ for all synthetic cases corresponding to homogeneous activation with added Perlin noise, except for the extreme case where the errors exceed $100\%$ as shown in Table \ref{tab:sum_MAE_noisy}.

For ``Type 2'' synthetic data, we plot the analytical displacements in both $X$ (horizontal) and $Y$ (vertical) against their tracked equivalents obtained by running our optical flow pipeline (Fig \ref{fig:type2_track}). The results reveal good agreement between the two with $R^2$ values of $0.998$ and $0.984$ for displacements in $X$ and $Y$, respectively. In Fig \ref{fig:type2_Ecc_track}, we compare subdomain-averaged $E_{cc}$ strain obtained via our tracking software to the known ground truth for a single beat. Overall, $R^{2}$ values (Fig \ref{fig:type2_Ecc_track}a, c \& d) reveal good agreement between tracked and ground truth subdomain-averaged $E_{cc}$ per subdomain, with $R^{2}$ values being greater than $0.9$. Furthermore, the mean absolute error in $E_{cc}$ per subdomain (Fig \ref{fig:type2_Ecc_track}b) indicates that the maximum error which occurs in subdomain {\fontfamily{pcr}\selectfont A4} is less than $7.4\%$ of the peak ground truth $E_{cc}$ in the respective subdomain.

\begin{table}[ht]
\begin{center}
\caption{\label{tab:sum_MAD_noisy}{\bf Summary of the mean absolute displacement validation errors for synthetic data with added Perlin noise.}} 
\includegraphics[width=0.5\textwidth]{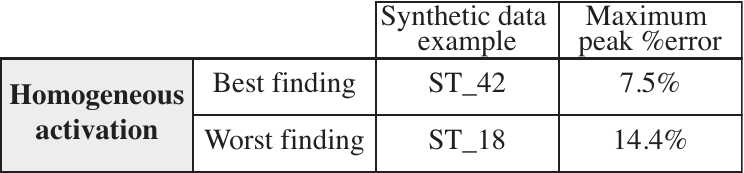}
\end{center}
\end{table}

\begin{table}[ht!]
\begin{center}
\caption{\label{tab:sum_MAE_noisy}\textbf{Summary of the mean absolute $\mathbf{E_{cc}}$ strain validation errors for synthetic data with added Perlin noise.}}
\includegraphics[width=1\textwidth]{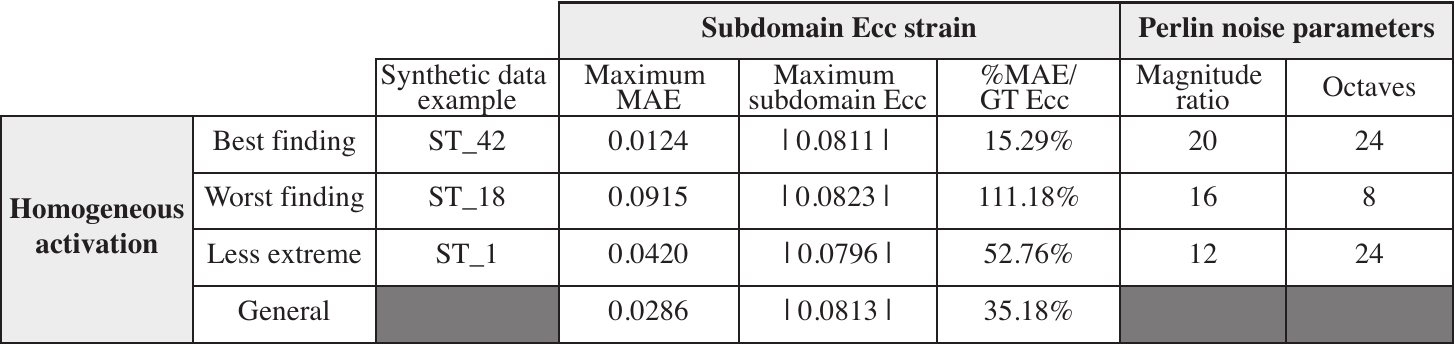} 
\end{center}
The ``Worst finding'' example corresponds to the largest obtained error whereas the ``Less extreme'' finding corresponds to the second largest error. For the ``General'' case, we consider the mean of the maximum MAE values for the remaining synthetic examples based on homogeneous activation excluding examples ST\_1, ST\_18, and ST\_42, as well as the mean of the maximum absolute ground truth subdomain $E_{cc}$.
\end{table}

\subsection*{Validation against manual tracking}
\label{subsubsi1:manual_res}
In Fig \ref{fig:man_track}, we show the comparison between the displacements in the $X$ (horizontal) and $Y$ (vertical) directions obtained via manual tracking to those obtained by ``MicroBundleCompute'' at the highlighted marker points. In both cases, the $R^2$ values indicate a good correlation between manual tracking results and the results from our computational tracking pipeline. Notably, the difference in tracking results between users $1$ and $2$ not only provides a range of values for comparison to our pipeline, but also indicates the fallibility of the manual tracking and the need for automated tools that lead to reproducible results.

\section*{Final remarks} 
\label{subsubsi1:final_rem}
Here, we have presented our two approaches to software validation: 1) comparison against synthetic data with known ground truth, and 2) comparison against manually labeled data (``\nameref{si1:methods}'' Section). From the results shown in the ``\nameref{si1:res}'' Section, we see that our computational pipeline produces fairly accurate displacement results and slightly less accurate Green-Lagrange strain results, with the peak errors for both outputs falling below $10\%$ for noiseless data exhibiting microbundle contractions that exceed a single pixel. For these examples with relatively large deformation, our software proves to be robust even against significant added noise. However, for \textit{sub}-pixel deformation examples, the results are less reliable, with added Perlin noise drastically affecting their accuracy. This is important context for highlighting the applicability and limitations of our pipeline.

\newpage
\FloatBarrier

\begin{figure}[p]
\begin{center}
\includegraphics[width=0.9\textwidth]{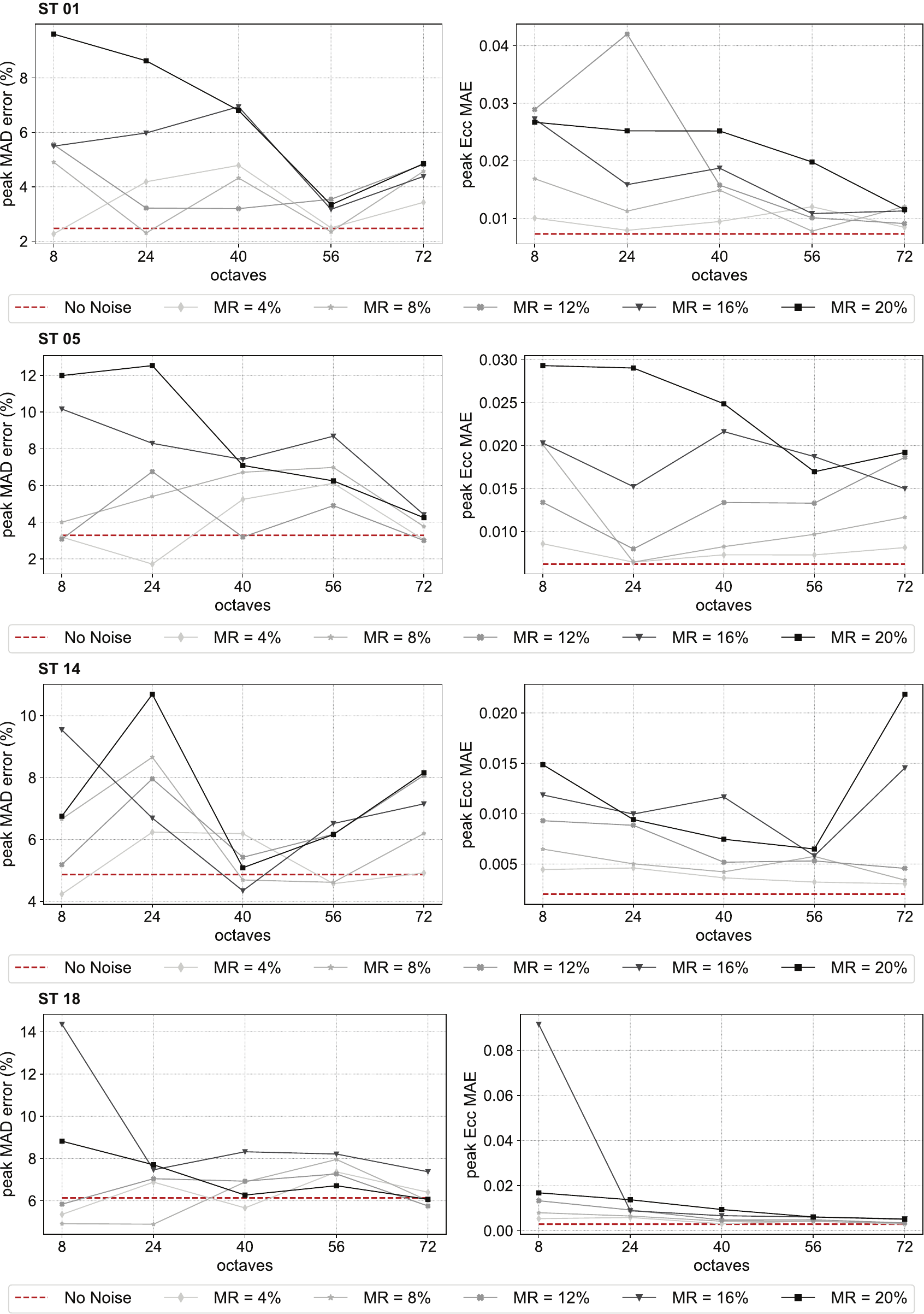}
   
\end{center}
\end{figure}

\begin{figure}[p]
\begin{center}
\includegraphics[width=0.9\textwidth]{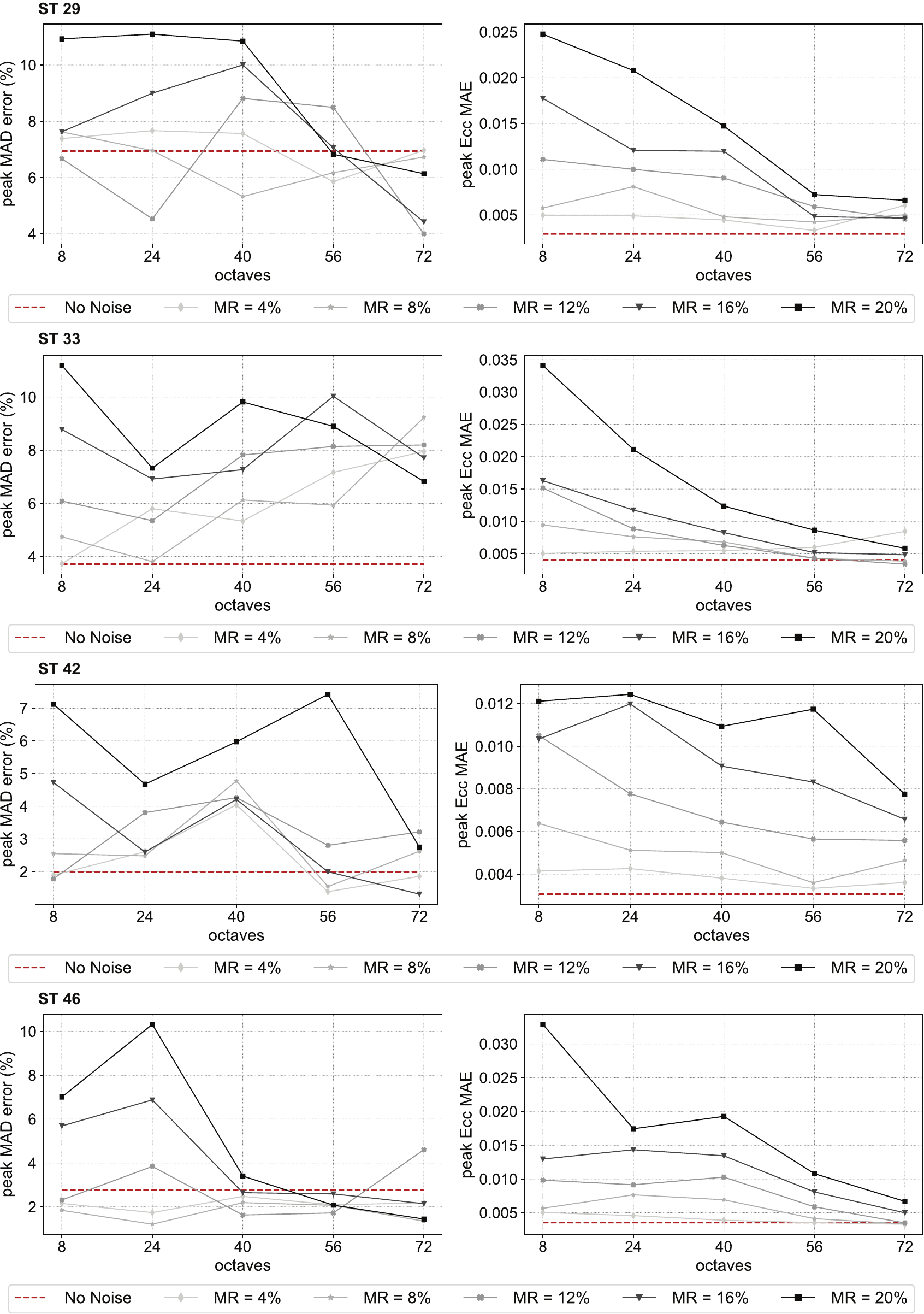}
\caption{\label{fig:homog_syn_res}Results of our pipeline validation against noisy synthetic data of ``Type 1'' based on FE simulations with homogeneous activation for different Perlin noise octaves and magnitude ratios (MR).}    
\end{center}
\end{figure}

\begin{figure}[ht]
\begin{center}
\includegraphics[width=0.9\textwidth]{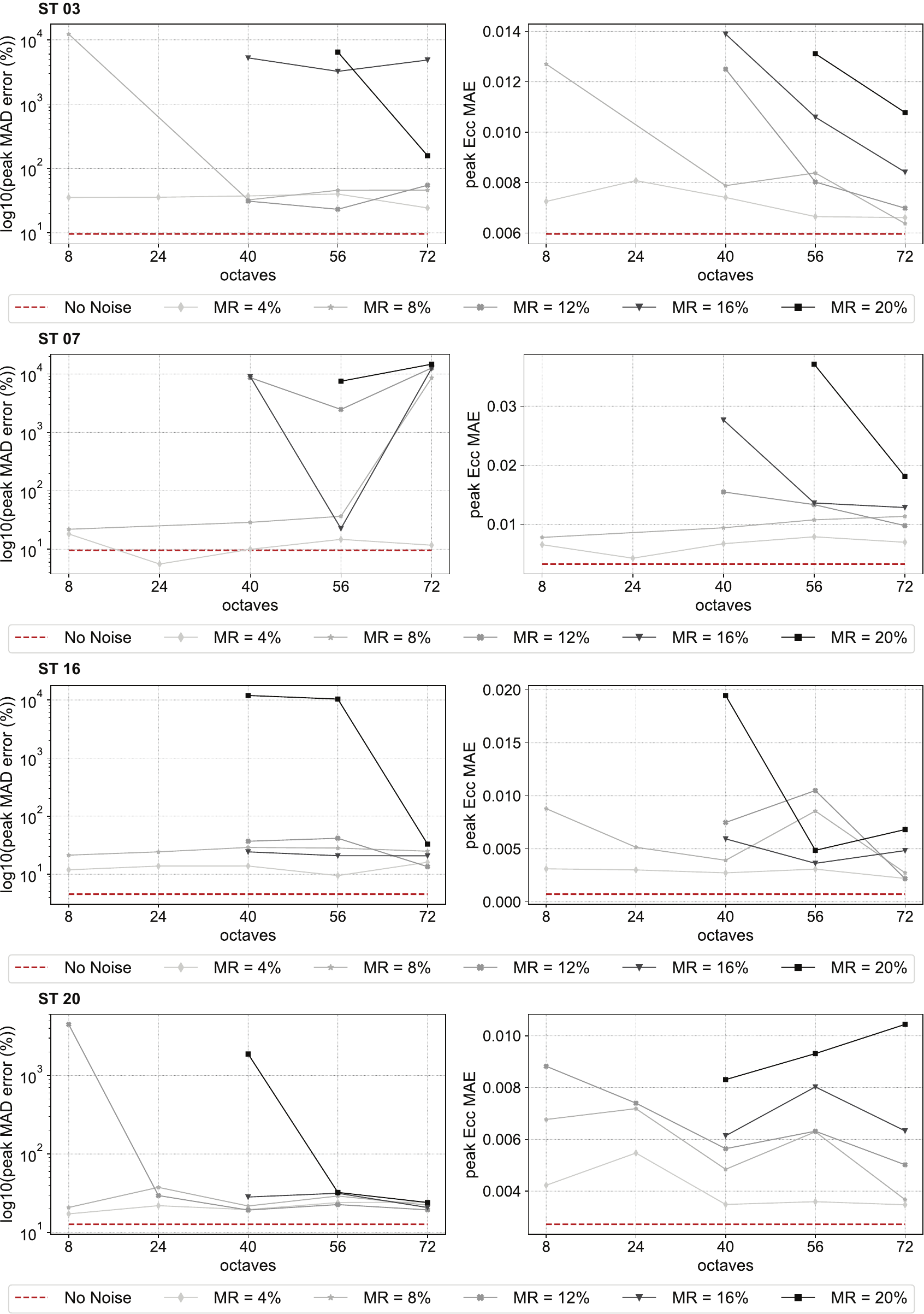}
\end{center}
\end{figure}

\begin{figure}[p]
\begin{center}
\includegraphics[width=0.9\textwidth]{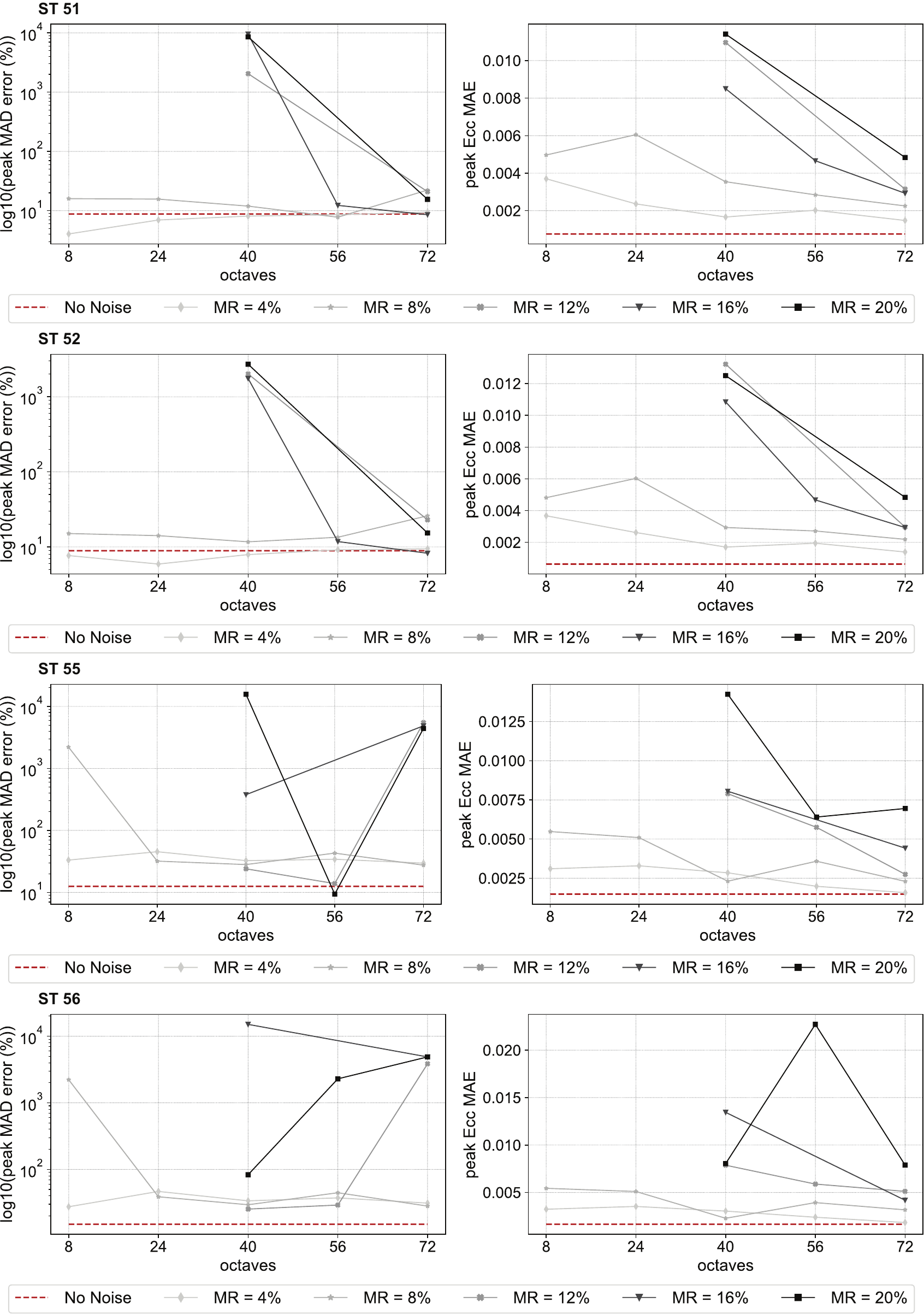}
\caption{\label{fig:hetero_syn_res}Results of our pipeline validation against noisy synthetic data of ``Type 1'' based on FE simulations with heterogeneous activation for different Perlin noise octaves and magnitude ratios (MR).}    
\end{center}
\end{figure}

\begin{figure}[p]
\begin{center}
\includegraphics[width=0.95\textwidth]{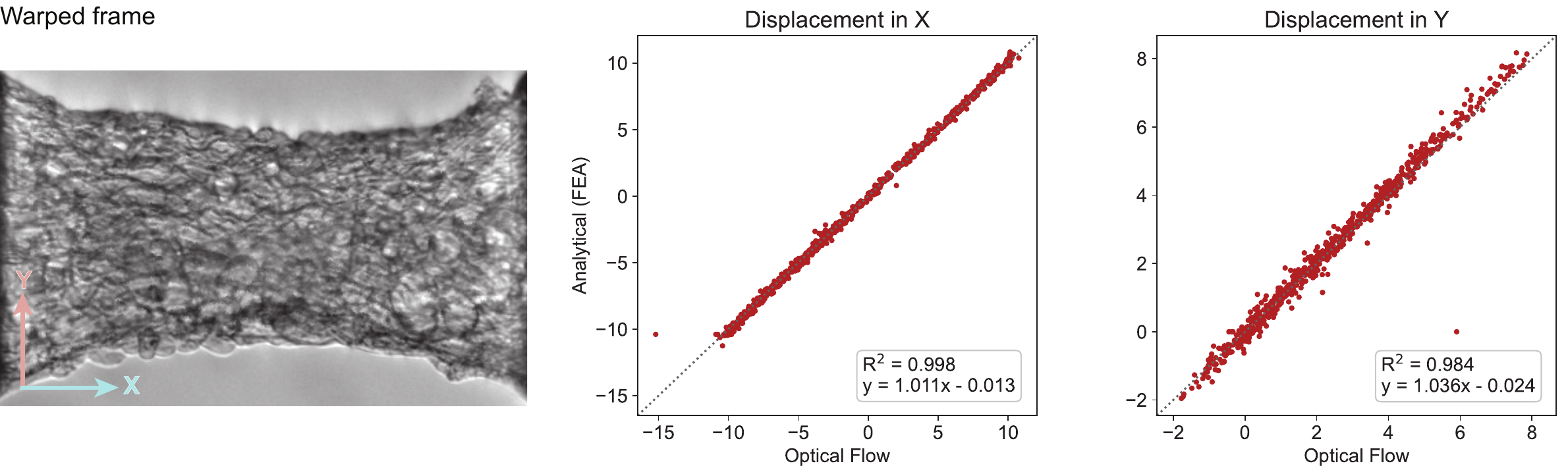}
\caption{\label{fig:type2_track}Results of our pipeline validation against synthetic data of ``Type 2'' for displacement outputs.}    
\end{center}
\end{figure}

\begin{figure}[p]
\begin{center}
\includegraphics[width=0.95\textwidth]{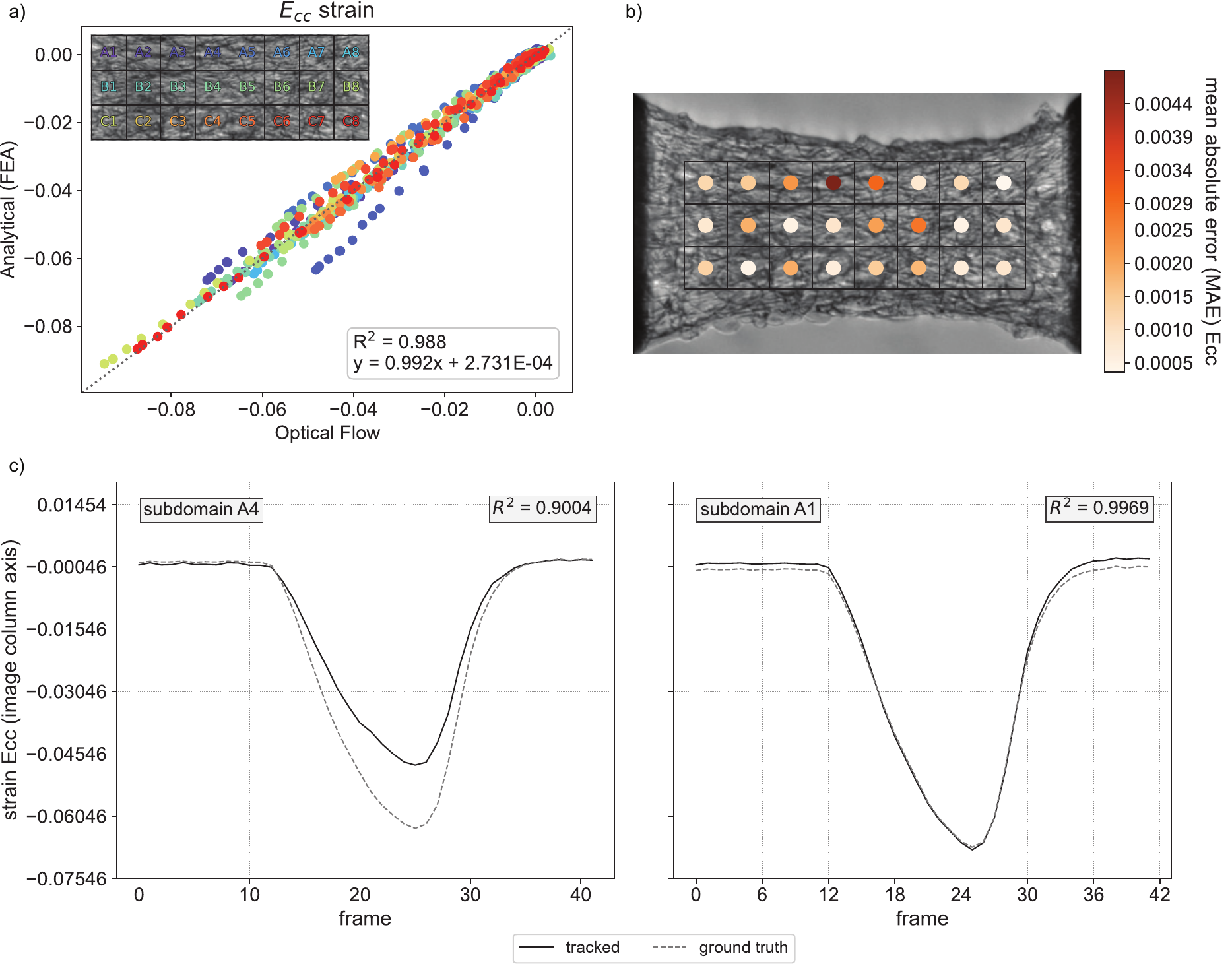}
\caption{\label{fig:type2_Ecc_track}Results of our pipeline validation against synthetic data of ``Type 2'' for $E_{cc}$ strain outputs: a) $E_{cc}$ strains obtained analytically and by tracking are in good agreement across all subdomains; (b) the mean absolute errors in $E_{cc}$ are below $0.0049$ in all subdomains; (c) with the maximum error occurring in subdomain {\fontfamily{pcr}\selectfont A4} and the median error occurring in subdomain {\fontfamily{pcr}\selectfont A1}.}    
\end{center}
\end{figure}

 \FloatBarrier
 \clearpage
\begin{figure}[ht!]
\begin{center}
\includegraphics[width=1\textwidth]{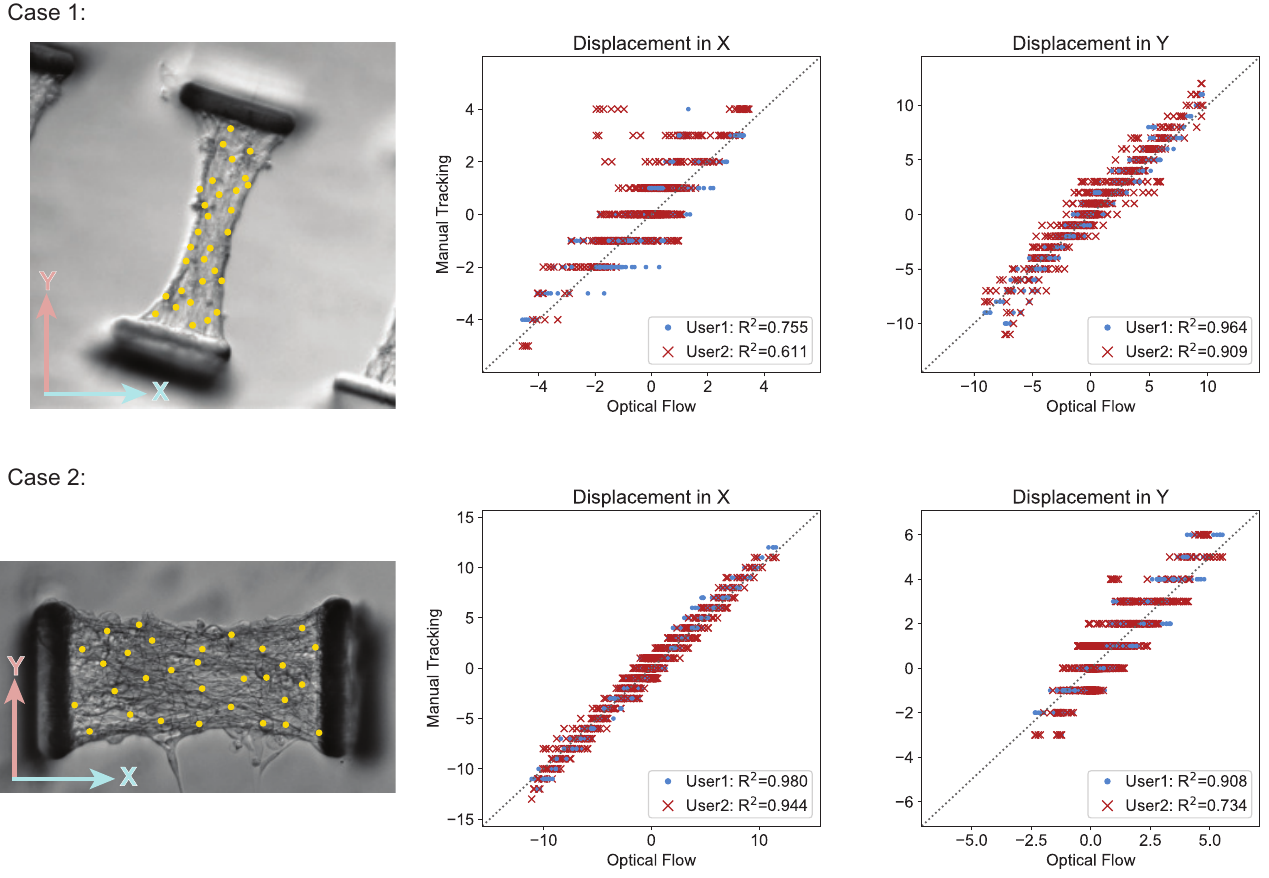}
\caption{\label{fig:man_track}Results of our pipeline validation against manually tracked points for $2$ different cases of ``Type 2'' data.}    
\end{center}
\end{figure}

%% file: S2_Appendix.tex
\begin{figure}[!htb]
\begin{center}
\includegraphics[width=1\textwidth]{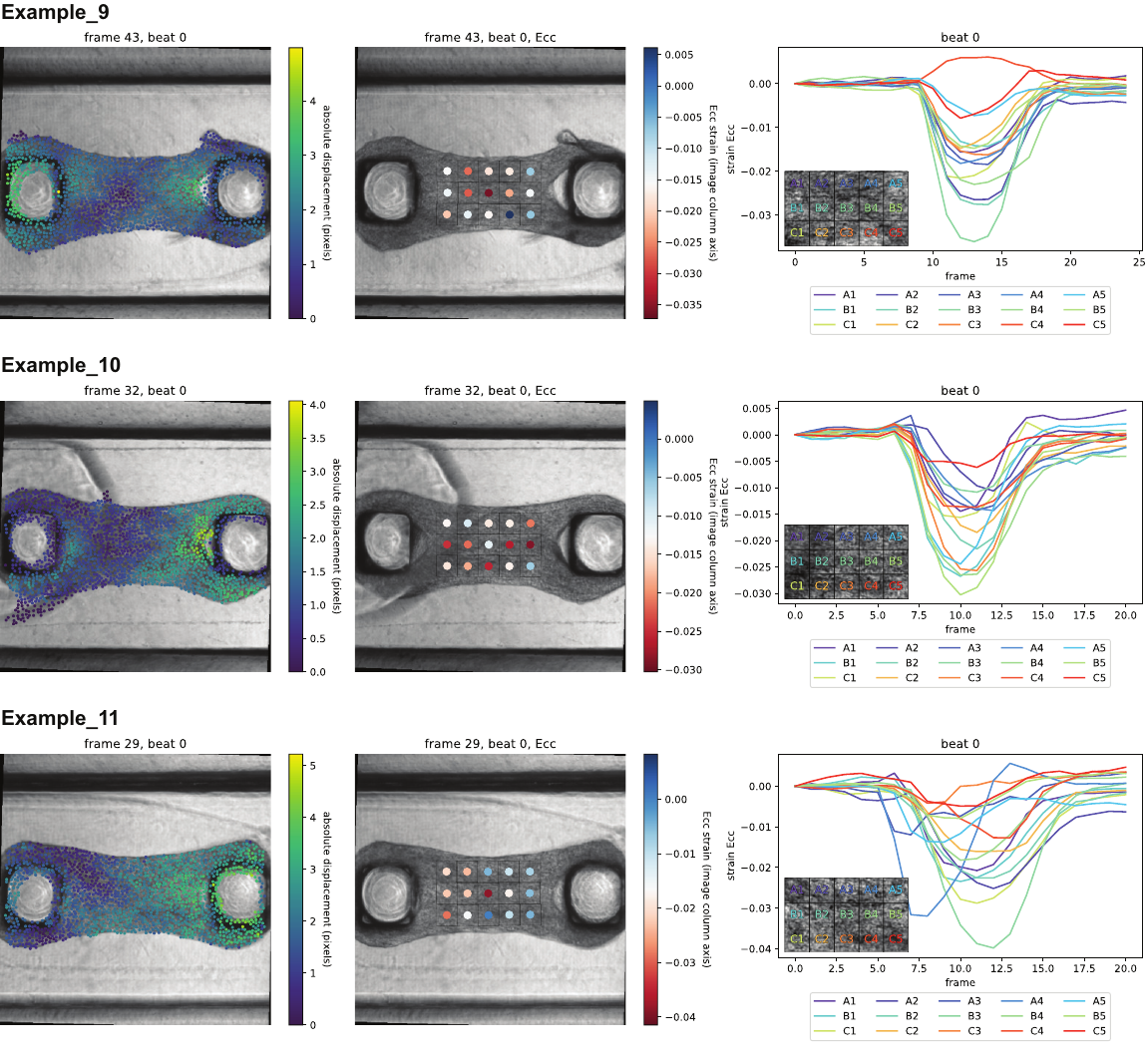}
\caption{\label{fig:res_type1}Example outputs of ``MicroBundleCompute'' run on ``Type 1'' experimental data. Here we show, in order from left to right, full-field mean absolute displacement and subdomain-averaged Green-Lagrange $E_{cc}$ strain at the first tracked peak, as well as a time series plot of $E_{cc}$ strain for the first tracked beat.}    
\end{center}
\end{figure}
\begin{figure}[!htb]
\begin{center}
\includegraphics[width=1\textwidth]{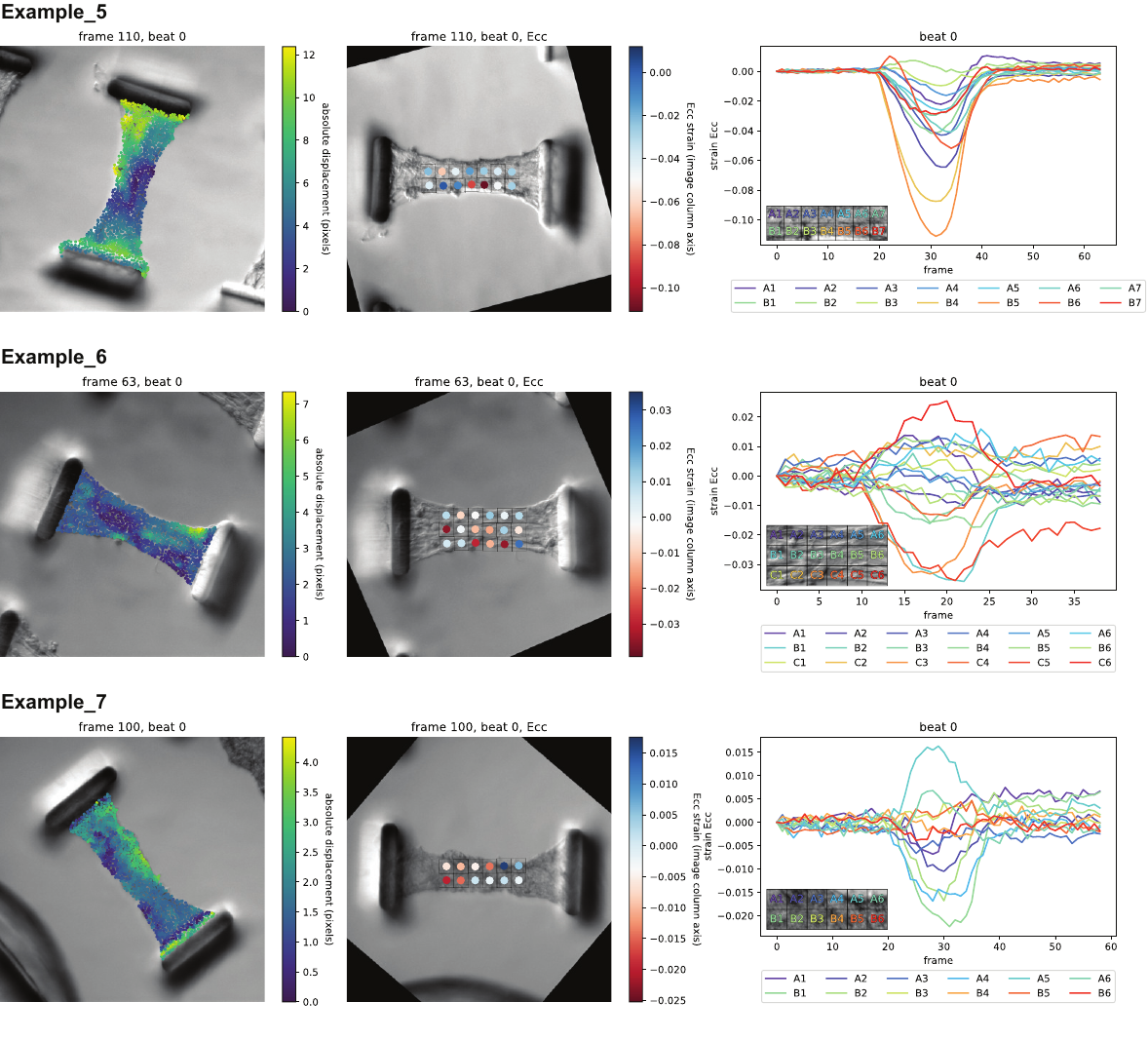}
\caption{\label{fig:res_type2}Example outputs of ``MicroBundleCompute'' run on ``Type 2'' experimental data. Here we show, in order from left to right, full-field mean absolute displacement and subdomain-averaged Green-Lagrange $E_{cc}$ strain at the first tracked peak, as well as a time series plot of $E_{cc}$ strain for the first tracked beat.}    
\end{center}
\end{figure}
\begin{figure}[!htb]
\begin{center}
\includegraphics[width=1\textwidth]{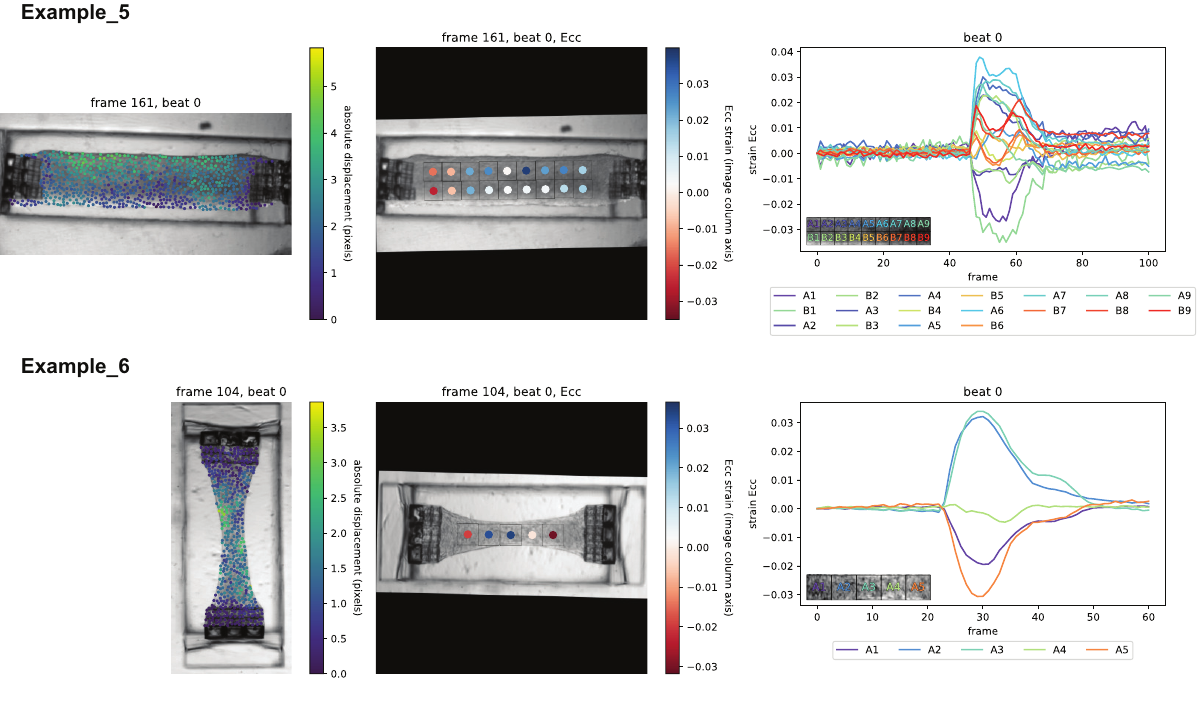}
\caption{\label{fig:res_type3}Example outputs of ``MicroBundleCompute'' run on ``Type 3'' experimental data. Here we show, in order from left to right, full-field mean absolute displacement and subdomain-averaged Green-Lagrange $E_{cc}$ strain at the first tracked peak, as well as a time series plot of $E_{cc}$ strain for the first tracked beat.}    
\end{center}
\end{figure}

In this Supplementary Document, we (1) provide the details necessary to reproduce our analysis on all examples in the ``Microbundle Time-lapse Dataset,'' and (2) show the results of running ``MicroBundleCompute'' on all of these data. In brief, the ``Microbundle Time-lapse Dataset'' contains $24$ experimental time-lapse images of cardiac microbundles. The dataset is hosted under a CC0 open-source license on the Dryad Digital Repository \cite{microbundle2023data}. Consistent with our description of the experimental dataset in the ``Experimental data'' Section of the main paper document, we categorize these data as ``Type 1'' ($11$ examples), ``Type 2'' ($7$ examples), and ``Type 3'' ($6$ examples). A brief metadata summary for each type is given in Table (Table \ref{tab:sum_exp_data}). These details are also elaborated on in the ``Experimental data'' Section, and documented as metadata for the ``Microbundle Time-lapse Dataset.'' In Tables \ref{tab:sum_type1}, \ref{tab:sum_type2}, and \ref{tab:sum_type3}, we provide the implementation details (mask type, first frame adjustment, and subdomain segmentation parameters) necessary to reproduce all results. For the subdomain segmentation parameters, we only include those that were changed from the default values provided within the ``run\_code.py'' file located in the ``MicroBundleCompute'' GitHub repository \href{Github} {https://github.com/HibaKob/MicroBundleCompute}. Finally, in Figs \ref{fig:res_type1}, \ref{fig:res_type2}, and \ref{fig:res_type3} we show representative outputs from running our code: full-field mean absolute displacement at the first tracked peak, subdomain averaged Green-Lagrange $E_{cc}$ (horizontal) strain at the first tracked peak, and time series plots of $E_{cc}$ strain for the first tracked beat. These results not only show the versatility of our framework, but also showcase the information that can be obtained from these data.  





\begin{table}[ht]
\begin{center}
\caption{\label{tab:sum_exp_data} {\bf A summary of the experimental conditions associated with each example movie.}}  
\includegraphics[width=0.75\textwidth]{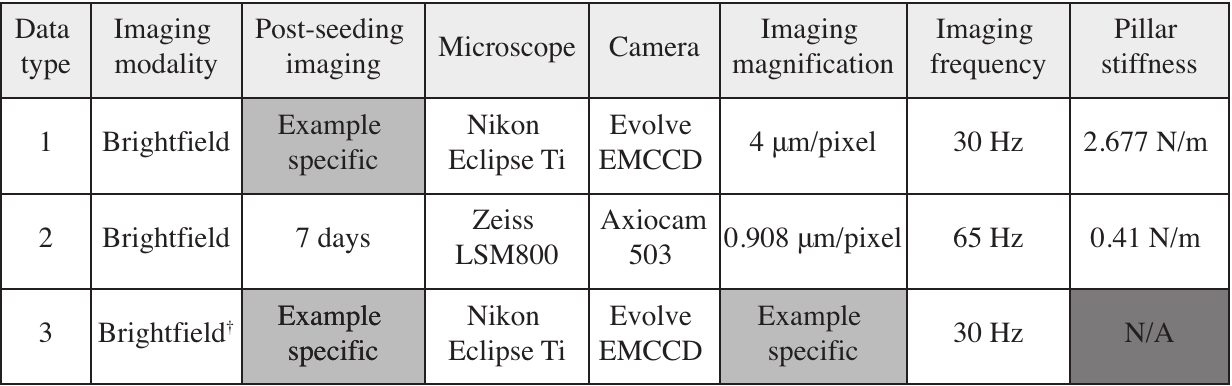}
\end{center}
We note that the symbol $^{\dagger}$ for ``Type 3'' ``Imaging modality'' is used to indicate that Example 1 is obtained via phase contrast microscopy, whereas the remainder of the examples are obtained via brightfield microscopy.
\end{table}

\clearpage
\begin{table}[ht]
\begin{center}
\caption{\label{tab:sum_type1} {\bf A summary of the example-specific information, code implementation details and subdomain segmentation parameters for each example of ``Type 1'' data.}} 
\includegraphics[width=0.6\textwidth]{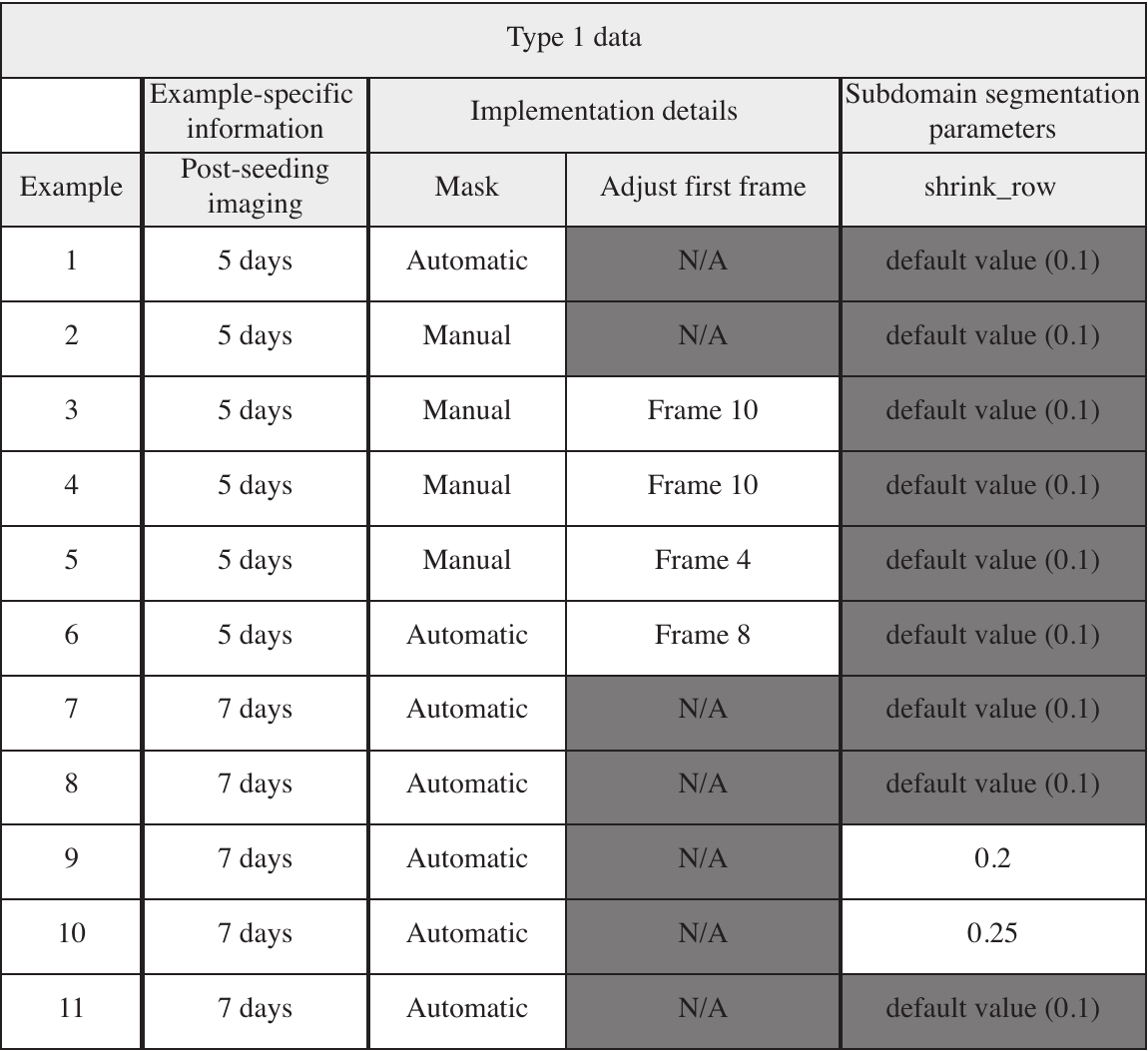}
\end{center}
\end{table}

\clearpage
\begin{table}[ht]
\begin{center}
\caption{\label{tab:sum_type2} {\bf A summary of the example-specific information, code implementation details, and subdomain segmentation parameters for each example of ``Type 2'' data.}} 
\includegraphics[width=1\textwidth]{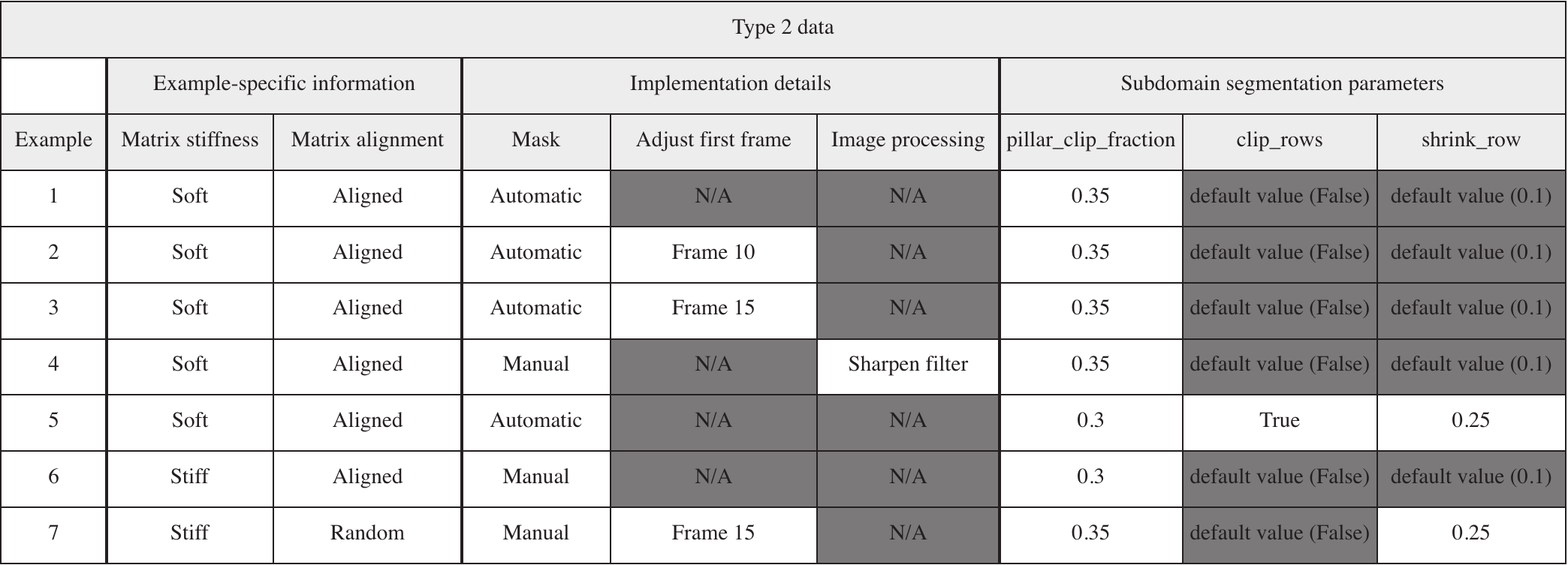}
\end{center}
\end{table}

\begin{table}[!htb]
\begin{center}
\caption{\label{tab:sum_type3} {\bf A summary of the example-specific information and subdomain segmentation parameters for each example of ``Type 3'' data.}} 
\includegraphics[width=1\textwidth]{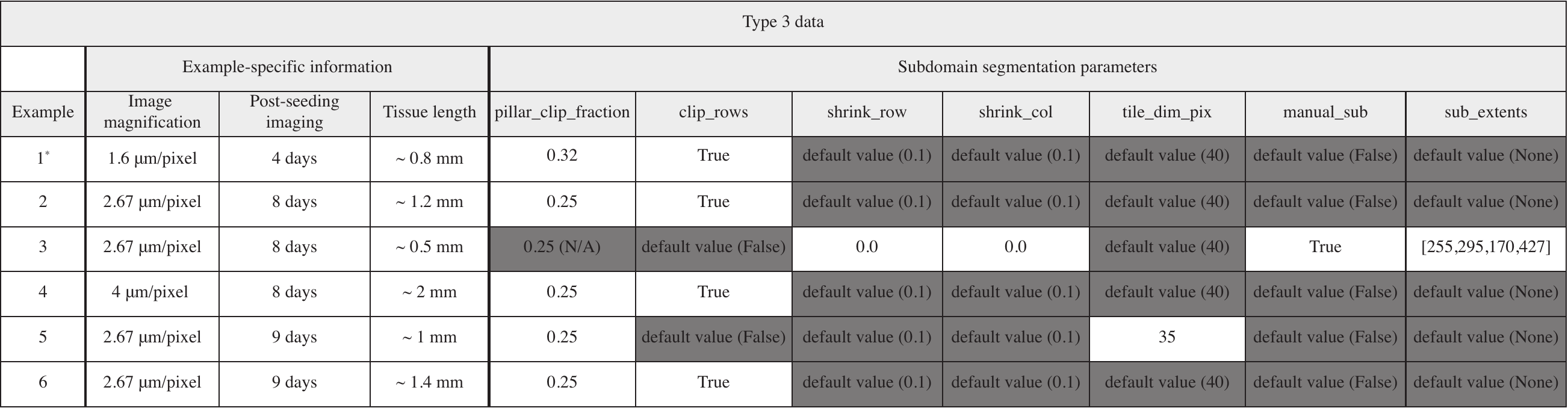}
\end{center}
We note that example $1$ (marked by an asterisk) is actuated by applying sawtooth pressure waves with $\sim$ - 6 kPa peak amplitude (equivalent to $\sim$ 2.5\% strain) using a microfluidic pump (Elveflow OB1) to stretch and release the tissue from one side at 0.5 Hz. All other examples are not subjected to any form of actuation. We also note that all masks for this data type were manually generated.
\end{table}

\clearpage
\begin{figure}[!htb]
\begin{center}
\includegraphics[width=1\textwidth]{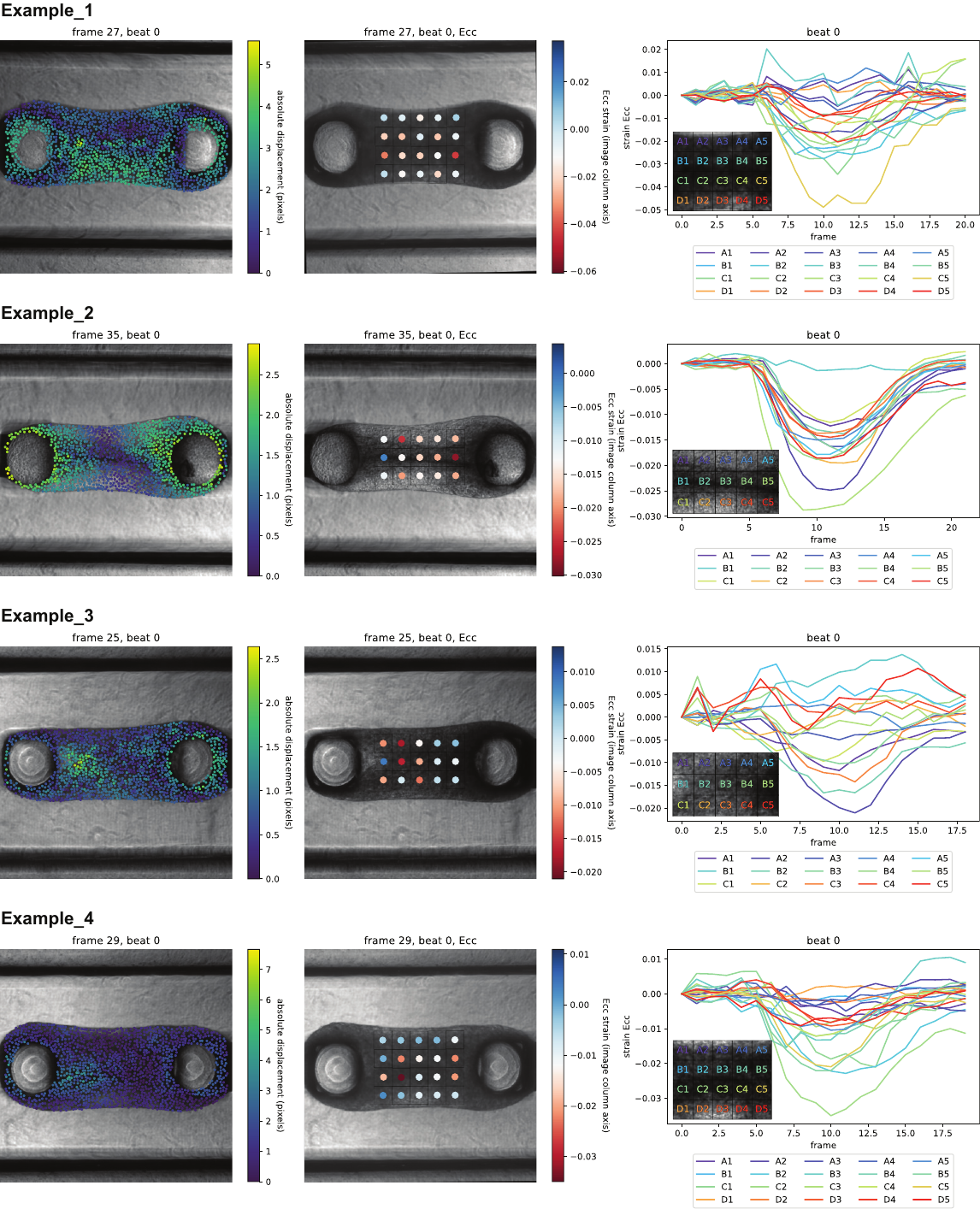}
\end{center}
\end{figure}

\clearpage
\begin{figure}[!htb]
\begin{center}
\includegraphics[width=1\textwidth]{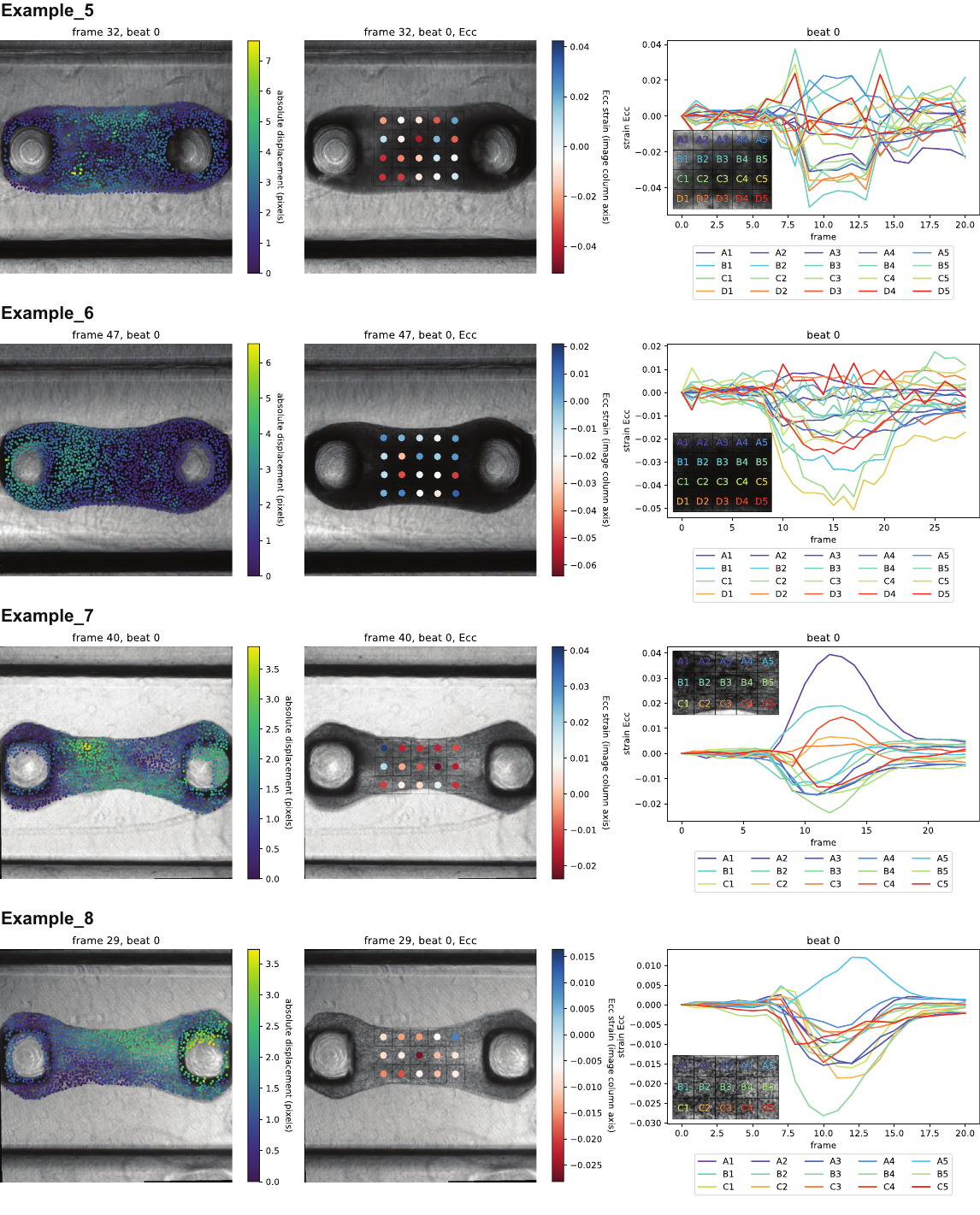}  
\end{center}
\end{figure}

\clearpage

\clearpage
\begin{figure}[!htb]
\begin{center}
\includegraphics[width=1\textwidth]{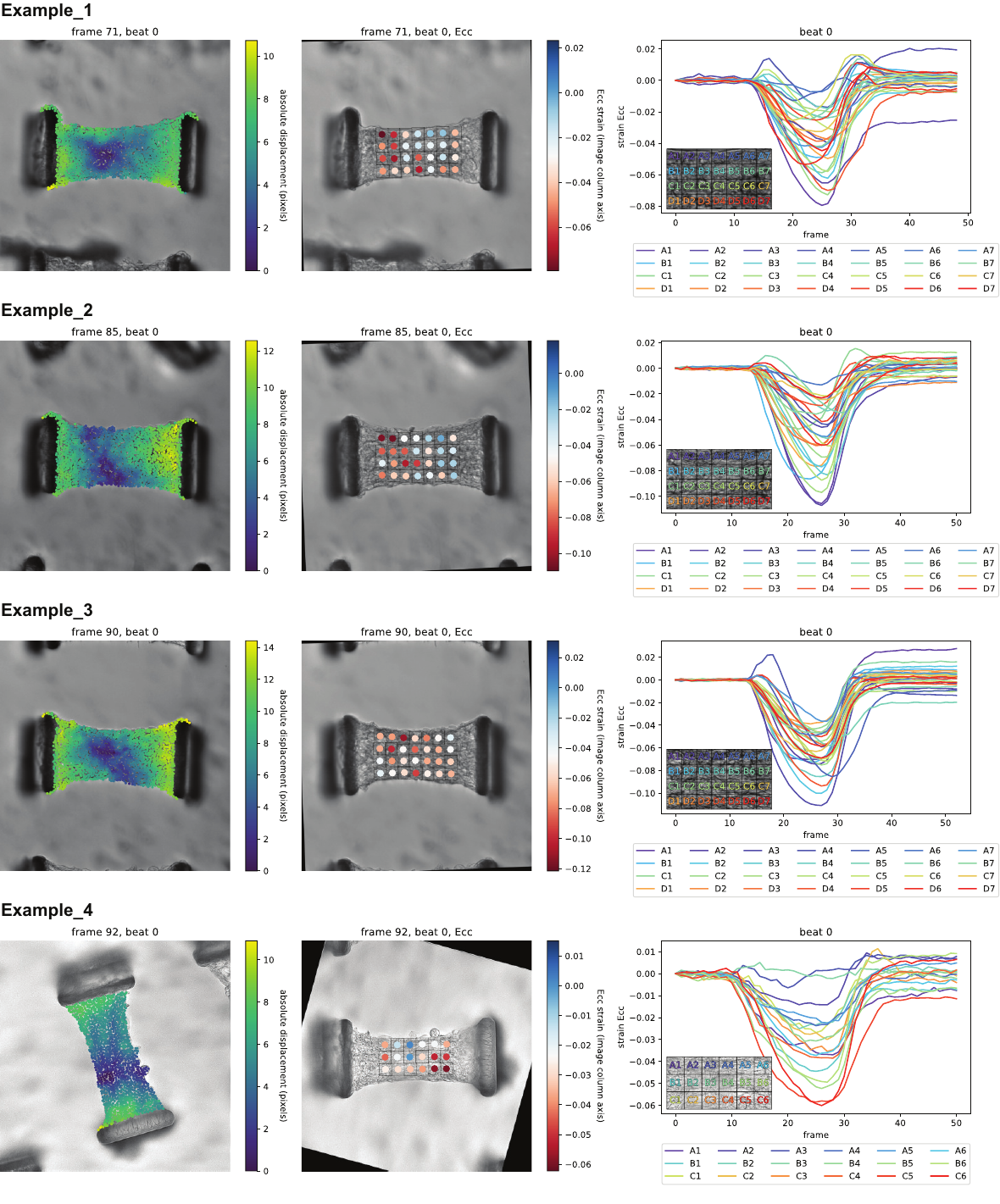}  
\end{center}
\end{figure}

\clearpage

\clearpage
\begin{figure}[!htb]
\begin{center}
\includegraphics[width=1\textwidth]{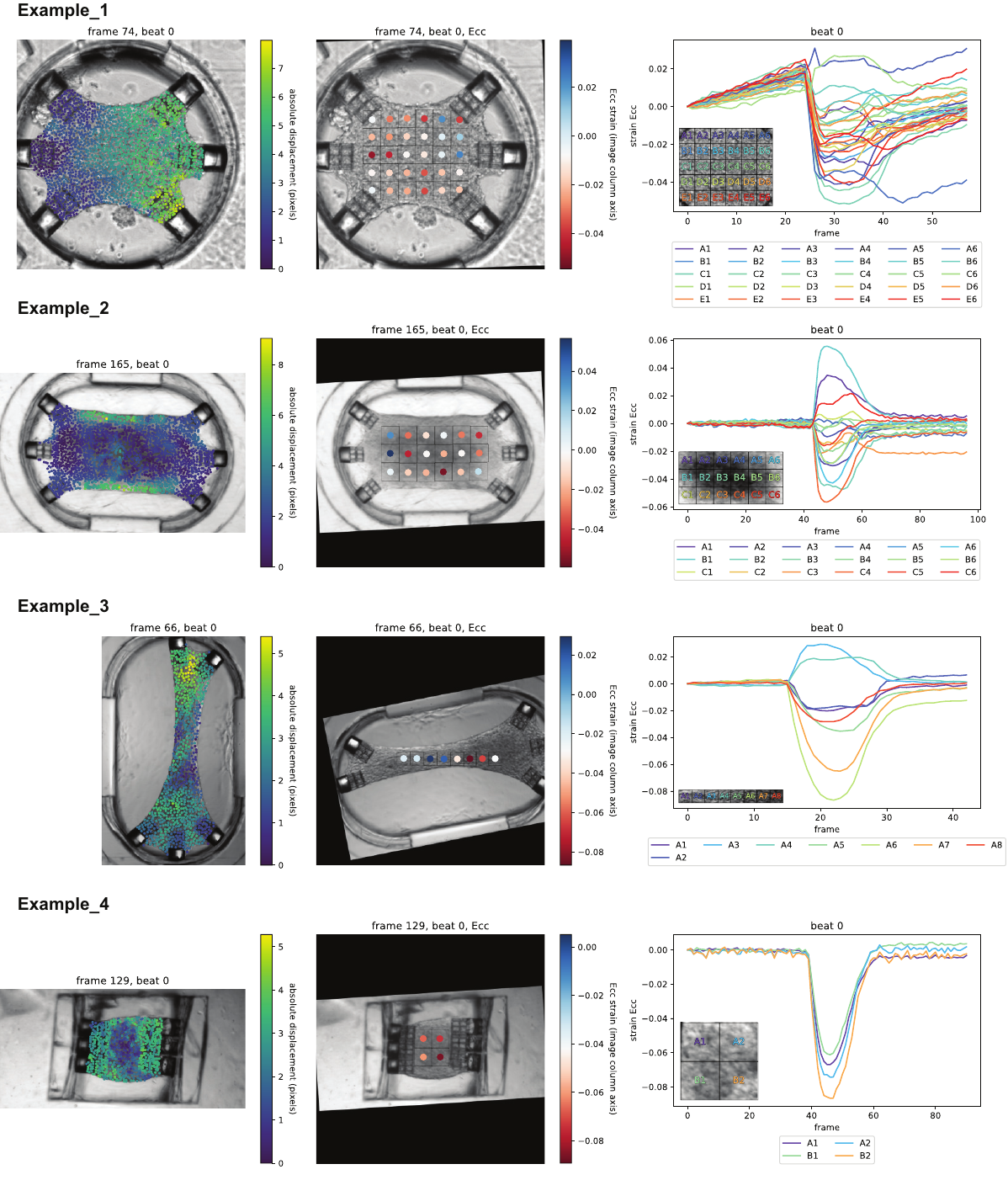}  
\end{center}
\end{figure}

\clearpage




%
%
%
%
%

%% file: S3_Appendix.tex
In this Supplementary Document, we elaborate on the basic pillar tracking feature that is included within our computational pipeline. 
In brief, we demonstrate how the functionality to track tissue deformation can be readily adapted and applied to compute the standard metrics of pillar displacement and subsequently cardiac microbundle twitch force. 
First, we describe our pipeline to extract pillar displacements, and then we describe our approach to convert these pillar displacements into twitch forces following standard equations from the literature (the ``\nameref{si3:methods}'' Section) \cite{legant2009microfabricated}. In the ``\nameref{si3:res}'' Section, we show the results obtained from running pillar tracking on our example data. Finally, in the ``\nameref{si3:future}'' Section, we briefly describe our future plans to advance our pillar tracking functionality beyond the scope of this project. 

 \section*{Methods} 
\label{si3:methods}
Here we describe our straightforward approach to tracking microbundle pillars and outputting twitch forces. Briefly, the pillar tracking functionality requires either manually or externally generating separate masks of each pillar, and follows a similar pipeline to the microbundle tissue tracking described in the ``Code'' Section of the main paper document with one main difference: temporal segmentation is by default skipped. Instead, we directly compute the mean position of all tracked points at every time step, and subsequently derive the mean absolute displacement relative to the first frame, considered to be a valley frame. However, we retain the temporal segmentation as an optional step to remove any drift present in the tracked results, as explained in the ``Temporal segmentation'' Section of the main paper document. We briefly note that this approach is possible because we do not output full-field pillar displacement results.

Following standard approaches in the literature \cite{legant2009microfabricated}, the pillar directional and absolute forces can be computed from the obtained mean directional and absolute displacement or deflection results. Specifically, an approximation of the pillar force $F$ can be found by applying Hooke's law: 

\begin{align}
F = k \delta
\label{hooke's}
\end{align}

where the deflection $\delta$ refers to the mean tracked pillar displacement and the combined geometric and material pillar stiffness $k$ is provided as an experimentally derived quantity.

Alternatively, if experimental testing data is not available, the poly(dimethylsiloxane) (PDMS) molded pillars can be approximated as cantilever beams \cite{legant2009microfabricated, das2022mechanical},
and the stiffness $k$ can be determined by the pillar geometry and material properties as:
\begin{align}
k = \frac {6 E I}{a^2 (3L - a)}
\label{stiffness}
\end{align}
where $E$ is the material's elastic modulus, $a$ is the location of force application, $L$ is the cantilever length, and finally $I$ is the moment of inertia defined as a function of the cantilever's geometry. For rectangular cross section beams, $I = \frac{wt^3}{12}$ where $w$ is the pillar width and $t$ is the pillar thickness while for circular cross section beams, $I = \frac{\pi D^4}{64}$ where $D$ is the cylindrical pillar diameter. To compute mean tissue stress, pillar force is divided by tissue cross sectional area, where the latter area is a function of tissue width computed from the tissue mask geometry, and experimentally measured tissue depth.

Experimental platforms that are based on pillar constructs (i.e., the ``Type 1'' and ``Type 2'' data introduced prior) are amenable to pillar tracking. We share our example implementation as well as general guidelines to use this feature on the ``MicroBundleCompute'' GitHub page (\href{Github} {https://github.com/HibaKob/MicroBundleCompute}). For our examples of ``Type 1'' data, we run the code with a pillar stiffness of $k = 2.677 \mu$N$/\mu$m and length scale ($ls$) $4 \mu$m/pixel. For the ``Type 2'' examples, we specify $k = 0.41 \mu$N$/\mu$m and $ls = 0.908 \mu$m/pixel. 

 \section*{Results}
\label{si3:res}
\begin{figure}[!htb]
\begin{center}
\includegraphics[width=0.85\textwidth]{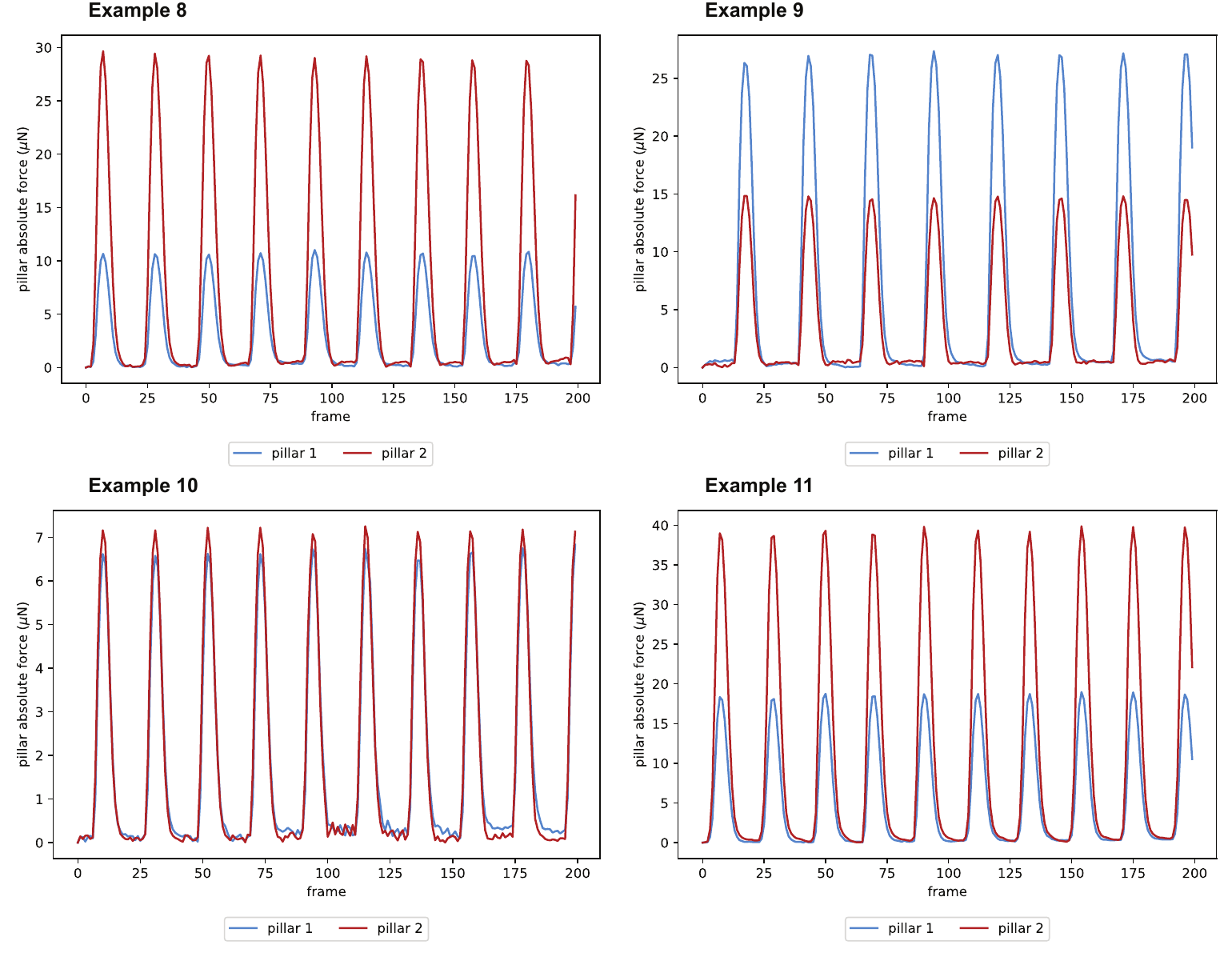}
\caption{\label{fig:type1_pillar_force} Pillar absolute force ($\mu$N) obtained by running the pillar tracking pipeline with ``Type 1'' data. We note that, in these cases,``pillar 1'' and ``pillar 2'' in the legends consistently refer to the left and right pillars, respectively. However, in general, ``pillar 1'' would refer to the user-defined pillar mask saved as ``pillar\_mask\_1.txt'' and ``pillar 2'' to the pillar whose mask is defined by ``pillar\_mask\_2.txt''.}
\end{center}
\end{figure}

\begin{figure}[htb!]
\begin{center}
\includegraphics[width=0.85\textwidth]{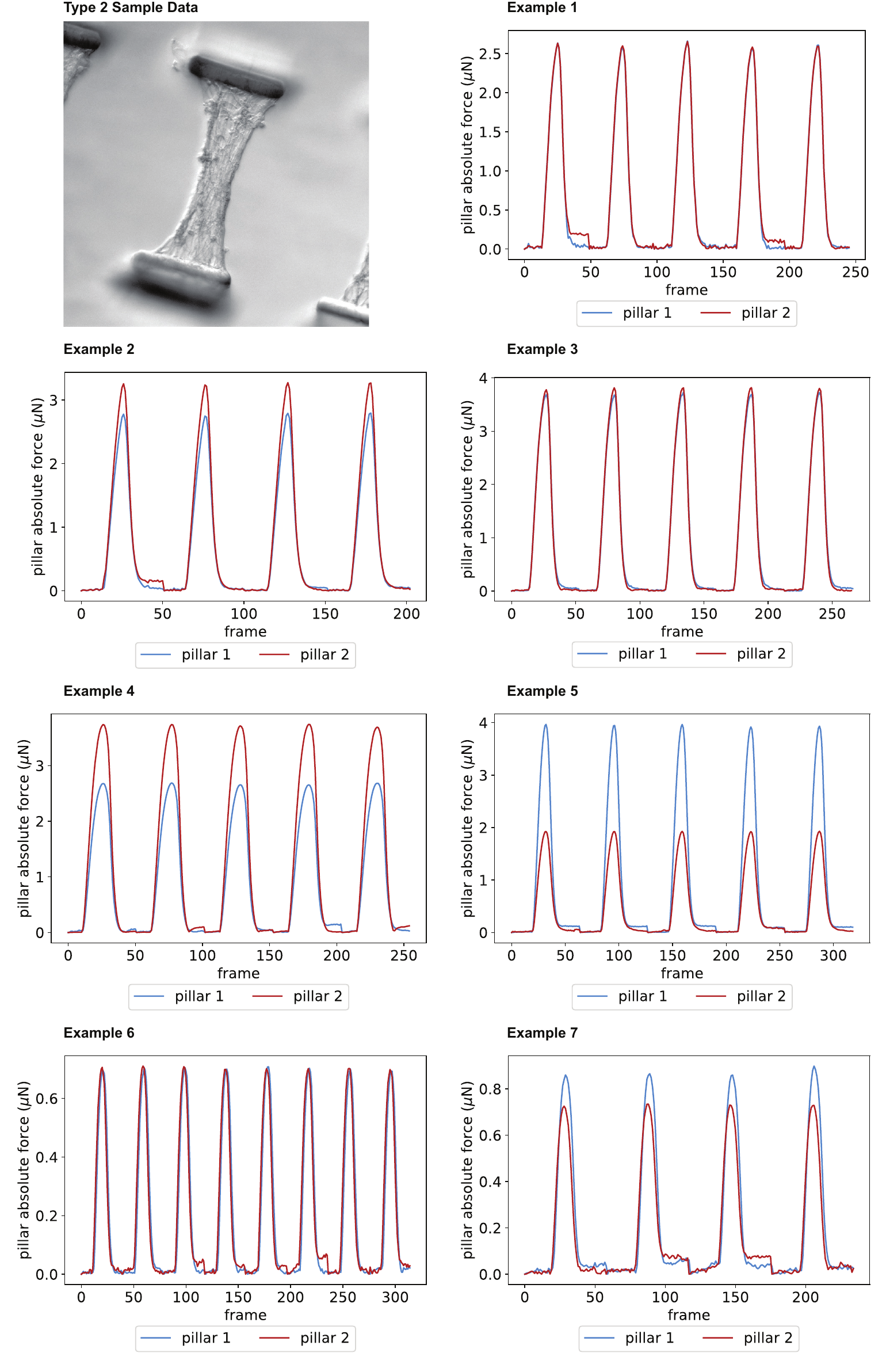}
\caption{\label{fig:type2_pillar_force} Pillar absolute force ($\mu$N) obtained by running the pillar tracking pipeline with ``Type 2'' data. We note that, in these cases, ``pillar 1'' refers to the user-defined pillar mask saved as ``pillar\_mask\_1.txt'' and ``pillar 2'' to the pillar whose mask is defined by ``pillar\_mask\_2.txt''.}
\end{center}
\end{figure}

Here, we present the results of implementing the pillar tracking pipeline with all $14$ experimental examples of ``Type 1'' and ``Type 2'' data. Results are shown in Figs \ref{fig:type1_pillar_force} and \ref{fig:type2_pillar_force} and reveal that the absolute force is not always consistent among the two pillars. While imaging artifacts, present in both types of data, might marginally contribute to the discrepancy in measured force outputs, we believe that there are a number of mechanistic reasons that we have yet to investigate, as we explain in the ``\nameref{si3:future}'' Section.
Of note, ``Type 1'' data was run without temporal segmentation while ``Type 2'' data displayed non-trivial drift and required the temporal segmentation step. We attribute this difference to the out-of-plane motion that is more significantly visible in the substantially thinner ``Type 2'' tissues.
\begin{figure}[!htb]
\begin{center}
\includegraphics[width=0.85\textwidth]{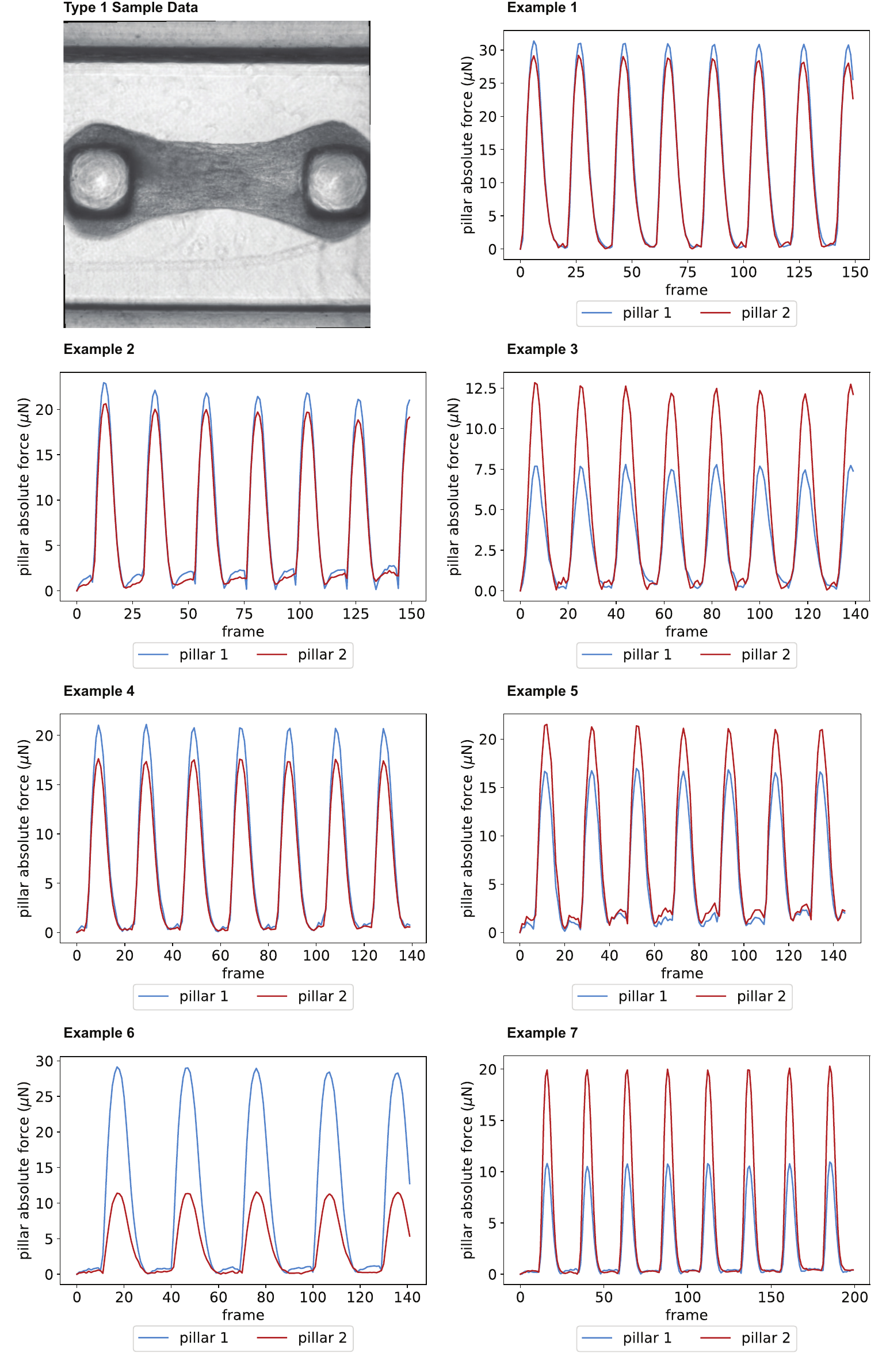}
\end{center}
\end{figure}

\clearpage

\FloatBarrier

 \section*{Future Work}
\label{si3:future}
In this document, we have provided a description of the pillar tracking functionality included within the ``MicroBundleCompute'' computational framework and have shared representative examples of implementing this functionality. We believe that, in its current state, the pillar tracking option would be useful to a number of researchers, given its ease of use. However, we have multiple plans to further improve our pipeline including: 

\begin{itemize}
    \item automatic pillar mask extraction
    \item validated outputs against synthetic data and against other tools currently available in the literature\cite{tamargo2021millipillar}
    \item generalized pillar tracking functionality that can be implemented on experimental setups beyond the scope of the ``Type 1'' and ``Type 2'' data examples shown here
\end{itemize}

In addition, by examining the time series plots of the pillar absolute force for both pillars in Figs \ref{fig:type1_pillar_force} and \ref{fig:type2_pillar_force}, we can see that the deflections of two pillars contained within one experimental sample do not always match. 
There are multiple potential mechanistic explanations of this observation. For example, the tissue may be at different vertical positions on each pillar, pillars may have variable material or geometric properties, or there may be an asymmetry in tissue-pillar interaction that we do not currently understand. Further investigations are needed to both formulate solid explanations of this finding, and develop the appropriate statistical tools to address the associated uncertainty as a companion to our software. 


%
%
%
%
%
%
%

%% file: Manuscript.bbl
\begin{thebibliography}{114}
\providecommand{\natexlab}[1]{#1}
\providecommand{\url}[1]{\texttt{#1}}
\expandafter\ifx\csname urlstyle\endcsname\relax
  \providecommand{\doi}[1]{doi: #1}\else
  \providecommand{\doi}{doi: \begingroup \urlstyle{rm}\Url}\fi

\bibitem[Brush et~al.(2022)Brush, Sherbino, and Norman]{brush2022diagnostic}
John~E Brush, Jonathan Sherbino, and Geoffrey~R Norman.
\newblock Diagnostic reasoning in cardiovascular medicine.
\newblock \emph{bmj}, 376, 2022.
\newblock \doi{https://doi.org/10.1136/bmj-2021-064389}.

\bibitem[Vardas et~al.(2022)Vardas, Asselbergs, van Smeden, and
  Friedman]{vardas2022year}
Panos~E Vardas, Folkert~W Asselbergs, Maarten van Smeden, and Paul Friedman.
\newblock The year in cardiovascular medicine 2021: digital health and
  innovation.
\newblock \emph{European heart journal}, 43\penalty0 (4):\penalty0 271--279,
  2022.
\newblock \doi{https://doi.org/10.1093/eurheartj/ehab874}.

\bibitem[Kaptoge et~al.(2019)Kaptoge, Pennells, De~Bacquer, Cooney, Kavousi,
  Stevens, Riley, Savin, Khan, Altay, et~al.]{kaptoge2019world}
Stephen Kaptoge, Lisa Pennells, Dirk De~Bacquer, Marie~Therese Cooney, Maryam
  Kavousi, Gretchen Stevens, Leanne~Margaret Riley, Stefan Savin, Taskeen Khan,
  Servet Altay, et~al.
\newblock World health organization cardiovascular disease risk charts: revised
  models to estimate risk in 21 global regions.
\newblock \emph{The Lancet Global Health}, 7\penalty0 (10):\penalty0
  e1332--e1345, 2019.
\newblock \doi{https://doi.org/10.1016/S2214-109X(19)30318-3}.

\bibitem[Nguyen et~al.(2021)Nguyen, de~Bakker, and Bakkers]{nguyen2021cardiac}
Phong~D Nguyen, Dennis~EM de~Bakker, and Jeroen Bakkers.
\newblock Cardiac regenerative capacity: an evolutionary afterthought?
\newblock \emph{Cellular and molecular life sciences}, 78\penalty0
  (12):\penalty0 5107--5122, 2021.
\newblock \doi{10.1007/s00018-021-03831-9}.

\bibitem[Bursill et~al.(2022)Bursill, Smith, Palpant, Tan, Sunde, Harvey,
  Lewis, Figtree, Vandenberg, and Alliance]{bursill2022don}
Christina~A Bursill, Nicola~J Smith, Nathan Palpant, Isabella Tan, Margaret
  Sunde, Richard~P Harvey, Benjamin Lewis, Gemma~A Figtree, Jamie~I Vandenberg,
  and Australian~Cardiovascular Alliance.
\newblock Don’t turn off the tap! the importance of discovery science to the
  australian cardiovascular sector and improving clinical outcomes into the
  future.
\newblock \emph{Heart, Lung and Circulation}, 2022.
\newblock \doi{https://doi.org/10.1016/j.hlc.2022.06.669}.

\bibitem[Thygesen et~al.(2007)Thygesen, Alpert, and
  White]{thygesen2007universal}
Kristian Thygesen, Joseph~S. Alpert, and Harvey~D. White.
\newblock Universal definition of myocardial infarction.
\newblock \emph{Circulation}, 116\penalty0 (22):\penalty0 2634--2653, 2007.
\newblock \doi{10.1161/CIRCULATIONAHA.107.187397}.
\newblock URL
  \url{https://www.ahajournals.org/doi/abs/10.1161/CIRCULATIONAHA.107.187397}.

\bibitem[Takahashi et~al.(2007)Takahashi, Tanabe, Ohnuki, Narita, Ichisaka,
  Tomoda, and Yamanaka]{takahashi2007induction}
Kazutoshi Takahashi, Koji Tanabe, Mari Ohnuki, Megumi Narita, Tomoko Ichisaka,
  Kiichiro Tomoda, and Shinya Yamanaka.
\newblock Induction of pluripotent stem cells from adult human fibroblasts by
  defined factors.
\newblock \emph{cell}, 131\penalty0 (5):\penalty0 861--872, 2007.
\newblock \doi{https://doi.org/10.1016/j.cell.2007.11.019}.

\bibitem[Brand{\~a}o et~al.(2017)Brand{\~a}o, Tabel, Atsma, Mummery, and
  Davis]{brandao2017human}
Karina~O Brand{\~a}o, Viola~A Tabel, Douwe~E Atsma, Christine~L Mummery, and
  Richard~P Davis.
\newblock Human pluripotent stem cell models of cardiac disease: from
  mechanisms to therapies.
\newblock \emph{Disease models \& mechanisms}, 10\penalty0 (9):\penalty0
  1039--1059, 2017.
\newblock \doi{https://doi.org/10.1242/dmm.030320}.

\bibitem[Nakao et~al.(2020)Nakao, Ihara, Hasegawa, and
  Kawamura]{nakao2020applications}
Shu Nakao, Dai Ihara, Koji Hasegawa, and Teruhisa Kawamura.
\newblock Applications for induced pluripotent stem cells in disease modelling
  and drug development for heart diseases.
\newblock \emph{European Cardiology Review}, 15, 2020.
\newblock \doi{doi:10.15420/ecr.2019.03}.

\bibitem[Daly et~al.(2021)Daly, Davidson, and Burdick]{daly2021bioprinting}
Andrew~C Daly, Matthew~D Davidson, and Jason~A Burdick.
\newblock 3d bioprinting of high cell-density heterogeneous tissue models
  through spheroid fusion within self-healing hydrogels.
\newblock \emph{Nature communications}, 12\penalty0 (1):\penalty0 753, 2021.
\newblock \doi{https://doi.org/10.1038/s41467-021-21029-2}.

\bibitem[Dou et~al.(2022)Dou, Malhi, Zhao, Wang, Huang, Law, Liu, Simmons,
  Maynes, and Sun]{dou2022microengineered}
Wenkun Dou, Manpreet Malhi, Qili Zhao, Li~Wang, Zongjie Huang, Junhui Law,
  Na~Liu, Craig~A Simmons, Jason~T Maynes, and Yu~Sun.
\newblock Microengineered platforms for characterizing the contractile function
  of in vitro cardiac models.
\newblock \emph{Microsystems \& Nanoengineering}, 8\penalty0 (1):\penalty0 26,
  2022.
\newblock \doi{https://doi.org/10.1038/s41378-021-00344-0}.

\bibitem[Hnatiuk et~al.(2021)Hnatiuk, Briganti, Staudt, and
  Mercola]{hnatiuk2021human}
Anna~P Hnatiuk, Francesca Briganti, David~W Staudt, and Mark Mercola.
\newblock Human i{PSC} modeling of heart disease for drug development.
\newblock \emph{Cell chemical biology}, 28\penalty0 (3):\penalty0 271--282,
  2021.
\newblock \doi{https://doi.org/10.1016/j.chembiol.2021.02.016}.

\bibitem[Ronaldson-Bouchard and Vunjak-Novakovic(2018)]{ronaldson2018organs}
Kacey Ronaldson-Bouchard and Gordana Vunjak-Novakovic.
\newblock Organs-on-a-chip: a fast track for engineered human tissues in drug
  development.
\newblock \emph{Cell stem cell}, 22\penalty0 (3):\penalty0 310--324, 2018.
\newblock \doi{https://doi.org/10.1016/j.stem.2018.02.011}.

\bibitem[Hirt et~al.(2014)Hirt, Hansen, and Eschenhagen]{hirt2014cardiac}
Marc~N Hirt, Arne Hansen, and Thomas Eschenhagen.
\newblock Cardiac tissue engineering: state of the art.
\newblock \emph{Circulation research}, 114\penalty0 (2):\penalty0 354--367,
  2014.
\newblock \doi{https://doi.org/10.1161/CIRCRESAHA.114.300522}.

\bibitem[Masumoto et~al.(2014)Masumoto, Ikuno, Takeda, Fukushima, Marui,
  Katayama, Shimizu, Ikeda, Okano, Sakata, et~al.]{masumoto2014human}
Hidetoshi Masumoto, Takeshi Ikuno, Masafumi Takeda, Hiroyuki Fukushima, Akira
  Marui, Shiori Katayama, Tatsuya Shimizu, Tadashi Ikeda, Teruo Okano, Ryuzo
  Sakata, et~al.
\newblock Human ips cell-engineered cardiac tissue sheets with cardiomyocytes
  and vascular cells for cardiac regeneration.
\newblock \emph{Scientific reports}, 4\penalty0 (1):\penalty0 1--7, 2014.
\newblock \doi{https://doi.org/10.1038/srep06716}.

\bibitem[Karbassi et~al.(2020)Karbassi, Fenix, Marchiano, Muraoka, Nakamura,
  Yang, and Murry]{karbassi2020cardiomyocyte}
Elaheh Karbassi, Aidan Fenix, Silvia Marchiano, Naoto Muraoka, Kenta Nakamura,
  Xiulan Yang, and Charles~E Murry.
\newblock Cardiomyocyte maturation: advances in knowledge and implications for
  regenerative medicine.
\newblock \emph{Nature Reviews Cardiology}, 17\penalty0 (6):\penalty0 341--359,
  2020.
\newblock \doi{https://doi.org/10.1038/s41569-019-0331-x}.

\bibitem[DePalma et~al.(2021)DePalma, Davidson, Stis, Helms, and
  Baker]{depalma2021microenvironmental}
Samuel~J. DePalma, Christopher~D. Davidson, Austin~E. Stis, Adam~S. Helms, and
  Brendon~M. Baker.
\newblock Microenvironmental determinants of organized i{PSC}-cardiomyocyte
  tissues on synthetic fibrous matrices.
\newblock \emph{Biomater. Sci.}, 9:\penalty0 93--107, 2021.
\newblock \doi{10.1039/D0BM01247E}.
\newblock URL \url{http://dx.doi.org/10.1039/D0BM01247E}.

\bibitem[Batalov et~al.(2021)Batalov, Jallerat, Kim, Bliley, and
  Feinberg]{batalov2021engineering}
Ivan Batalov, Quentin Jallerat, Sean Kim, Jacqueline Bliley, and Adam~W
  Feinberg.
\newblock Engineering aligned human cardiac muscle using developmentally
  inspired fibronectin micropatterns.
\newblock \emph{Scientific Reports}, 11\penalty0 (1):\penalty0 1--14, 2021.
\newblock \doi{https://doi.org/10.1038/s41598-021-87550-y}.

\bibitem[Jayne et~al.(2021)Jayne, Karakan, Zhang, Pierce, Michas, Bishop, Chen,
  Ekinci, and White]{jayne2021direct}
Rachael~K Jayne, M~{\c{C}}a{\u{g}}atay Karakan, Kehan Zhang, Noelle Pierce,
  Christos Michas, David~J Bishop, Christopher~S Chen, Kamil~L Ekinci, and
  Alice~E White.
\newblock Direct laser writing for cardiac tissue engineering: a microfluidic
  heart on a chip with integrated transducers.
\newblock \emph{Lab on a Chip}, 21\penalty0 (9):\penalty0 1724--1737, 2021.
\newblock \doi{10.1039/D0LC01078B}.

\bibitem[Karakan(2023)]{karakan2023direct}
M~{\c{C}}a{\u{g}}atay Karakan.
\newblock \emph{A Direct-Laser-Written Heart-on-a-Chip Platform for Generation
  and Stimulation of Engineered Heart Tissues}.
\newblock Ph.d. thesis, Boston University, 2023.

\bibitem[Ronaldson-Bouchard et~al.(2019)Ronaldson-Bouchard, Yeager, Teles,
  Chen, Ma, Song, Morikawa, Wobma, Vasciaveo, Ruiz,
  et~al.]{ronaldson2019engineering}
Kacey Ronaldson-Bouchard, Keith Yeager, Diogo Teles, Timothy Chen, Stephen Ma,
  LouJin Song, Kumi Morikawa, Holly~M Wobma, Alessandro Vasciaveo, Edward~C
  Ruiz, et~al.
\newblock Engineering of human cardiac muscle electromechanically matured to an
  adult-like phenotype.
\newblock \emph{Nature protocols}, 14\penalty0 (10):\penalty0 2781--2817, 2019.
\newblock \doi{https://doi.org/10.1038/s41596-019-0189-8}.

\bibitem[Javor et~al.(2021)Javor, Sundaram, Chen, and Bishop]{javor2021pillar}
Josh Javor, Subramanian Sundaram, Christopher~S. Chen, and David~J. Bishop.
\newblock A microtissue platform to simultaneously actuate and detect
  mechanical forces via non-contact magnetic approach.
\newblock \emph{Journal of Microelectromechanical Systems}, 30\penalty0
  (1):\penalty0 96--104, 2021.
\newblock \doi{10.1109/JMEMS.2020.3036978}.

\bibitem[Zhao et~al.(2019)Zhao, Rafatian, Feric, Cox, Aschar-Sobbi, Wang,
  Aggarwal, Zhang, Conant, Ronaldson-Bouchard, et~al.]{zhao2019platform}
Yimu Zhao, Naimeh Rafatian, Nicole~T Feric, Brian~J Cox, Roozbeh Aschar-Sobbi,
  Erika~Yan Wang, Praful Aggarwal, Boyang Zhang, Genevieve Conant, Kacey
  Ronaldson-Bouchard, et~al.
\newblock A platform for generation of chamber-specific cardiac tissues and
  disease modeling.
\newblock \emph{Cell}, 176\penalty0 (4):\penalty0 913--927, 2019.
\newblock \doi{https://doi.org/10.1016/j.cell.2018.11.042}.

\bibitem[Ruan et~al.(2016)Ruan, Tulloch, Razumova, Saiget, Muskheli, Pabon,
  Reinecke, Regnier, and Murry]{ruan2016mechanical}
Jia-Ling Ruan, Nathaniel~L Tulloch, Maria~V Razumova, Mark Saiget, Veronica
  Muskheli, Lil Pabon, Hans Reinecke, Michael Regnier, and Charles~E Murry.
\newblock Mechanical stress conditioning and electrical stimulation promote
  contractility and force maturation of induced pluripotent stem cell-derived
  human cardiac tissue.
\newblock \emph{Circulation}, 134\penalty0 (20):\penalty0 1557--1567, 2016.
\newblock \doi{https://doi.org/10.1161/CIRCULATIONAHA.114.014998}.

\bibitem[Feric et~al.(2019)Feric, Pallotta, Singh, Bogdanowicz, Gustilo,
  Chaudhary, Willette, Chendrimada, Xu, Graziano, et~al.]{feric2019engineered}
Nicole~T Feric, Isabella Pallotta, Rishabh Singh, Danielle~R Bogdanowicz,
  Marietta~M Gustilo, Khuram~W Chaudhary, Robert~N Willette, Tim~P Chendrimada,
  Xiaoping Xu, Michael~P Graziano, et~al.
\newblock Engineered cardiac tissues generated in the biowire ii: a platform
  for human-based drug discovery.
\newblock \emph{Toxicological sciences}, 172\penalty0 (1):\penalty0 89--97,
  2019.
\newblock \doi{https://doi.org/10.1093/toxsci/kfz168}.

\bibitem[Huang et~al.(2020)Huang, Maia-Joca, Ong, Wilson, DiSilvestre,
  Tomaselli, and Reich]{huang2020enhancement}
Chen~Yu Huang, Rebeca Peres~Moreno Maia-Joca, Chin~Siang Ong, Ijala Wilson,
  Deborah DiSilvestre, Gordon~F Tomaselli, and Daniel~H Reich.
\newblock Enhancement of human i{PSC}-derived cardiomyocyte maturation by
  chemical conditioning in a 3{D} environment.
\newblock \emph{Journal of molecular and cellular cardiology}, 138:\penalty0
  1--11, 2020.
\newblock \doi{https://doi.org/10.1016/j.yjmcc.2019.10.001}.

\bibitem[Lee et~al.(2019)Lee, Manoharan, Cheung, Lee, Cha, Newman, Farzad,
  Mehrotra, Zhang, Khan, et~al.]{lee2019nanoparticle}
Junmin Lee, Vijayan Manoharan, Louis Cheung, Seungkyu Lee, Byung-Hyun Cha,
  Peter Newman, Razieh Farzad, Shreya Mehrotra, Kaizhen Zhang, Fazal Khan,
  et~al.
\newblock Nanoparticle-based hybrid scaffolds for deciphering the role of
  multimodal cues in cardiac tissue engineering.
\newblock \emph{ACS nano}, 13\penalty0 (11):\penalty0 12525--12539, 2019.
\newblock \doi{https://doi.org/10.1021/acsnano.9b03050}.

\bibitem[Huebsch et~al.(2022)Huebsch, Charrez, Neiman, Siemons, Boggess, Wall,
  Charwat, J{\ae}ger, Cleres, Telle, et~al.]{huebsch2022metabolically}
Nathaniel Huebsch, Berenice Charrez, Gabriel Neiman, Brian Siemons, Steven~C
  Boggess, Samuel Wall, Verena Charwat, Karoline~H J{\ae}ger, David Cleres,
  {\AA}shild Telle, et~al.
\newblock Metabolically driven maturation of
  human-induced-pluripotent-stem-cell-derived cardiac microtissues on
  microfluidic chips.
\newblock \emph{Nature Biomedical Engineering}, 6\penalty0 (4):\penalty0
  372--388, 2022.
\newblock \doi{https://doi.org/10.1038/s41551-022-00884-4}.

\bibitem[Pointon et~al.(2017)Pointon, Pilling, Dorval, Wang, Archer, and
  Pollard]{pointon2017cover}
Amy Pointon, James Pilling, Thierry Dorval, Yinhai Wang, Caroline Archer, and
  Christopher Pollard.
\newblock From the cover: high-throughput imaging of cardiac microtissues for
  the assessment of cardiac contraction during drug discovery.
\newblock \emph{Toxicological Sciences}, 155\penalty0 (2):\penalty0 444--457,
  2017.
\newblock \doi{https://doi.org/10.1093/toxsci/kfw227}.

\bibitem[Huebsch et~al.(2015)Huebsch, Loskill, Mandegar, Marks, Sheehan, Ma,
  Mathur, Nguyen, Yoo, Judge, et~al.]{huebsch2015automated}
Nathaniel Huebsch, Peter Loskill, Mohammad~A Mandegar, Natalie~C Marks, Alice~S
  Sheehan, Zhen Ma, Anurag Mathur, Trieu~N Nguyen, Jennie~C Yoo, Luke~M Judge,
  et~al.
\newblock Automated video-based analysis of contractility and calcium flux in
  human-induced pluripotent stem cell-derived cardiomyocytes cultured over
  different spatial scales.
\newblock \emph{Tissue Engineering Part C: Methods}, 21\penalty0 (5):\penalty0
  467--479, 2015.
\newblock \doi{https://doi.org/10.1089/ten.tec.2014.0283}.

\bibitem[Ronaldson-Bouchard et~al.(2018)Ronaldson-Bouchard, Ma, Yeager, Chen,
  Song, Sirabella, Morikawa, Teles, Yazawa, and
  Vunjak-Novakovic]{ronaldson2018advanced}
Kacey Ronaldson-Bouchard, Stephen~P Ma, Keith Yeager, Timothy Chen, LouJin
  Song, Dario Sirabella, Kumi Morikawa, Diogo Teles, Masayuki Yazawa, and
  Gordana Vunjak-Novakovic.
\newblock Advanced maturation of human cardiac tissue grown from pluripotent
  stem cells.
\newblock \emph{Nature}, 556\penalty0 (7700):\penalty0 239--243, 2018.
\newblock \doi{https://doi.org/10.1038/s41586-018-0016-3}.

\bibitem[Hansen et~al.(2010)Hansen, Eder, Bönstrup, Flato, Mewe, Schaaf,
  Aksehirlioglu, Schwörer, Uebeler, and Eschenhagen]{hansen2010development}
Arne Hansen, Alexandra Eder, Marlene Bönstrup, Marianne Flato, Marco Mewe,
  Sebastian Schaaf, Bülent Aksehirlioglu, Alexander Schwörer, June Uebeler,
  and Thomas Eschenhagen.
\newblock Development of a drug screening platform based on engineered heart
  tissue.
\newblock \emph{Circulation research}, 107\penalty0 (1):\penalty0 35--44, 2010.
\newblock \doi{https://doi.org/10.1161/CIRCRESAHA.109.211458}.

\bibitem[Tamargo et~al.(2021)Tamargo, Nash, Fleischer, Kim, Vila, Yeager,
  Summers, Zhao, Lock, Chavez, et~al.]{tamargo2021millipillar}
Manuel~Alejandro Tamargo, Trevor~Ray Nash, Sharon Fleischer, Youngbin Kim,
  Olaia~Fernandez Vila, Keith Yeager, Max Summers, Yimu Zhao, Roberta Lock,
  Miguel Chavez, et~al.
\newblock millipillar: a platform for the generation and real-time assessment
  of human engineered cardiac tissues.
\newblock \emph{ACS biomaterials science \& engineering}, 7\penalty0
  (11):\penalty0 5215--5229, 2021.
\newblock \doi{https://doi.org/10.1021/acsbiomaterials.1c01006}.

\bibitem[Rivera-Arbel{\'a}ez et~al.(2022)Rivera-Arbel{\'a}ez,
  Cofi{\~n}o-Fabres, Schwach, Boonen, Ten~Den, Vermeul, van~den Berg, Segerink,
  Ribeiro, and Passier]{rivera2022contractility}
Jos{\'e}~M Rivera-Arbel{\'a}ez, Carla Cofi{\~n}o-Fabres, Verena Schwach, Tom
  Boonen, Simone~A Ten~Den, Kim Vermeul, Albert van~den Berg, Loes~I Segerink,
  Marcelo~C Ribeiro, and Robert Passier.
\newblock Contractility analysis of human engineered 3d heart tissues by an
  automatic tracking technique using a standalone application.
\newblock \emph{Plos one}, 17\penalty0 (4):\penalty0 e0266834, 2022.
\newblock \doi{https://doi.org/10.1371/journal.pone.0266834}.

\bibitem[Oyunbaatar et~al.(2016)Oyunbaatar, Lee, Patil, Kim, and
  Lee]{oyunbaatar2016biomechanical}
Nomin-Erdene Oyunbaatar, Deok-Hyu Lee, Swati~J Patil, Eung-Sam Kim, and
  Dong-Weon Lee.
\newblock Biomechanical characterization of cardiomyocyte using pdms pillar
  with microgrooves.
\newblock \emph{Sensors}, 16\penalty0 (8):\penalty0 1258, 2016.
\newblock \doi{10.3390/s16081258}.

\bibitem[Dostanić et~al.(2020)Dostanić, Windt, Stein, van Meer, Bellin,
  Orlova, Mastrangeli, Mummery, and Sarro]{dostanic2020miniaturized}
Milica Dostanić, Laura~M. Windt, Jeroen~M. Stein, Berend~J. van Meer, Milena
  Bellin, Valeria Orlova, Massimo Mastrangeli, Christine~L. Mummery, and
  Pasqualina~M. Sarro.
\newblock A miniaturized eht platform for accurate measurements of tissue
  contractile properties.
\newblock \emph{Journal of Microelectromechanical Systems}, 29\penalty0
  (5):\penalty0 881--887, 2020.
\newblock \doi{10.1109/JMEMS.2020.3011196}.

\bibitem[Thavandiran et~al.(2020)Thavandiran, Hale, Blit, Sandberg, McElvain,
  Gagliardi, Sun, Witty, Graham, Do, et~al.]{thavandiran2020functional}
Nimalan Thavandiran, Christopher Hale, Patrick Blit, Mark~L Sandberg, Michele~E
  McElvain, Mark Gagliardi, Bo~Sun, Alec Witty, George Graham, Van~TH Do,
  et~al.
\newblock Functional arrays of human pluripotent stem cell-derived cardiac
  microtissues.
\newblock \emph{Scientific reports}, 10\penalty0 (1):\penalty0 6919, 2020.
\newblock \doi{https://doi.org/10.1038/s41598-020-62955-3}.

\bibitem[Tsan et~al.(2021)Tsan, DePalma, Zhao, Capilnasiu, Wu, Elder, Panse,
  Ufford, Matera, Friedline, et~al.]{tsan2021physiologic}
Yao-Chang Tsan, Samuel~J DePalma, Yan-Ting Zhao, Adela Capilnasiu, Yu-Wei Wu,
  Brynn Elder, Isabella Panse, Kathryn Ufford, Daniel~L Matera, Sabrina
  Friedline, et~al.
\newblock Physiologic biomechanics enhance reproducible contractile development
  in a stem cell derived cardiac muscle platform.
\newblock \emph{Nature Communications}, 12\penalty0 (1):\penalty0 6167, 2021.
\newblock \doi{https://doi.org/10.1038/s41467-021-26496-1}.

\bibitem[M{\'e}ry et~al.(2023)M{\'e}ry, Ruppel, Revilloud, Balland, Cappello,
  and Boudou]{mery2023light}
Adrien M{\'e}ry, Artur Ruppel, Jean Revilloud, Martial Balland, Giovanni
  Cappello, and Thomas Boudou.
\newblock Light-driven biological actuators to probe the rheology of 3d
  microtissues.
\newblock \emph{Nature Communications}, 14\penalty0 (1):\penalty0 717, 2023.
\newblock \doi{https://doi.org/10.1038/s41467-023-36371-w}.

\bibitem[Steadman et~al.(1988)Steadman, Moore, Spitzer, and
  Bridge]{steadman1988video}
BW~Steadman, KB~Moore, KW~Spitzer, and JHB Bridge.
\newblock A video system for measuring motion in contracting heart cells.
\newblock \emph{IEEE Transactions on Biomedical Engineering}, 35\penalty0
  (4):\penalty0 264--272, 1988.
\newblock \doi{doi: 10.1109/10.1375}.

\bibitem[Lim et~al.(2008)Lim, Yang, Yang, Wang, Shi, Berg, Pimentel, Gwathmey,
  Hajjar, Helmes, et~al.]{lim2008novel}
Chee~Chew Lim, Haijun Yang, Mingfeng Yang, Chien-Kao Wang, Jianru Shi, Eric~A
  Berg, David~R Pimentel, Judith~K Gwathmey, Roger~J Hajjar, Michiel Helmes,
  et~al.
\newblock A novel mutant cardiac troponin c disrupts molecular motions critical
  for calcium binding affinity and cardiomyocyte contractility.
\newblock \emph{Biophysical journal}, 94\penalty0 (9):\penalty0 3577--3589,
  2008.
\newblock \doi{https://doi.org/10.1529/biophysj.107.112896}.

\bibitem[Hossain et~al.(2010)Hossain, Shimizu, Saito, Rao, Yamaguchi, and
  Tamiya]{hossain2010non}
Mohammad~Mosharraf Hossain, Eiichi Shimizu, Masato Saito, Sathuluri~Ramachandra
  Rao, Yoshinori Yamaguchi, and Eiichi Tamiya.
\newblock Non-invasive characterization of mouse embryonic stem cell derived
  cardiomyocytes based on the intensity variation in digital beating video.
\newblock \emph{Analyst}, 135\penalty0 (7):\penalty0 1624--1630, 2010.
\newblock \doi{10.1039/C0AN00208A}.

\bibitem[Hayakawa et~al.(2012)Hayakawa, Kunihiro, Dowaki, Uno, Matsui, Uchida,
  Kobayashi, Yasuda, Shimizu, and Okano]{hayakawa2012noninvasive}
Tomohiro Hayakawa, Takeshi Kunihiro, Suguru Dowaki, Hatsume Uno, Eriko Matsui,
  Masashi Uchida, Seiji Kobayashi, Akio Yasuda, Tatsuya Shimizu, and Teruo
  Okano.
\newblock Noninvasive evaluation of contractile behavior of cardiomyocyte
  monolayers based on motion vector analysis.
\newblock \emph{Tissue Engineering Part C: Methods}, 18\penalty0 (1):\penalty0
  21--32, 2012.
\newblock \doi{https://doi-org.ezproxy.bu.edu/10.1089/ten.tec.2011.0273}.

\bibitem[Ahola et~al.(2014)Ahola, Kiviaho, Larsson, Honkanen,
  Aalto-Set{\"a}l{\"a}, and Hyttinen]{ahola2014video}
Antti Ahola, Anna~L Kiviaho, Kim Larsson, Markus Honkanen, Katriina
  Aalto-Set{\"a}l{\"a}, and Jari Hyttinen.
\newblock Video image-based analysis of single human induced pluripotent stem
  cell derived cardiomyocyte beating dynamics using digital image correlation.
\newblock \emph{Biomedical engineering online}, 13\penalty0 (1):\penalty0
  1--18, 2014.
\newblock \doi{https://doi.org/10.1186/1475-925X-13-39}.

\bibitem[Hayakawa et~al.(2014)Hayakawa, Kunihiro, Ando, Kobayashi, Matsui,
  Yada, Kanda, Kurokawa, and Furukawa]{hayawaka2014image}
Tomohiro Hayakawa, Takeshi Kunihiro, Tomoko Ando, Seiji Kobayashi, Eriko
  Matsui, Hiroaki Yada, Yasunari Kanda, Junko Kurokawa, and Tetsushi Furukawa.
\newblock Image-based evaluation of contraction–relaxation kinetics of
  human-induced pluripotent stem cell-derived cardiomyocytes: Correlation and
  complementarity with extracellular electrophysiology.
\newblock \emph{Journal of Molecular and Cellular Cardiology}, 77:\penalty0
  178--191, 2014.
\newblock ISSN 0022-2828.
\newblock \doi{https://doi.org/10.1016/j.yjmcc.2014.09.010}.
\newblock URL
  \url{https://www.sciencedirect.com/science/article/pii/S0022282814002910}.

\bibitem[Rajasingh et~al.(2015)Rajasingh, Thangavel, Czirok, Samanta, Roby,
  Dawn, and Rajasingh]{rajasingh2015generation}
Sheeja Rajasingh, Jayakumar Thangavel, Andras Czirok, Saheli Samanta,
  Katherine~F Roby, Buddhadeb Dawn, and Johnson Rajasingh.
\newblock Generation of functional cardiomyocytes from efficiently generated
  human i{PSC}s and a novel method of measuring contractility.
\newblock \emph{PLoS One}, 10\penalty0 (8):\penalty0 e0134093, 2015.
\newblock \doi{https://doi.org/10.1371/journal.pone.0134093}.

\bibitem[Czirok et~al.(2017)Czirok, Isai, Kosa, Rajasingh, Kinsey, Neufeld, and
  Rajasingh]{czirok2017optical}
Andras Czirok, Dona~Greta Isai, Edina Kosa, Sheeja Rajasingh, William Kinsey,
  Zoltan Neufeld, and Johnson Rajasingh.
\newblock Optical-flow based non-invasive analysis of cardiomyocyte
  contractility.
\newblock \emph{Scientific reports}, 7\penalty0 (1):\penalty0 10404, 2017.
\newblock \doi{10.1038/s41598-017-10094-7}.

\bibitem[Shradhanjali et~al.(2019)Shradhanjali, Riehl, Duan, Yang, and
  Lim]{shradhanjali2019spatiotemporal}
Akankshya Shradhanjali, Brandon~D Riehl, Bin Duan, Ruiguo Yang, and Jung~Yul
  Lim.
\newblock Spatiotemporal characterizations of spontaneously beating
  cardiomyocytes with adaptive reference digital image correlation.
\newblock \emph{Scientific reports}, 9\penalty0 (1):\penalty0 1--10, 2019.
\newblock \doi{https://doi.org/10.1038/s41598-019-54768-w}.

\bibitem[Scalzo et~al.(2021)Scalzo, Afonso, da~Fonseca, Jesus, Alves,
  Mendon{\c{c}}a, Teixeira, Biagi, Cruvinel, Santos, et~al.]{scalzo2021dense}
S{\'e}rgio Scalzo, Marcelo~QL Afonso, N{\'e}li~J da~Fonseca, Itamar~CG Jesus,
  Ana~Paula Alves, Carolina~ATF Mendon{\c{c}}a, Vanessa~P Teixeira, Diogo
  Biagi, Estela Cruvinel, Anderson~K Santos, et~al.
\newblock Dense optical flow software to quantify cellular contractility.
\newblock \emph{Cell Reports Methods}, 1\penalty0 (4), 2021.
\newblock \doi{https://doi.org/10.1016/j.crmeth.2021.100044}.

\bibitem[Cheng et~al.(2023)Cheng, Yang, Jiang, Nie, Yang, Tu, Liang, and
  Xiang]{cheng2023quantification}
Zhiyang Cheng, Yuxin Yang, Kai Jiang, Hongyi Nie, Xingbo Yang, Zizhuo Tu, Jiayi
  Liang, and Yaozu Xiang.
\newblock Quantification of cardiomyocyte contraction in vitro and drug
  screening by myocytobeats.
\newblock \emph{Journal of Cardiovascular Translational Research}, pages 1--10,
  2023.
\newblock \doi{10.1007/s12265-023-10357-x}.

\bibitem[Sala et~al.(2018)Sala, Van~Meer, Tertoolen, Bakkers, Bellin, Davis,
  Denning, Dieben, Eschenhagen, Giacomelli, et~al.]{sala2018musclemotion}
Luca Sala, Berend~J Van~Meer, Leon~GJ Tertoolen, Jeroen Bakkers, Milena Bellin,
  Richard~P Davis, Chris Denning, Michel~AE Dieben, Thomas Eschenhagen, Elisa
  Giacomelli, et~al.
\newblock Musclemotion: a versatile open software tool to quantify
  cardiomyocyte and cardiac muscle contraction in vitro and in vivo.
\newblock \emph{Circulation research}, 122\penalty0 (3):\penalty0 e5--e16,
  2018.
\newblock \doi{https://doi.org/10.1161/CIRCRESAHA.117.312067}.

\bibitem[Bouguet et~al.(2001)]{bouguet2001pyramidal}
Jean-Yves Bouguet et~al.
\newblock Pyramidal implementation of the affine lucas kanade feature tracker
  description of the algorithm.
\newblock \emph{Intel corporation}, 5\penalty0 (1-10):\penalty0 4, 2001.

\bibitem[Das et~al.(2022)Das, Sutherland, Lejeune, Eyckmans, and
  Chen]{das2022mechanical}
Shoshana~L Das, Bryan~P Sutherland, Emma Lejeune, Jeroen Eyckmans, and
  Christopher~S Chen.
\newblock Mechanical response of cardiac microtissues to acute localized
  injury.
\newblock \emph{American Journal of Physiology-Heart and Circulatory
  Physiology}, 323\penalty0 (4):\penalty0 H738--H748, 2022.
\newblock \doi{https://doi.org/10.1152/ajpheart.00305.2022}.

\bibitem[Raghupathy(2011)]{raghupathy2011form}
Ramesh Raghupathy.
\newblock \emph{Form from function: generalized anisotropic inverse mechanics
  for soft tissues}.
\newblock University of Minnesota, 2011.

\bibitem[Genovese et~al.(2014)Genovese, Casaletto, Humphrey, and
  Lu]{genovese2014digital}
Katia Genovese, Luciana Casaletto, Jay~D Humphrey, and Jia Lu.
\newblock Digital image correlation-based point-wise inverse characterization
  of heterogeneous material properties of gallbladder in vitro.
\newblock \emph{Proceedings of the Royal Society A: Mathematical, Physical and
  Engineering Sciences}, 470\penalty0 (2167):\penalty0 20140152, 2014.
\newblock \doi{https://doi.org/10.1098/rspa.2014.0152}.

\bibitem[Narayanan et~al.(2021)Narayanan, Olender, Marlevi, Edelman, and
  Nezami]{narayanan2021inverse}
Bharath Narayanan, Max~L Olender, David Marlevi, Elazer~R Edelman, and Farhad~R
  Nezami.
\newblock An inverse method for mechanical characterization of heterogeneous
  diseased arteries using intravascular imaging.
\newblock \emph{Scientific Reports}, 11\penalty0 (1):\penalty0 22540, 2021.
\newblock \doi{https://doi.org/10.1038/s41598-021-01874-3}.

\bibitem[Wang et~al.(2021)Wang, Estrada, Arruda, and
  Garikipati]{wang2021inference}
Zhenlin Wang, Jonathan~B Estrada, Ellen~M Arruda, and Krishna Garikipati.
\newblock Inference of deformation mechanisms and constitutive response of soft
  material surrogates of biological tissue by full-field characterization and
  data-driven variational system identification.
\newblock \emph{Journal of the Mechanics and Physics of Solids}, 153:\penalty0
  104474, 2021.
\newblock \doi{10.1016/j.jmps.2021.104474}.

\bibitem[Weiss et~al.(2020)Weiss, Cavinato, Gray, Ramachandra, Avril, Humphrey,
  and Latorre]{weiss2020mechanics}
Dar Weiss, Cristina Cavinato, Authia Gray, Abhay~B Ramachandra, Stephane Avril,
  Jay~D Humphrey, and Marcos Latorre.
\newblock Mechanics-driven mechanobiological mechanisms of arterial tortuosity.
\newblock \emph{Science advances}, 6\penalty0 (49):\penalty0 eabd3574, 2020.
\newblock \doi{10.1126/sciadv.abd3574}.

\bibitem[Sree et~al.(2019)Sree, Rausch, and Tepole]{Sree2019linking}
Vivek~D Sree, Manuel~K Rausch, and Adrian~B Tepole.
\newblock Linking microvascular collapse to tissue hypoxia in a multiscale
  model of pressure ulcer initiation.
\newblock \emph{Biomechanics and modeling in mechanobiology}, 18\penalty0
  (6):\penalty0 1947--1964, 2019.
\newblock \doi{https://doi.org/10.1007/s10237-019-01187-5}.

\bibitem[Kong et~al.(2018)Kong, Pham, Martin, McKay, Primiano, Hashim, Kodali,
  and Sun]{kong2018finite}
Fanwei Kong, Thuy Pham, Caitlin Martin, Raymond McKay, Charles Primiano, Sabet
  Hashim, Susheel Kodali, and Wei Sun.
\newblock Finite element analysis of tricuspid valve deformation from
  multi-slice computed tomography images.
\newblock \emph{Annals of biomedical engineering}, 46\penalty0 (8):\penalty0
  1112--1127, 2018.
\newblock \doi{https://doi.org/10.1007/s10439-018-2024-8}.

\bibitem[Joldes et~al.(2019)Joldes, Bourantas, Zwick, Chowdhury, Wittek,
  Agrawal, Mountris, Hyde, Warfield, and Miller]{Joldes2019Suite}
Grand Joldes, George Bourantas, Benjamin Zwick, Habib Chowdhury, Adam Wittek,
  Sudip Agrawal, Konstantinos Mountris, Damon Hyde, Simon~K. Warfield, and
  Karol Miller.
\newblock Suite of meshless algorithms for accurate computation of soft tissue
  deformation for surgical simulation.
\newblock \emph{Medical Image Analysis}, 56:\penalty0 152--171, 2019.
\newblock ISSN 1361-8415.
\newblock \doi{https://doi.org/10.1016/j.media.2019.06.004}.

\bibitem[de~Z{\'e}licourt et~al.(2010)de~Z{\'e}licourt, Marsden, Fogel, and
  Yoganathan]{de2010imaging}
Diane~A de~Z{\'e}licourt, Alison Marsden, Mark~A Fogel, and Ajit~P Yoganathan.
\newblock Imaging and patient-specific simulations for the fontan surgery:
  current methodologies and clinical applications.
\newblock \emph{Progress in Pediatric Cardiology}, 30\penalty0 (1-2):\penalty0
  31--44, 2010.
\newblock \doi{https://doi.org/10.1016/j.ppedcard.2010.09.005}.

\bibitem[Wang et~al.(2015)Wang, Kieu, Nguyen, and Le]{wang2015digital}
Zhaoyang Wang, Hien Kieu, Hieu Nguyen, and Minh Le.
\newblock Digital image correlation in experimental mechanics and image
  registration in computer vision: Similarities, differences and complements.
\newblock \emph{Optics and Lasers in Engineering}, 65:\penalty0 18--27, 2015.
\newblock \doi{https://doi.org/10.1016/j.optlaseng.2014.04.002}.

\bibitem[Chu et~al.(1985)Chu, Ranson, and Sutton]{chu1985applications}
TC~Chu, WF~Ranson, and Michael~A Sutton.
\newblock Applications of digital-image-correlation techniques to experimental
  mechanics.
\newblock \emph{Experimental mechanics}, 25:\penalty0 232--244, 1985.
\newblock \doi{https://doi.org/10.1007/BF02325092}.

\bibitem[Zhang and Arola(2004)]{zhang2004applications}
Dongsheng Zhang and Dwayne~D Arola.
\newblock Applications of digital image correlation to biological tissues.
\newblock \emph{Journal of Biomedical Optics}, 9\penalty0 (4):\penalty0
  691--699, 2004.
\newblock \doi{https://doi.org/10.1117/1.1753270}.

\bibitem[Annaidh et~al.(2012)Annaidh, Bruy{\`e}re, Destrade, Gilchrist, and
  Ott{\'e}nio]{annaidh2012characterization}
Aisling~N{\'\i} Annaidh, Karine Bruy{\`e}re, Michel Destrade, Michael~D
  Gilchrist, and M{\'e}lanie Ott{\'e}nio.
\newblock Characterization of the anisotropic mechanical properties of excised
  human skin.
\newblock \emph{Journal of the mechanical behavior of biomedical materials},
  5\penalty0 (1):\penalty0 139--148, 2012.
\newblock \doi{https://doi.org/10.1016/j.jmbbm.2011.08.016}.

\bibitem[Blaber et~al.(2015)Blaber, Adair, and Antoniou]{blaber2015ncorr}
J~Blaber, B~Adair, and Antonia Antoniou.
\newblock Ncorr: open-source 2d digital image correlation matlab software.
\newblock \emph{Experimental Mechanics}, 55\penalty0 (6):\penalty0 1105--1122,
  2015.
\newblock \doi{https://doi.org/10.1007/s11340-015-0009-1}.

\bibitem[Yang and Bhattacharya(2021)]{yang2021fast}
Jin Yang and Kaushik Bhattacharya.
\newblock Fast adaptive mesh augmented lagrangian digital image correlation.
\newblock \emph{Experimental Mechanics}, 61\penalty0 (4):\penalty0 719--735,
  2021.
\newblock \doi{https://doi.org/10.1007/s11340-021-00695-9}.

\bibitem[Solav et~al.(2018)Solav, Moerman, Jaeger, Genovese, and
  Herr]{solav2018MultiDIC}
Dana Solav, Kevin~M. Moerman, Aaron~M. Jaeger, Katia Genovese, and Hugh~M.
  Herr.
\newblock Multidic: An open-source toolbox for multi-view 3d digital image
  correlation.
\newblock \emph{IEEE Access}, 6:\penalty0 30520--30535, 2018.
\newblock \doi{10.1109/ACCESS.2018.2843725}.

\bibitem[Murienne and Nguyen(2016)]{murienne2016comparison}
Barbara~J Murienne and Thao~D Nguyen.
\newblock A comparison of 2d and 3d digital image correlation for a membrane
  under inflation.
\newblock \emph{Optics and lasers in engineering}, 77:\penalty0 92--99, 2016.
\newblock \doi{https://doi.org/10.1016/j.optlaseng.2015.07.013}.

\bibitem[Bay et~al.(1999)Bay, Smith, Fyhrie, and Saad]{bay1999digital}
Brian~K Bay, Tait~S Smith, David~P Fyhrie, and Malik Saad.
\newblock Digital volume correlation: three-dimensional strain mapping using
  x-ray tomography.
\newblock \emph{Experimental mechanics}, 39:\penalty0 217--226, 1999.
\newblock \doi{https://doi.org/10.1007/BF02323555}.

\bibitem[Dall’Ara et~al.(2014)Dall’Ara, Barber, and
  Viceconti]{dall2014inevitable}
E~Dall’Ara, D~Barber, and M~Viceconti.
\newblock About the inevitable compromise between spatial resolution and
  accuracy of strain measurement for bone tissue: A 3d zero-strain study.
\newblock \emph{Journal of biomechanics}, 47\penalty0 (12):\penalty0
  2956--2963, 2014.
\newblock \doi{https://doi.org/10.1016/j.jbiomech.2014.07.019}.

\bibitem[Acosta~Santamar{\'\i}a et~al.(2018)Acosta~Santamar{\'\i}a,
  Flechas~Garc{\'\i}a, Molimard, and Avril]{acosta2018three}
V{\'\i}ctor~A Acosta~Santamar{\'\i}a, Mar{\'\i}a Flechas~Garc{\'\i}a,
  J{\'e}r{\^o}me Molimard, and Stephane Avril.
\newblock Three-dimensional full-field strain measurements across a whole
  porcine aorta subjected to tensile loading using optical coherence
  tomography--digital volume correlation.
\newblock \emph{Frontiers in Mechanical Engineering}, 4:\penalty0 3, 2018.
\newblock \doi{https://doi.org/10.3389/fmech.2018.00003}.

\bibitem[Bersi et~al.(2020)Bersi, Acosta~Santamar{\'\i}a, Marback, Di~Achille,
  Phillips, Goergen, Humphrey, and Avril]{bersi2020multimodality}
Matthew~R Bersi, V{\'\i}ctor~A Acosta~Santamar{\'\i}a, Karl Marback, Paolo
  Di~Achille, Evan~H Phillips, Craig~J Goergen, Jay~D Humphrey, and
  St{\'e}phane Avril.
\newblock Multimodality imaging-based characterization of regional material
  properties in a murine model of aortic dissection.
\newblock \emph{Scientific reports}, 10\penalty0 (1):\penalty0 9244, 2020.
\newblock \doi{https://doi.org/10.1038/s41598-020-65624-7}.

\bibitem[Lucas and Kanade(1981)]{lucas1981iterative}
Bruce~D. Lucas and Takeo Kanade.
\newblock An iterative image registration technique with an application to
  stereo vision.
\newblock In \emph{Proceedings of the 7th International Joint Conference on
  Artificial Intelligence - Volume 2}, IJCAI'81, page 674–679, San Francisco,
  CA, USA, 1981. Morgan Kaufmann Publishers Inc.

\bibitem[Boyle et~al.(2014)Boyle, Kume, Wyczalkowski, Taber, Pless, Xia, Genin,
  and Thomopoulos]{boyle2014simple}
John~J Boyle, Maiko Kume, Matthew~A Wyczalkowski, Larry~A Taber, Robert~B
  Pless, Younan Xia, Guy~M Genin, and Stavros Thomopoulos.
\newblock Simple and accurate methods for quantifying deformation, disruption,
  and development in biological tissues.
\newblock \emph{Journal of the Royal Society Interface}, 11\penalty0
  (100):\penalty0 20140685, 2014.
\newblock \doi{http://dx.doi.org/10.1098/rsif.2014.0685}.

\bibitem[Boyle et~al.(2019)Boyle, Soepriatna, Damen, Rowe, Pless, Kovacs,
  Goergen, Thomopoulos, and Genin]{boyle2019regularization}
John~J Boyle, Arvin Soepriatna, Frederick Damen, Roger~A Rowe, Robert~B Pless,
  Attila Kovacs, Craig~J Goergen, Stavros Thomopoulos, and Guy~M Genin.
\newblock Regularization-free strain mapping in three dimensions, with
  application to cardiac ultrasound.
\newblock \emph{Journal of biomechanical engineering}, 141\penalty0
  (1):\penalty0 011010, 2019.
\newblock \doi{10.1115/1.4041576}.

\bibitem[Bourne and Bourne(2010)]{bourne2010imagej}
Roger Bourne and Roger Bourne.
\newblock Imagej.
\newblock \emph{Fundamentals of digital imaging in medicine}, pages 185--188,
  2010.

\bibitem[Inc.(2022)]{MATLAB}
The~MathWorks Inc.
\newblock Matlab, 2022.
\newblock URL \url{https://www.mathworks.com}.

\bibitem[Ghanbari(1990)]{ghanbari1990cross}
M.~Ghanbari.
\newblock The cross-search algorithm for motion estimation (image coding).
\newblock \emph{IEEE Transactions on Communications}, 38\penalty0 (7):\penalty0
  950--953, 1990.
\newblock \doi{10.1109/26.57512}.

\bibitem[Geuzaine and Remacle(2009)]{geuzaine2009gmsh}
Christophe Geuzaine and Jean-Fran{\c{c}}ois Remacle.
\newblock Gmsh: A 3-d finite element mesh generator with built-in pre-and
  post-processing facilities.
\newblock \emph{International journal for numerical methods in engineering},
  79\penalty0 (11):\penalty0 1309--1331, 2009.
\newblock \doi{https://doi.org/10.1002/nme.2579}.

\bibitem[Aln{\ae}s et~al.(2015)Aln{\ae}s, Blechta, Hake, Johansson, Kehlet,
  Logg, Richardson, Ring, Rognes, and Wells]{alnaes2015fenics}
Martin Aln{\ae}s, Jan Blechta, Johan Hake, August Johansson, Benjamin Kehlet,
  Anders Logg, Chris Richardson, Johannes Ring, Marie~E Rognes, and Garth~N
  Wells.
\newblock The {FEniCS} project version 1.5.
\newblock \emph{Archive of Numerical Software}, 3\penalty0 (100):\penalty0
  9--23, 2015.

\bibitem[Logg et~al.(2012)Logg, Mardal, and Wells]{logg2012automated}
Anders Logg, Kent-Andre Mardal, and Garth Wells.
\newblock \emph{Automated solution of differential equations by the finite
  element method: The {FEniCS} book}, volume~84.
\newblock Springer Science \& Business Media, Germany, 2012.

\bibitem[Pezzuto et~al.(2014)Pezzuto, Ambrosi, and
  Quarteroni]{pezzuto2014orthotropic}
S.~Pezzuto, D.~Ambrosi, and A.~Quarteroni.
\newblock An orthotropic active–strain model for the myocardium mechanics and
  its numerical approximation.
\newblock \emph{European Journal of Mechanics - A/Solids}, 48:\penalty0 83--96,
  2014.
\newblock ISSN 0997-7538.
\newblock \doi{https://doi.org/10.1016/j.euromechsol.2014.03.006}.
\newblock URL
  \url{https://www.sciencedirect.com/science/article/pii/S0997753814000527}.
\newblock Frontiers in Finite-Deformation Electromechanics.

\bibitem[Gurev et~al.(2015)Gurev, Pathmanathan, Fattebert, Wen, Magerlein,
  Gray, Richards, and Rice]{gurev2015high}
Viatcheslav Gurev, Pras Pathmanathan, Jean-Luc Fattebert, Hui-Fang Wen, John
  Magerlein, Richard~A Gray, David~F Richards, and J~Jeremy Rice.
\newblock A high-resolution computational model of the deforming human heart.
\newblock \emph{Biomechanics and modeling in mechanobiology}, 14:\penalty0
  829--849, 2015.
\newblock \doi{https://doi.org/10.1007/s10237-014-0639-8}.

\bibitem[Finsberg et~al.(2018)Finsberg, Xi, Tan, Zhong, Genet, Sundnes, Lee,
  and Wall]{finsberg2018efficient}
Henrik Finsberg, Ce~Xi, Ju~Le Tan, Liang Zhong, Martin Genet, Joakim Sundnes,
  Lik~Chuan Lee, and Samuel~T Wall.
\newblock Efficient estimation of personalized biventricular mechanical
  function employing gradient-based optimization.
\newblock \emph{International journal for numerical methods in biomedical
  engineering}, 34\penalty0 (7):\penalty0 e2982, 2018.

\bibitem[Land et~al.(2015)Land, Gurev, Arens, Augustin, Baron, Blake, Bradley,
  Castro, Crozier, Favino, et~al.]{land2015verification}
Sander Land, Viatcheslav Gurev, Sander Arens, Christoph~M Augustin, Lukas
  Baron, Robert Blake, Chris Bradley, Sebastian Castro, Andrew Crozier, Marco
  Favino, et~al.
\newblock Verification of cardiac mechanics software: benchmark problems and
  solutions for testing active and passive material behaviour.
\newblock \emph{Proceedings of the Royal Society A: Mathematical, Physical and
  Engineering Sciences}, 471\penalty0 (2184):\penalty0 20150641, 2015.
\newblock \doi{https://doi.org/10.1098/rspa.2015.0641}.

\bibitem[van~der Walt et~al.(2014)van~der Walt, {S}ch\"onberger,
  {Nunez-Iglesias}, {B}oulogne, {W}arner, {Y}ager, {G}ouillart, {Y}u, and the
  scikit-image contributors]{scikit_image}
{S}t\'efan van~der Walt, {J}ohannes~{L}. {S}ch\"onberger, {J}uan
  {Nunez-Iglesias}, {F}ran\c{c}ois {B}oulogne, {J}oshua~{D}. {W}arner, {N}eil
  {Y}ager, {E}mmanuelle {G}ouillart, {T}ony {Y}u, and the scikit-image
  contributors.
\newblock scikit-image: image processing in {P}ython.
\newblock \emph{PeerJ}, 2:\penalty0 e453, 6 2014.
\newblock ISSN 2167-8359.
\newblock \doi{10.7717/peerj.453}.
\newblock URL \url{https://doi.org/10.7717/peerj.453}.

\bibitem[Jilberto et~al.(2023(in press))Jilberto, DePalma, Lo, Kobeissi,
  Lejeune, Baker, and Nordsletten]{jilberto2023computational}
J.~Jilberto, S.~J. DePalma, J.~Lo, H.~Kobeissi, E.~Lejeune, B.~M. Baker, and
  D.~Nordsletten.
\newblock A data--driven computational modeling for engineered cardiac
  microtissues.
\newblock \emph{Acta Biomaterialia}, 2023(in press).

\bibitem[Perlin(1985)]{perlin1985noise}
Ken Perlin.
\newblock An image synthesizer.
\newblock \emph{Siggraph Computer Graphics}, 19\penalty0 (3):\penalty0
  287–296, jul 1985.
\newblock ISSN 0097-8930.
\newblock \doi{10.1145/325165.325247}.
\newblock URL \url{https://doi.org/10.1145/325165.325247}.

\bibitem[Boudou et~al.(2012)Boudou, Legant, Mu, Borochin, Thavandiran, Radisic,
  Zandstra, Epstein, Margulies, and Chen]{boudou2012microfabricated}
Thomas Boudou, Wesley~R Legant, Anbin Mu, Michael~A Borochin, Nimalan
  Thavandiran, Milica Radisic, Peter~W Zandstra, Jonathan~A Epstein, Kenneth~B
  Margulies, and Christopher~S Chen.
\newblock A microfabricated platform to measure and manipulate the mechanics of
  engineered cardiac microtissues.
\newblock \emph{Tissue Engineering Part A}, 18\penalty0 (9-10):\penalty0
  910--919, 2012.
\newblock \doi{10.1089/ten.tea.2011.0341}.

\bibitem[Xu et~al.(2015)Xu, Zhao, Liu, Metz, Shi, Bose, and
  Reich]{xu2015microfabricated}
Fan Xu, Ruogang Zhao, Alan~S Liu, Tristin Metz, Yu~Shi, Prasenjit Bose, and
  Daniel~H Reich.
\newblock A microfabricated magnetic actuation device for mechanical
  conditioning of arrays of 3d microtissues.
\newblock \emph{Lab on a Chip}, 15\penalty0 (11):\penalty0 2496--2503, 2015.
\newblock \doi{10.1039/C4LC01395F}.

\bibitem[Bielawski et~al.(2016)Bielawski, Leonard, Bhandari, Murry, and
  Sniadecki]{bielawski2016real}
Kevin~S Bielawski, Andrea Leonard, Shiv Bhandari, Chuck~E Murry, and Nathan~J
  Sniadecki.
\newblock Real-time force and frequency analysis of engineered human heart
  tissue derived from induced pluripotent stem cells using magnetic sensing.
\newblock \emph{Tissue Engineering Part C: Methods}, 22\penalty0 (10):\penalty0
  932--940, 2016.
\newblock \doi{10.1089/ten.TEC.2016.0257}.

\bibitem[DePalma et~al.(2023(under review))DePalma, Jilberto, Stis, Huang, Lo,
  Davidson, Jewett, Kobeissi, Chowdhury, Chen, Lejeune, Helms, Nordsletten, and
  Baker]{depalma2023microenvironmental}
S.~J. DePalma, J.~Jilberto, A.~E. Stis, D.~D. Huang, J.~Lo, C.~D. Davidson,
  M.~E. Jewett, H.~Kobeissi, A.~Chowdhury, C.~S. Chen, E.~Lejeune, A.~S. Helms,
  D.~Nordsletten, and B.~M. Baker.
\newblock Microenvironmental mechanical inputs dictate the assembly and
  function of i{PSC}-derived cardiomyocyte microtissues.
\newblock \emph{Nature Communications}, 2023(under review).
\newblock \doi{https://doi.org/10.1101/2023.10.20.563346}.

\bibitem[Lian et~al.(2013)Lian, Zhang, Azarin, Zhu, Hazeltine, Bao, Hsiao,
  Kamp, and Palecek]{lian2013directed}
Xiaojun Lian, Jianhua Zhang, Samira~M Azarin, Kexian Zhu, Laurie~B Hazeltine,
  Xiaoping Bao, Cheston Hsiao, Timothy~J Kamp, and Sean~P Palecek.
\newblock Directed cardiomyocyte differentiation from human pluripotent stem
  cells by modulating wnt/$\beta$-catenin signaling under fully defined
  conditions.
\newblock \emph{Nature protocols}, 8\penalty0 (1):\penalty0 162--175, 2013.
\newblock \doi{https://doi.org/10.1038/nprot.2012.150}.

\bibitem[Kobeissi et~al.(2023)Kobeissi, Jilberto, Karakan, Gao, DePalma, Das,
  Quach, Urquia, Baker, Chen, Nordsletten, and Lejeune]{microbundle2023data}
Hiba Kobeissi, Javiera Jilberto, M~{\c{C}}a{\u{g}}atay Karakan, Xining Gao,
  Samuel~J DePalma, Shoshana~L Das, Lani Quach, Jonathan Urquia, Brendon~M
  Baker, Christopher~S Chen, David Nordsletten, and Emma Lejeune.
\newblock Microbundle {T}ime-lapse {D}ataset, 2023.

\bibitem[Bradski and Kaehler(2008)]{bradski2008learning}
Gary Bradski and Adrian Kaehler.
\newblock \emph{Learning OpenCV: {Computer} vision with the OpenCV library}.
\newblock O'Reilly Media, Inc., Sebastopol, CA, 2008.

\bibitem[Shi et~al.(1994)]{shi1994good}
Jianbo Shi et~al.
\newblock Good features to track.
\newblock In \emph{1994 Proceedings of IEEE conference on computer vision and
  pattern recognition}, pages 593--600. IEEE, 1994.
\newblock \doi{10.1109/CVPR.1994.323794}.

\bibitem[Virtanen et~al.(2020)Virtanen, Gommers, Oliphant, Haberland, Reddy,
  Cournapeau, Burovski, Peterson, Weckesser, Bright, {van der Walt}, Brett,
  Wilson, Millman, Mayorov, Nelson, Jones, Kern, Larson, Carey, Polat, Feng,
  Moore, {VanderPlas}, Laxalde, Perktold, Cimrman, Henriksen, Quintero, Harris,
  Archibald, Ribeiro, Pedregosa, {van Mulbregt}, and {SciPy 1.0
  Contributors}]{2020SciPy}
Pauli Virtanen, Ralf Gommers, Travis~E. Oliphant, Matt Haberland, Tyler Reddy,
  David Cournapeau, Evgeni Burovski, Pearu Peterson, Warren Weckesser, Jonathan
  Bright, St{\'e}fan~J. {van der Walt}, Matthew Brett, Joshua Wilson, K.~Jarrod
  Millman, Nikolay Mayorov, Andrew R.~J. Nelson, Eric Jones, Robert Kern, Eric
  Larson, C~J Carey, {\.I}lhan Polat, Yu~Feng, Eric~W. Moore, Jake
  {VanderPlas}, Denis Laxalde, Josef Perktold, Robert Cimrman, Ian Henriksen,
  E.~A. Quintero, Charles~R. Harris, Anne~M. Archibald, Ant{\^o}nio~H. Ribeiro,
  Fabian Pedregosa, Paul {van Mulbregt}, and {SciPy 1.0 Contributors}.
\newblock {{SciPy} 1.0: Fundamental Algorithms for Scientific Computing in
  Python}.
\newblock \emph{Nature Methods}, 17:\penalty0 261--272, 2020.
\newblock \doi{10.1038/s41592-019-0686-2}.

\bibitem[Stout et~al.(2016)Stout, Bar-Kochba, Estrada, Toyjanova, Kesari,
  Reichner, and Franck]{Franck2016mean_def}
David~A. Stout, Eyal Bar-Kochba, Jonathan~B. Estrada, Jennet Toyjanova, Haneesh
  Kesari, Jonathan~S. Reichner, and Christian Franck.
\newblock Mean deformation metrics for quantifying 3d cell–matrix
  interactions without requiring information about matrix material properties.
\newblock \emph{Proceedings of the National Academy of Sciences}, 113\penalty0
  (11):\penalty0 2898--2903, 2016.
\newblock \doi{10.1073/pnas.1510935113}.
\newblock URL \url{https://www.pnas.org/doi/abs/10.1073/pnas.1510935113}.

\bibitem[Zhao et~al.(2021)Zhao, Zhang, Chen, and Lejeune]{Lejeune2021sarcgraph}
Bill Zhao, Kehan Zhang, Christopher~S. Chen, and Emma Lejeune.
\newblock Sarc-graph: Automated segmentation, tracking, and analysis of
  sarcomeres in hi{PSC}-derived cardiomyocytes.
\newblock \emph{PLOS Computational Biology}, 17\penalty0 (10):\penalty0 1--27,
  10 2021.
\newblock \doi{10.1371/journal.pcbi.1009443}.
\newblock URL \url{https://doi.org/10.1371/journal.pcbi.1009443}.

\bibitem[Holzapfel(2002)]{holzapfel2002nonlinear}
Gerhard~A Holzapfel.
\newblock Nonlinear solid mechanics: a continuum approach for engineering
  science, 2002.

\bibitem[Hunter(2007)]{hunter2007matplotlib}
John~D. Hunter.
\newblock Matplotlib: A 2d graphics environment.
\newblock \emph{Computing in Science \& Engineering}, 9\penalty0 (3):\penalty0
  90--95, 2007.
\newblock \doi{10.1109/MCSE.2007.55}.

\bibitem[Legant et~al.(2009)Legant, Pathak, Yang, Deshpande, McMeeking, and
  Chen]{legant2009microfabricated}
Wesley~R Legant, Amit Pathak, Michael~T Yang, Vikram~S Deshpande, Robert~M
  McMeeking, and Christopher~S Chen.
\newblock Microfabricated tissue gauges to measure and manipulate forces from
  3d microtissues.
\newblock \emph{Proceedings of the National Academy of Sciences}, 106\penalty0
  (25):\penalty0 10097--10102, 2009.
\newblock \doi{https://doi.org/10.1073/pnas.0900174106}.

\bibitem[Kabadi et~al.(2015)Kabadi, Vantangoli, Rodd, Leary, Madnick, Morgan,
  Kane, and Boekelheide]{kabadi2015into}
Pranita~K Kabadi, Marguerite~M Vantangoli, April~L Rodd, Elizabeth Leary,
  Samantha~J Madnick, Jeffrey~R Morgan, Agnes Kane, and Kim Boekelheide.
\newblock Into the depths: Techniques for in vitro three-dimensional
  microtissue visualization.
\newblock \emph{Biotechniques}, 59\penalty0 (5):\penalty0 279--286, 2015.
\newblock \doi{10.2144/000114353}.

\bibitem[Gilbert and Pollak(1960)]{shot1960}
E.~N. Gilbert and H.~O. Pollak.
\newblock Amplitude distribution of shot noise.
\newblock \emph{The Bell System Technical Journal}, 39\penalty0 (2):\penalty0
  333--350, 1960.
\newblock \doi{10.1002/j.1538-7305.1960.tb01603.x}.

\bibitem[Wang et~al.(2013)Wang, Svoronos, Boudou, Sakar, Schell, Morgan, Chen,
  and Shenoy]{wang2013necking}
Hailong Wang, Alexander~A Svoronos, Thomas Boudou, Mahmut~Selman Sakar,
  Jacquelyn~Youssef Schell, Jeffrey~R Morgan, Christopher~S Chen, and Vivek~B
  Shenoy.
\newblock Necking and failure of constrained 3d microtissues induced by
  cellular tension.
\newblock \emph{Proceedings of the National Academy of Sciences}, 110\penalty0
  (52):\penalty0 20923--20928, 2013.
\newblock \doi{https://doi.org/10.1073/pnas.1313662110}.

\bibitem[Guatimosim et~al.(2011)Guatimosim, Guatimosim, and
  Song]{guatimosim2011imaging}
Silvia Guatimosim, Cristina Guatimosim, and Long-Sheng Song.
\newblock Imaging calcium sparks in cardiac myocytes.
\newblock \emph{Light Microscopy: Methods and Protocols}, pages 205--214, 2011.
\newblock \doi{https://doi.org/10.1007/978-1-60761-950-5\_12}.

\bibitem[Mohammadzadeh and Lejeune(2023)]{Mohammadzadeh2023}
Saeed Mohammadzadeh and Emma Lejeune.
\newblock Sarcgraph: A python package for analyzing the contractile behavior of
  pluripotent stem cell-derived cardiomyocytes.
\newblock \emph{Journal of Open Source Software}, 8\penalty0 (85):\penalty0
  5322, 2023.
\newblock \doi{10.21105/joss.05322}.
\newblock URL \url{https://doi.org/10.21105/joss.05322}.

\bibitem[Rios et~al.(2023)Rios, Bu, Sheehan, Kobeissi, Kohli, Shah, Lejeune,
  and Raman]{rios2023mechanically}
Brandon Rios, Angel Bu, Tara Sheehan, Hiba Kobeissi, Sonika Kohli, Karina Shah,
  Emma Lejeune, and Ritu Raman.
\newblock Mechanically programming anisotropy in engineered muscle with
  actuating extracellular matrices.
\newblock \emph{Device}, 1\penalty0 (4), 2023.
\newblock \doi{https://doi.org/10.1016/j.device.2023.100097}.

\bibitem[Weiss et~al.(1996)Weiss, Maker, and Govindjee]{weiss1996finite}
Jeffrey~A Weiss, Bradley~N Maker, and Sanjay Govindjee.
\newblock Finite element implementation of incompressible, transversely
  isotropic hyperelasticity.
\newblock \emph{Computer Methods in Applied Mechanics and Engineering},
  135\penalty0 (1):\penalty0 107--128, 1996.
\newblock ISSN 0045-7825.
\newblock \doi{https://doi.org/10.1016/0045-7825(96)01035-3}.
\newblock URL
  \url{https://www.sciencedirect.com/science/article/pii/0045782596010353}.

\bibitem[Hood and Taylor(1974)]{hood1974navier}
P~Hood and C~Taylor.
\newblock Navier-stokes equations using mixed interpolation.
\newblock \emph{Finite element methods in flow problems}, pages 121--132, 1974.

\bibitem[Ambrosi and Pezzuto(2012)]{ambrosi2012active}
D~Ambrosi and Pezzuto.
\newblock Active stress vs. active strain in mechanobiology: constitutive
  issues.
\newblock \emph{Journal of Elasticity}, 107:\penalty0 199--212, 2012.
\newblock \doi{https://doi.org/10.1007/s10659-011-9351-4}.

\bibitem[Zhivomirov(2018)]{zhivomirov2018method}
Hristo Zhivomirov.
\newblock A method for colored noise generation.
\newblock \emph{Romanian journal of acoustics and vibration}, 15\penalty0
  (1):\penalty0 14--19, 2018.

\end{thebibliography}


\begin{thebibliography}{15}
\providecommand{\natexlab}[1]{#1}
\providecommand{\url}[1]{\texttt{#1}}
\expandafter\ifx\csname urlstyle\endcsname\relax
  \providecommand{\doi}[1]{doi: #1}\else
  \providecommand{\doi}{doi: \begingroup \urlstyle{rm}\Url}\fi

\bibitem[Geuzaine and Remacle(2009)]{geuzaine2009gmsh}
Christophe Geuzaine and Jean-Fran{\c{c}}ois Remacle.
\newblock Gmsh: A 3-d finite element mesh generator with built-in pre-and
  post-processing facilities.
\newblock \emph{International journal for numerical methods in engineering},
  79\penalty0 (11):\penalty0 1309--1331, 2009.
\newblock \doi{https://doi.org/10.1002/nme.2579}.

\bibitem[Javor et~al.(2021)Javor, Sundaram, Chen, and Bishop]{javor2021pillar}
Josh Javor, Subramanian Sundaram, Christopher~S. Chen, and David~J. Bishop.
\newblock A microtissue platform to simultaneously actuate and detect
  mechanical forces via non-contact magnetic approach.
\newblock \emph{Journal of Microelectromechanical Systems}, 30\penalty0
  (1):\penalty0 96--104, 2021.
\newblock \doi{10.1109/JMEMS.2020.3036978}.

\bibitem[Aln{\ae}s et~al.(2015)Aln{\ae}s, Blechta, Hake, Johansson, Kehlet,
  Logg, Richardson, Ring, Rognes, and Wells]{alnaes2015fenics}
Martin Aln{\ae}s, Jan Blechta, Johan Hake, August Johansson, Benjamin Kehlet,
  Anders Logg, Chris Richardson, Johannes Ring, Marie~E Rognes, and Garth~N
  Wells.
\newblock The {FEniCS} project version 1.5.
\newblock \emph{Archive of Numerical Software}, 3\penalty0 (100):\penalty0
  9--23, 2015.

\bibitem[Logg et~al.(2012)Logg, Mardal, and Wells]{logg2012automated}
Anders Logg, Kent-Andre Mardal, and Garth Wells.
\newblock \emph{Automated solution of differential equations by the finite
  element method: The {FEniCS} book}, volume~84.
\newblock Springer Science \& Business Media, Germany, 2012.

\bibitem[Pezzuto et~al.(2014)Pezzuto, Ambrosi, and
  Quarteroni]{pezzuto2014orthotropic}
S.~Pezzuto, D.~Ambrosi, and A.~Quarteroni.
\newblock An orthotropic active–strain model for the myocardium mechanics and
  its numerical approximation.
\newblock \emph{European Journal of Mechanics - A/Solids}, 48:\penalty0 83--96,
  2014.
\newblock ISSN 0997-7538.
\newblock \doi{https://doi.org/10.1016/j.euromechsol.2014.03.006}.
\newblock URL
  \url{https://www.sciencedirect.com/science/article/pii/S0997753814000527}.
\newblock Frontiers in Finite-Deformation Electromechanics.

\bibitem[Gurev et~al.(2015)Gurev, Pathmanathan, Fattebert, Wen, Magerlein,
  Gray, Richards, and Rice]{gurev2015high}
Viatcheslav Gurev, Pras Pathmanathan, Jean-Luc Fattebert, Hui-Fang Wen, John
  Magerlein, Richard~A Gray, David~F Richards, and J~Jeremy Rice.
\newblock A high-resolution computational model of the deforming human heart.
\newblock \emph{Biomechanics and modeling in mechanobiology}, 14:\penalty0
  829--849, 2015.
\newblock \doi{https://doi.org/10.1007/s10237-014-0639-8}.

\bibitem[Finsberg et~al.(2018)Finsberg, Xi, Tan, Zhong, Genet, Sundnes, Lee,
  and Wall]{finsberg2018efficient}
Henrik Finsberg, Ce~Xi, Ju~Le Tan, Liang Zhong, Martin Genet, Joakim Sundnes,
  Lik~Chuan Lee, and Samuel~T Wall.
\newblock Efficient estimation of personalized biventricular mechanical
  function employing gradient-based optimization.
\newblock \emph{International journal for numerical methods in biomedical
  engineering}, 34\penalty0 (7):\penalty0 e2982, 2018.

\bibitem[Holzapfel(2002)]{holzapfel2002nonlinear}
Gerhard~A Holzapfel.
\newblock Nonlinear solid mechanics: a continuum approach for engineering
  science, 2002.

\bibitem[Weiss et~al.(1996)Weiss, Maker, and Govindjee]{weiss1996finite}
Jeffrey~A Weiss, Bradley~N Maker, and Sanjay Govindjee.
\newblock Finite element implementation of incompressible, transversely
  isotropic hyperelasticity.
\newblock \emph{Computer Methods in Applied Mechanics and Engineering},
  135\penalty0 (1):\penalty0 107--128, 1996.
\newblock ISSN 0045-7825.
\newblock \doi{https://doi.org/10.1016/0045-7825(96)01035-3}.
\newblock URL
  \url{https://www.sciencedirect.com/science/article/pii/0045782596010353}.

\bibitem[Hood and Taylor(1974)]{hood1974navier}
P~Hood and C~Taylor.
\newblock Navier-stokes equations using mixed interpolation.
\newblock \emph{Finite element methods in flow problems}, pages 121--132, 1974.

\bibitem[Ambrosi and Pezzuto(2012)]{ambrosi2012active}
D~Ambrosi and Pezzuto.
\newblock Active stress vs. active strain in mechanobiology: constitutive
  issues.
\newblock \emph{Journal of Elasticity}, 107:\penalty0 199--212, 2012.
\newblock \doi{https://doi.org/10.1007/s10659-011-9351-4}.

\bibitem[Zhivomirov(2018)]{zhivomirov2018method}
Hristo Zhivomirov.
\newblock A method for colored noise generation.
\newblock \emph{Romanian journal of acoustics and vibration}, 15\penalty0
  (1):\penalty0 14--19, 2018.

\bibitem[van~der Walt et~al.(2014)van~der Walt, {S}ch\"onberger,
  {Nunez-Iglesias}, {B}oulogne, {W}arner, {Y}ager, {G}ouillart, {Y}u, and the
  scikit-image contributors]{scikit_image}
{S}t\'efan van~der Walt, {J}ohannes~{L}. {S}ch\"onberger, {J}uan
  {Nunez-Iglesias}, {F}ran\c{c}ois {B}oulogne, {J}oshua~{D}. {W}arner, {N}eil
  {Y}ager, {E}mmanuelle {G}ouillart, {T}ony {Y}u, and the scikit-image
  contributors.
\newblock scikit-image: image processing in {P}ython.
\newblock \emph{PeerJ}, 2:\penalty0 e453, 6 2014.
\newblock ISSN 2167-8359.
\newblock \doi{10.7717/peerj.453}.
\newblock URL \url{https://doi.org/10.7717/peerj.453}.

\bibitem[Perlin(1985)]{perlin1985noise}
Ken Perlin.
\newblock An image synthesizer.
\newblock \emph{Siggraph Computer Graphics}, 19\penalty0 (3):\penalty0
  287–296, jul 1985.
\newblock ISSN 0097-8930.
\newblock \doi{10.1145/325165.325247}.
\newblock URL \url{https://doi.org/10.1145/325165.325247}.

\bibitem[Jilberto et~al.(2023(in press))Jilberto, DePalma, Lo, Kobeissi,
  Lejeune, Baker, and Nordsletten]{jilberto2023computational}
J.~Jilberto, S.~J. DePalma, J.~Lo, H.~Kobeissi, E.~Lejeune, B.~M. Baker, and
  D.~Nordsletten.
\newblock A data--driven computational modeling for engineered cardiac
  microtissues.
\newblock \emph{Acta Biomaterialia}, 2023(in press).

\end{thebibliography}


\begin{thebibliography}{1}
\providecommand{\natexlab}[1]{#1}
\providecommand{\url}[1]{\texttt{#1}}
\expandafter\ifx\csname urlstyle\endcsname\relax
  \providecommand{\doi}[1]{doi: #1}\else
  \providecommand{\doi}{doi: \begingroup \urlstyle{rm}\Url}\fi

\bibitem[Kobeissi et~al.(2023)Kobeissi, Jilberto, Karakan, Gao, DePalma, Das,
  Quach, Urquia, Baker, Chen, Nordsletten, and Lejeune]{microbundle2023data}
Hiba Kobeissi, Javiera Jilberto, M~{\c{C}}a{\u{g}}atay Karakan, Xining Gao,
  Samuel~J DePalma, Shoshana~L Das, Lani Quach, Jonathan Urquia, Brendon~M
  Baker, Christopher~S Chen, David Nordsletten, and Emma Lejeune.
\newblock Microbundle {T}ime-lapse {D}ataset, 2023.

\end{thebibliography}


\begin{thebibliography}{3}
\providecommand{\natexlab}[1]{#1}
\providecommand{\url}[1]{\texttt{#1}}
\expandafter\ifx\csname urlstyle\endcsname\relax
  \providecommand{\doi}[1]{doi: #1}\else
  \providecommand{\doi}{doi: \begingroup \urlstyle{rm}\Url}\fi

\bibitem[Legant et~al.(2009)Legant, Pathak, Yang, Deshpande, McMeeking, and
  Chen]{legant2009microfabricated}
Wesley~R Legant, Amit Pathak, Michael~T Yang, Vikram~S Deshpande, Robert~M
  McMeeking, and Christopher~S Chen.
\newblock Microfabricated tissue gauges to measure and manipulate forces from
  3d microtissues.
\newblock \emph{Proceedings of the National Academy of Sciences}, 106\penalty0
  (25):\penalty0 10097--10102, 2009.
\newblock \doi{https://doi.org/10.1073/pnas.0900174106}.

\bibitem[Das et~al.(2022)Das, Sutherland, Lejeune, Eyckmans, and
  Chen]{das2022mechanical}
Shoshana~L Das, Bryan~P Sutherland, Emma Lejeune, Jeroen Eyckmans, and
  Christopher~S Chen.
\newblock Mechanical response of cardiac microtissues to acute localized
  injury.
\newblock \emph{American Journal of Physiology-Heart and Circulatory
  Physiology}, 323\penalty0 (4):\penalty0 H738--H748, 2022.
\newblock \doi{https://doi.org/10.1152/ajpheart.00305.2022}.

\bibitem[Tamargo et~al.(2021)Tamargo, Nash, Fleischer, Kim, Vila, Yeager,
  Summers, Zhao, Lock, Chavez, et~al.]{tamargo2021millipillar}
Manuel~Alejandro Tamargo, Trevor~Ray Nash, Sharon Fleischer, Youngbin Kim,
  Olaia~Fernandez Vila, Keith Yeager, Max Summers, Yimu Zhao, Roberta Lock,
  Miguel Chavez, et~al.
\newblock millipillar: a platform for the generation and real-time assessment
  of human engineered cardiac tissues.
\newblock \emph{ACS biomaterials science \& engineering}, 7\penalty0
  (11):\penalty0 5215--5229, 2021.
\newblock \doi{https://doi.org/10.1021/acsbiomaterials.1c01006}.

\end{thebibliography}
